\documentclass[useAMS,usenatbib,twocolumn]{mnras}
\usepackage{epsfig,amssymb}
\usepackage{graphicx,amsmath,multicol}
\usepackage{multirow}
\usepackage{caption}
\usepackage{subcaption}
\usepackage{placeins}
\usepackage{physics}
\usepackage[T1]{fontenc}
\usepackage{aecompl}

\usepackage[utf8]{inputenc} 
\usepackage{setspace}
\usepackage{graphicx}
\usepackage{mathtools}
\usepackage{grffile}
\usepackage{amsmath}
\usepackage{longtable}
\usepackage{graphics}
\usepackage{verbatim}
\usepackage[usenames]{color}
\newcommand\numberthis{\addtocounter{equation}{1}\tag{\theequation}}

\usepackage{float}
\usepackage{enumitem}

\title[Mergers of BHNS binaries and rates of EM counterparts]{Mergers of black hole--neutron star binaries and rates of associated electromagnetic counterparts}
\author[Bhattacharya, Kumar \& Smoot]
{Mukul Bhattacharya,$^{1}$\thanks{E-mail: mukul.b@utexas.edu (MB)} 
Pawan Kumar$^2$
and George Smoot$^{3,4,5}$\\  
$^{1}$ Department of Physics, University of Texas at Austin, Austin, TX 78712, USA\\
$^{2}$ Department of Astronomy, University of Texas at Austin, Austin, TX 78712, USA\\
$^{3}$ IAS TT \& WF Chao Professor, Institute for Advanced Study, Hong Kong University of Science and Technology,\\
Clear Water Bay, Kowloon, 999077 Hong Kong, China\\
$^{4}$ PCCP; APC, Universit\`e Paris Diderot, Universit\`e Sorbonne Paris Cit\`e, 75013, France\\
$^{5}$ BCCP; LBNL \& Physics Dept. University of California at Berkeley, CA 94720, USA}

\begin{document}

\date{Accepted . Received ; in original form }

\pagerange{\pageref{firstpage}--\pageref{lastpage}} \pubyear{2019}

\maketitle

\label{firstpage}

\begin{abstract} 
Black hole-neutron star (BHNS) binaries are amongst promising candidates for the joint detection of electromagnetic (EM) signals with gravitational waves (GWs) and are expected to be detected in the near future. Here we study the effect of the BHNS binary parameters on the merger ejecta properties and associated EM signals. 
We estimate the remnant disk and unbound ejecta masses for BH mass and spin distributions motivated from the observations of transient low-mass X-ray binaries (LMXBs) and specific NS equation of state (EoS).
The amount of r-process elements synthesised in BHNS mergers is estimated to be 
a factor of $\sim 10^{2}-10^{4}$ smaller than BNS mergers, due to the smaller dynamical ejecta and merger rates for the former.
We compute the EM luminosities and light curves for the early- and late-time emissions from the ultra-relativistic jet, sub-relativistic dynamical ejecta and wind, and the mildly-relativistic cocoon for typical ejecta parameters. We then evaluate the low-latency EM follow-up rates of the GW triggers in terms of the GW detection rate $\dot{N}_{GW}$ for current telescope sensitivities and typical BHNS binary parameters 
to find that most of the EM counterparts are detectable for high BH spin, small BH mass and stiffer NS EoS when NS disruption is significant. 
Based on the relative detection rates for given binary parameters, we find the ease of EM follow-up to be: ejecta afterglow $>$ cocoon afterglow $\gtrsim$ jet prompt $>$ ejecta macronova $>$ cocoon prompt $>$ jet afterglow $>>$ wind macronova $>>$ wind afterglow.

\end{abstract}


\section{Introduction}
\label{Intro}

In addition to being one of the most likely candidates for the progenitor of short gamma-ray bursts (sGRB; \citealt{Eichler89,Nakar07}), the coalescence of BNS and BHNS compact binaries are also considered to be promising sources of GWs detectable by the current generation of ground-based interferometers such as Advanced LIGO (aLIGO, \citealt{LIGO}), Advanced Virgo \citep{Virgo} and KAGRA \citep{KAGRA}. At present, six BBH and one BNS mergers have been detected and BHNS merger detections are anticipated in the upcoming aLIGO/Virgo runs at a rate of more than one per year.
The observations from binaries with at least one neutron star will help us constrain the EoS of dense nuclear matter in these compact objects \citep{Takami15,Clark16}. Furthermore, the detection of accompanying EM counterparts is a potent prospect \citep{MB12}, as it will improve source localisation and redshift estimate, and provide more information about the merger process as well as the properties of the compact binary. 

Many numerical-relativity simulations of compact binary mergers have been carried out in the recent past to study the merger dynamics and properties of the ejected material as well as the post-merger remnant \citep{Foucart12,Fou13,Fou14,Fou18,Kyu15,Kaw15,Kaw16}. The outcome of the BHNS merger critically depends on both the binary separation at tidal disruption of the NS ($R_{td}$) and the location of the innermost stable circular orbit (ISCO) of the BH ($R_{{\rm ISCO}}$). 
The most important parameters that determine the BHNS merger dynamics and ejecta properties are the mass ratio of the binary system \citep{Fou12,Kyu16}, the BH spin magnitude and orientation \citep{Pan11,Foucart12}, and the EoS of the NS \citep{Duez10,Kyu10}. 

If the NS is tidally disrupted, most of the in-falling matter is accreted by the BH, however, some material remains outside for longer timescales in the form of a hot accretion disk, eccentric bound orbits (tails) and/or unbound (dynamical) ejecta around the BH. A considerable fraction of the remnant accretion disk mass can also get unbound over long timescales due to winds driven by magnetic fields \citep{Kiuchi15}, viscous flows \citep{FM13} and/or neutrinos \citep{Dess09}. The ejected material from a BHNS merger can be classified into four different components: a sGRB jet, dynamical ejected mass, magnetic/viscous/neutrino-driven winds, and a cocoon from jet interacting with other ejecta - see Figure \ref{fig0} for a schematic sketch and Table \ref{Table1} for the properties of these ejecta components and their associated emission. The dynamical ejected mass is found to be much more anisotropic for BHNS mergers compared to BNS mergers with most of the matter being concentrated around the orbital plane \citep{Kyu15}.
Recent studies \citep{Bar19,CD19} have shown that joint multi-messenger analysis of the GW and EM signals, especially those arising from the sGRB jet and the dynamical ejected mass, can break degeneracies in the GW parameter space to provide better constraints on the BHNS binary parameters.

The BNS merger GW170817 is the only known astrophysical source at present with both GW and multi-wavelength EM radiation observations (for instance, \citealt{Abbott17a,Abbott17b,Mooley18a}). The steadily increasing afterglow luminosity observations \citep{Ghirlanda18,Margutti18,Mooley18a} are better explained with a jet-cocoon system whereby a highly relativistic narrow jet propagates through a environment of denser and mildly-relativistic material \citep{IN17,Lazzati18}, as opposed to a mildly relativistic isotropic fireball \citep{Haggard17} and/or top-hat jet seen off-axis \citep{Alex17}. The VLBI observations of the rapid turnover around the radio light curve peak and the very fast subsequent decline suggests that the early-time afterglow is most likely due to the emission from the wider-angle cocoon whereas the late-time emission is powered by a collimated relativistic jet with an opening angle $< 5^{\circ}$ viewed from an angle of $\sim 20^{\circ}$ \citep{Ghirlanda18,Mooley18b}.

There can be several possible EM counterparts for BHNS mergers: optical-infrared (IR) transients known as macronova driven by radioactive decay of neutron-rich dynamical ejecta on timescales of $t \sim {\rm few\ hours-day}$ \citep{LP98,Kul05,Metzger10}, synchrotron radio afterglow from ejecta components decelerated by the surrounding interstellar medium (ISM) with $t \sim {\rm week-year}$ \citep{NP11,MB12,PNR13}, relativistic sGRB jet prompt emission on short timescales $t \lesssim 1\ {\rm sec}$ \citep{Eichler89,Nakar07} and gamma/X-ray prompt emission from cocoon with $t \lesssim 1\ {\rm sec}$ \citep{Lazzati18}. Even though all four ejecta components contribute to synchrotron afterglow emission directly, different ejecta component afterglows peak at different timescales and with different intensities due to their separate masses and kinetic energies (KEs). 

In this paper, we explore the dependence of BHNS merger ejecta properties and associated EM signals on the mass ratio of the binary, the magnitude of the BH spin and the unknown NS EoS. For simplicity, we will only consider the cases where BH spin is completely aligned with the orbital angular momentum and the binary orbit is of low eccentricity. We also evaluate the detection rates for low-latency EM follow-up of GW triggers restricted to the nearby universe, determined by the aLIGO sensitivity at a given frequency.
The structure of this paper is as follows. In Section 2, we evaluate the dynamical/unbound ejecta mass, the disk mass and the dynamical ejecta velocity for a given BH mass and spin distribution and NS EoS. In Section 3, we calculate the horizon distance from aLIGO sensitivity to find the GW event detection rate using BHNS merger rates from population synthesis models. In Section 4, we discuss all the possible EM counterparts for BHNS mergers and further estimate their rates for EM follow-up of GW triggers in Section 5. Finally, we summarize our results in Section 6. 


\section{BHNS merger ejecta properties from binary parameters}
In this section, we estimate the mass of bound ejecta ($M_{ej,b}$) along with the dynamical ejecta mass ($M_{ej,dyn}$) and velocity ($v_{ej,dyn}$) for BHNS binaries with a given BH mass ($M_{BH}$) and spin ($a_{BH}$) distributions, and NS EoS. 
We consider different distributions for $M_{BH}$ and $a_{BH}$ in order to find the range of BHNS binary parameters for which the disruption of NS and subsequent mass ejection is more likely. 
The tidal disruption of NS by a BH is facilitated by a larger NS radius ($R_{NS}$), a smaller $M_{BH}$ and a larger $a_{BH}$. The BHNS merger ejecta calculations are done for a single NS mass $M_{NS}$ and therefore $R_{NS}$ is directly related to the NS EoS (as discussed in Section 2.2). For the rest of this paper, we use units in which $G = c = 1$ and $M_{BH}/M_{NS}$ is the Arnowitt-Deser-Misner (ADM) mass of the BH/NS at infinite separation.

\begin{table*}
\begin{center}
\caption{\small Properties of different ejecta components and their associated EM signals} 
\label{Table1}
\begin{tabular}{| c | c | c | c | c | c | c | c |}
\hline
\hline
\centering
\emph{Component} & \emph{Mass ($M_{\odot}$)} & \emph{Average $\Gamma$} & \emph{KE (erg)} & \emph{EM signal} & \emph{Band} & \emph{$L_{peak}$ (erg/s)} & \emph{$t_{peak}$} \\ \hline \hline
sGRB jet & $10^{-4}-10^{-3}$ & 10--30 & $10^{51}-10^{52}$ & prompt & gamma/X-ray & $10^{48}-10^{49}$ & $<$ second \\ \hline
               & & & & afterglow & optical/radio & $10^{36}-10^{38}$ & $\sim$ day-week \\ \hline
Cocoon & $10^{-5}-10^{-3}$ & 3--10 & $10^{49}-10^{51}$ & prompt & gamma/X-ray & $10^{47}-10^{48}$ & $\sim$ second-minute \\ \hline
	     & & & & afterglow & optical/radio & $10^{36}-10^{39}$ & $\sim$ week-month \\ \hline
Dynamical ejecta & $10^{-4}-10^{-1}$ & $\sim$ 1 & $10^{48}-10^{51}$ & macronova & optical/IR & $10^{40}-10^{42}$ & $\sim$ day-week \\ \hline
			   & & & & afterglow & optical/radio & $10^{35}-10^{38}$ & $\sim$ year-decade \\ \hline
Wind & $10^{-4}-10^{-3}$ & $\sim$ 1 & $10^{50}-10^{51}$ & macronova & optical/IR & $10^{39}-10^{40}$ & $\sim$ hour-day \\ \hline
			   & & & & afterglow & optical/radio & $10^{32}-10^{34}$ & $\sim$ year-decade \\ \hline
\hline
\end{tabular}
\end{center}
\end{table*}

\subsection{Bound and dynamical ejecta mass} 
The outcome of any BHNS merger is two-fold: either the NS plunges directly into the BH before it can get tidally disrupted, or the tidal forces on the NS are sufficiently strong to disrupt it before reaching the $R_{ISCO}$ of the BH. The binary parameters that essentially determine this outcome are the mass ratio of the compact objects $q = M_{BH}/M_{NS}$, the orientation and magnitude of $a_{BH}$, and $R_{NS}$ based on the EoS of the NS. The mass of the disrupted NS present outside the BH at late times $\sim 10\ {\rm ms}$ is almost entirely determined by the relative positions of $R_{td}$ and $R_{ISCO}$. While $R_{td} \sim R_{NS}(M_{BH}/M_{NS})^{1/3}$ in Newtonian theory is obtained by equating the self gravity of the NS with the tidal force from the BH, the ISCO radius is given by 
\begin{equation}
\frac{R_{ISCO}}{M_{BH}} = 3 + Z_{2} - {\rm sign}(a_{BH})\sqrt{(3-Z_{1})(3+Z_{1}+2\ Z_{2})}
\label{r_isco}
\end{equation}
where
\begin{eqnarray*}
&Z_{1} = 1 + (1-a_{BH}^{2})^{1/3}[(1+a_{BH})^{1/3} + (1-a_{BH})^{1/3}] \nonumber \\
&Z_{2} = \sqrt{3a_{BH}^{2} + Z_{1}^{2}}
\end{eqnarray*}
Here, $a_{BH} = a/M_{BH}$ is the dimensionless spin parameter of the BH. 

Using the numerical simulation results obtained from three different evolutionary codes \citep{Etienne09,Foucart11,Kyu11,Fou12,Fou13,Lovelace13,Fou14,Kyu15,Brege18}, \citet{Fou18} proposed a model to determine the combined mass $M_{ej,tot}$ of the remnant disk, the tidal tails and the unbound ejecta remaining outside the BH $t \sim$ 10 ms after the merger has occurred,
\begin{align}
\frac{M_{ej,tot}}{M_{b,NS}} &= {\rm Max}\biggl[\biggl(0.406\ q^{0.333}(1 - 2 C_{NS}) \nonumber \\ 
&- 0.139\ \frac{R_{ISCO}}{R_{NS}} + 0.255\biggr)^{1.761},0\biggr]
\label{M_ej_tot}
\end{align}
where $C_{NS} = M_{NS}/R_{NS}$ is the NS compactness parameter and $M_{b,NS}$ is the baryon mass of the NS. The error in $M_{ej,tot}$ is estimated to be $[(M_{ej,tot}/10)^{2} + (1/100)^{2}]^{1/2}$. 
The NS mass is fixed to be $M_{NS} = 1.35\ M_{\odot}$ whereas $M_{b,NS}$ depends on the EoS considered for the NS (see Section 2.2). \citet{Kaw16} used the numerical relativity simulation results of the Kyoto group \citep{Kaw15,Kyu15} to extend the formalism developed by \citet{Foucart12}, and determine the mass and average velocity of the dynamical ejecta from BHNS mergers with the least-squares fit 
\begin{align}
&\frac{M_{ej,dyn}}{M_{b,NS}} = {\rm Max}\biggl[4.464\times10^{-2}q^{0.250}\left(\frac{1 - 2 C_{NS}}{C_{NS}}\right) - 2.269\nonumber \\ 
& \times10^{-3}q^{1.352}\frac{R_{ISCO}}{M_{BH}} + 2.431\left(1 - \frac{M_{NS}}{M_{b,NS}}\right) -0.4159,0\biggr] \label{M_ej_dyn} \\
&v_{avg,dyn} = (1.533\times10^{-2}\ q + 1.907\times10^{-1})\ c 
\end{align}

Here, $M_{ej,dyn}$ is also estimated $\sim 10$ ms after the merger and does not include the possible ejecta component from the remnant BH accretion disk. 
Similar to $M_{ej,tot}$, the error in $M_{ej,dyn}$ is estimated as $[(M_{ej,dyn}/10)^{2} + (M_{\odot}/200)^{2}]^{1/2}$.
The mass of the bound ejecta component (in the remnant disk and tidal tails) is calculated as, $M_{ej,b} = M_{ej,tot} - M_{ej,dyn}$. In general, the majority of the bound ejecta mass is contained in the accretion disk ($M_{disk} \approx M_{ej,b}$). It should also be noted that these estimates are limited to low-eccentricity BHNS binary orbits for BH spins aligned with the orbital angular momentum and in the range of parameters: $M_{BH} = 1-7\ M_{NS}$, $a_{BH} = -0.5-0.97$, and $C_{NS} = 0.13-0.18$ ($R_{NS} = 11-16\ {\rm km}$). 

From equations \ref{M_ej_tot} and \ref{M_ej_dyn}, the ejecta masses $M_{ej,tot}$ and $M_{ej,dyn}$ are both larger for a smaller $M_{BH}$, a larger $a_{BH}$ and a stiffer NS EoS (smaller $C_{NS}$), which is expected from the physics of tidal disruption. While a smaller $M_{BH}$ means a larger tidal force from the BH on the infalling NS between $R_{td}$ and $R_{ISCO}$ resulting in a larger NS disruption, the increase in $a_{BH}$ reduces $R_{ISCO}$ - thereby causing a larger amount of NS matter to be stripped. For a stiffer NS EoS, the increase in NS tidal deformability is expected to give off larger ejecta material during its disruption. 

The ratio of the r-process elements synthesised from BHNS mergers to BNS mergers is approximately the product of the relative ratios of the dynamical ejecta produced and the merger rate. While $M_{ej,dyn} \sim 10^{-4}-10^{-3}\ M_{\odot}$ for BHNS mergers \citep{Kaw16} is about an order of magnitude smaller than that for typical BNS merger events such as GW170817 \citep{Abbott17c}, the relative merger rate $R_{BHNS}/R_{BNS}$ is a fairly uncertain quantity and ranges from $\sim 10^{-3}$ to $\sim 10^{-1}$ with a most probable value of $\sim 10^{-2}$ \citep{Abadie10}. Therefore, BHNS mergers are expected to synthesise only about 0.01-1\% of the r-process elements found in the universe.

\begin{figure*}
    \begin{subfigure}[tp]{0.95\linewidth}
    \centering
    \includegraphics[height=0.9\linewidth,width=0.97\linewidth]{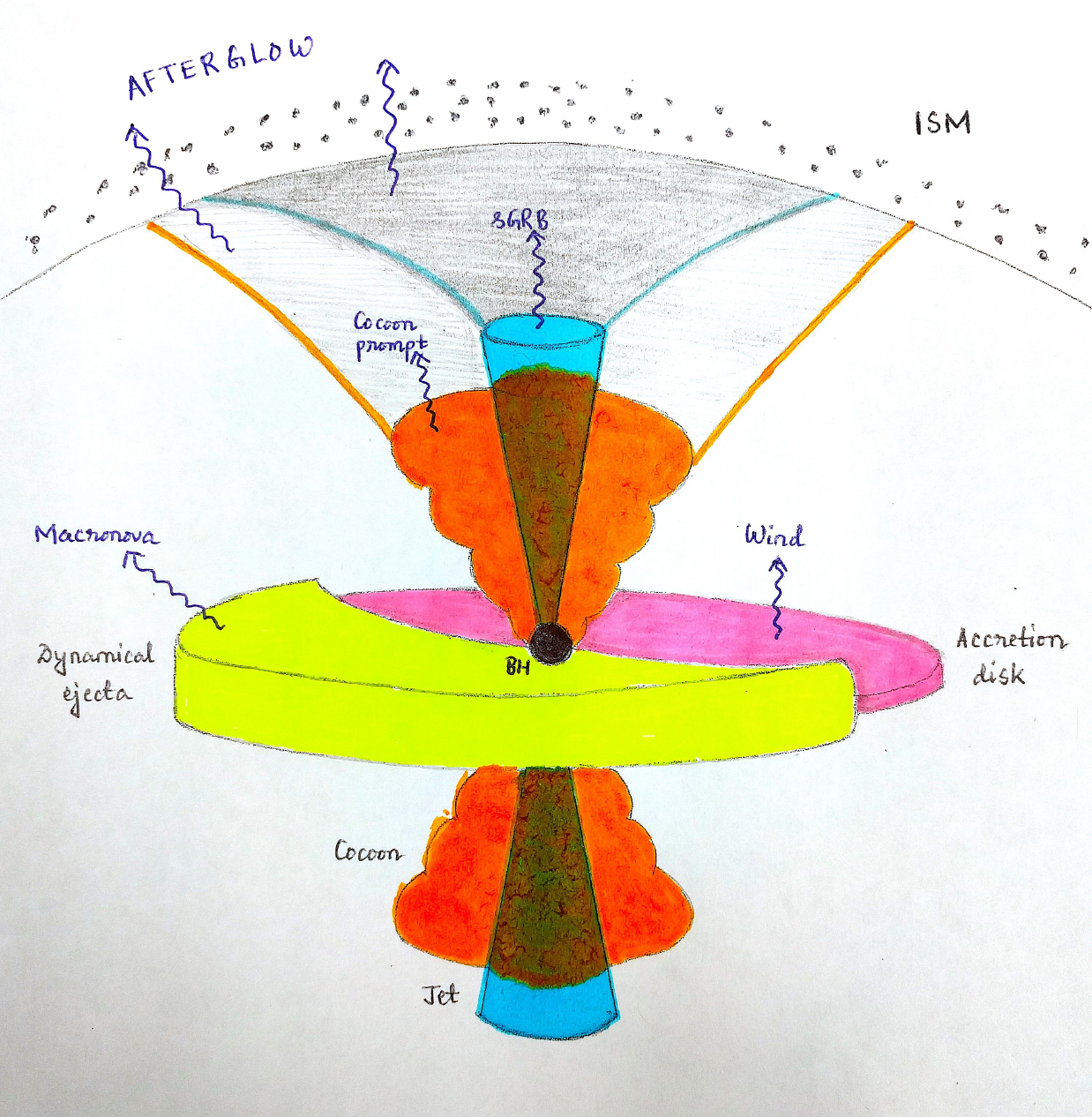} 
  \end{subfigure}
  \caption{A schematic diagram for the various ejecta components from BHNS mergers: \emph{ultra-relativistic} sGRB jet, \emph{mildly relativistic} cocoon, and \emph{sub-relativistic} dynamical ejecta and magnetic/viscous/neutrino-driven winds. The early time emissions are (optical/IR) macronova, (gamma/X-ray) sGRB jet prompt and (gamma/X-ray) cocoon prompt while the interaction of the ejecta components with the ambient ISM results in the late time radio and optical afterglows from the dynamical ejecta, jet, wind and cocoon.
 }
  \label{fig0} 
\end{figure*}

\subsection{Distributions for BHNS binary parameters}
Here, we consider different sets of binary parameters in order to estimate the ejecta properties for typical BHNS mergers. 
We perform Monte Carlo simulations to estimate the ejecta masses $M_{ej,tot}$ and $M_{ej,b} \approx M_{disk}$ - see Appendix \ref{dist_sims} for the results. We consider BHNS mergers with a given NS EoS as well as specific $M_{BH}$ and $a_{BH}$ distributions as described below:

\begin{itemize}

\item {\it BH mass distribution:} We consider gaussian, power-law and exponential $M_{BH}$ distributions inferred from transient LMXB observations. 
\citet{Ozel10} compiled the dynamical BH mass measurements in transient LMXBs \citep{MR06,RM06} to find that most of the BH masses were clustered within $\sim 6-10\ M_{\odot}$ whereas there were no observed BHs with $M_{BH} \sim 2-5\ M_{\odot}$. The narrow mass distribution of the 16 observed BHs used in their sample was found to be consistent with a Gaussian at $(7.8\pm1.2)\ M_{\odot}$. 
\citet{Farr11} then obtained a best fit power-law distribution, $P(M_{BH}) \propto M_{BH}^{-6.4}$ for $6.1\ M_{\odot} < M_{BH} < 23\ M_{\odot}$, with the same data as \citet{Ozel10}. The lower-end cut-off of this distribution accounts for the paucity of BHs within the mass range $\sim 2-5\ M_{\odot}$.
The BH masses in transient LMXBs have also been modelled using an exponentially decaying distribution with a sharp cut-off \citep{Ozel10,Steiner12}, $P(M_{BH}) = M_{scale}^{-1}\ {\rm exp}[-(M_{BH} - M_c)/M_{scale}]$ for $M_{BH} > M_c$, where $M_{c} = 6.30\ M_{\odot}$ and $M_{scale} = 1.57\ M_{\odot}$. This parametric mass distribution is motivated by the mass distributions of pre-SNe stars as well as the energetics of SNe explosions.
While we also consider $M_{BH}$ distributions with peak around $10-25\ M_{\odot}$ for low metallicity simulated BHNS systems \citep{Kruckow18}, such large BH masses do not produce almost any post-merger ejecta material (both bound and dynamical ejecta from equations \ref{M_ej_tot} and \ref{M_ej_dyn}) for the associated EM counterparts to be detected.\\




\item {\it BH spin distribution:} 
The magnitude of the BH spin in a BHNS binary can be modified by mass transfer via accretion from its NS companion either through an accretion disk or a common-envelope phase \citep{Shaug08}. A significant fraction of the intrinsic BH spin is expected to be attained during its formation via the collapse of the progenitor star. At later times, the BH can further spin-up due to repeated accretion episodes. At present, we do not have any direct observations of the BH spin which makes it very difficult to confidently constrain the BH spins in merging BHNS systems. Estimates of $a_{BH}$ from quasi-periodic oscillations and/or iron line profiles are largely dependent on how the disk accretion is modelled for XRBs \citep{Rey99,Miller04,Narayan05,OS05,Tor05} and can vary within a large range based on the assumptions used. Due to these uncertainties in the determination of $a_{BH}$, here we only consider two simplistic distributions for the BH spin: (a) $a_{BH} = {\rm Uniform}(0,0.97)$, and (b) $a_{BH} = A$, where $0 \leq A \leq 0.97$ is a constant. Note that the upper limit of $a_{BH}=0.97$ is assumed as the ejecta mass estimate fits are only valid up to that spin magnitude.\\

\item {\it NS equation of state:} Along with the mass and spin magnitude of the BH, the size of the NS specified by its EoS plays an important role in determining the amount of matter ejected in any BHNS merger event. The disruption of NS by BH results in larger ejected mass for a stiffer NS EoS i.e. larger NS radius. The NS parameters $M_{b,NS}$, $R_{NS}$ and $C_{NS} = M_{NS}/R_{NS}$ are closely associated with the NS EoS adopted. In this work, we consider two piecewise polytropic EoSs \citep{Read09} that are commonly used in the numerical-relativity simulations.

\begin{enumerate}
\item \emph{APR4 EoS model:} Soft NS EoS with $R_{NS} = 11.1\ {\rm km}$, $M_{b,NS} = 1.50\ M_{\odot}$ and $C_{NS} = M_{NS}/R_{NS} = 0.180$ for $M_{NS} = 1.35\ M_{\odot}$.

\item \emph{H4 EoS model:} Stiff NS EoS with $R_{NS} = 13.6\ {\rm km}$, $M_{b,NS} = 1.47\ M_{\odot}$ and $C_{NS} = M_{NS}/R_{NS} = 0.147$ for $M_{NS} = 1.35\ M_{\odot}$.

\end{enumerate} 

\end{itemize}

\section{GW event detection rate from LIGO sensitivity}
In this section, we first calculate the detection horizon distance ($D_{hor}$) for GW observations based on the aLIGO strain sensitivity curve and signal-to-noise ratio (SNR). We further evaluate the event detection rate of BHNS mergers within the observable volume using the plausible estimates for the BHNS coalescence rates from population synthesis models. The design strain sensitivity of the initial LIGO detectors were improved by a factor of $\sim 10$ for aLIGO in the entire operating frequency range $\sim 10\ {\rm Hz}-10\ {\rm kHz}$ \citep{LIGO}, thereby increasing the observable volume of the universe significantly by $\sim 1000$ times for the first observing run (O1) in September 2015. The source orientation- and sky location-averaged range for observing a binary coalescence event with aLIGO design sensitivity and at a SNR of 8 will be: $\sim 300\ {\rm Mpc}$ for  $\sim 1.4\ M_{\odot}$ BNS inspirals, $\sim 650\ {\rm Mpc}$ for  $\sim 1.4\ M_{\odot}-8\ M_{\odot}$ BHNS inspirals and $\sim 1.6\ {\rm Gpc}$ for  $\sim 10\ M_{\odot}$ BBH inspirals. 

Here we consider a well-separated BHNS binary with point mass constituents moving in a circular orbit to estimate the GW strain amplitude at a distance $r$ from the source. Even though the orbit of the BHNS system is elliptical initially, it can be shown that it gets quickly circularised due to GW emission \citep{Peters64}. The lowest order quadrupolar contribution gives the GW strain amplitude \citep{PY14}
\begin{align}
h_{GW} &= \left(\frac{32}{5}\right)^{1/2} \frac{(GM_{ch})^{5/3}}{rc^4} (\pi f_{GW})^{2/3} \nonumber \\
&= (7.56\times10^{-23}) M_{tot,M_{\odot}}^{2/3} \mu_{red,M_{\odot}} f_{GW,{\rm Hz}}^{2/3} r_{{\rm Mpc}}^{-1}
\label{hGW}
\end{align}
where, $M_{ch} = \mu_{red}^{3/5}M_{tot}^{2/5}$ is the chirp mass, $\mu_{red}=\mu_{red,M_{\odot}}\times M_{\odot}$ is the reduced mass, $M_{tot}=M_{tot,M_{\odot}}\times M_{\odot}$ is the total mass of the binary, $f_{GW}= f_{GW,{\rm Hz}}\times {1\ {\rm Hz}} = (1/\pi)\sqrt{GM_{tot}/a^{3}}$ is the GW frequency that is twice of the orbital frequency, 
$a$ is the orbital separation and $r_{{\rm Mpc}} = r/(1\: {\rm Mpc})$. The GW frequency is maximum, $f_{GW,max} \sim f_{ISCO} = (1/\pi)\sqrt{GM_{tot}/R_{ISCO}^{3}}$, at $a=R_{ISCO}$. 
It should be noted that we have excluded geometrical factors of order unity in the expression of $h_{GW}$ and equation (\ref{hGW}) corresponds to the case where the observer line-of-sight is along the orbital angular momentum axis of the binary.

We use the aLIGO design sensitivity curve \citep{LIGO}, to estimate the GW strain amplitude spectral density (ASD) as a function of detection frequency. As aLIGO is expected to reach its design sensitivity (lowest achievable noise level with current detector capabilities) by 2019, we fit the design strain ASD in the entire detector frequency range $f \sim 10\ {\rm Hz} - 10\ {\rm kHz}$ to obtain
\begin{eqnarray}
\sqrt{S_{n}(f)} = 2.6\times10^{-23} f^{-0.44} + 1.3\times10^{-19} f^{-3.06}\nonumber \\ 
+\ 3.8\times10^{-27} f
\label{ASD}
\end{eqnarray}
which is better by a factor of $\sim 3-4$ in the most sensitive frequency band $\sim 100-300\ {\rm Hz}$ compared to the aLIGO sensitivity during the detection of GW170814. In equation (\ref{ASD}), $S_{n}(f)$ is the strain power spectral density (PSD) of the interferometer. The noise level of the interferometer is mainly determined by the thermal noise and quantum noise.  

The SNR $\rho$ for a matched-filter search is determined by the strain PSD $S_{n}(f)$, Fourier transform of the GW strain amplitude $\tilde{h}(f)$ and $f_{ISCO}$,
\begin{eqnarray}
\rho = \sqrt{4Z(i)\int_0^{f_{isco}}\frac{|\tilde{h}(f)|^{2}}{S_{n}(f)}df}
\label{snr}
\end{eqnarray}
where $Z(i) = (1 + 6{\rm cos}^{2}i + {\rm cos}^{4}i)/8$ accounts for the variation of the radiation power emitted for a binary inclination angle $i$ \citep{SS09,Schutz11}.
The Fourier transform of the GW strain amplitude is \citep{Abadie10}
\begin{eqnarray}
|\tilde{h}(f)| = \frac{2c}{D} \left(\frac{5G\mu_{red}}{96c^3}\right)^{1/2}\left(\frac{GM_{tot}}{\pi^2 c^3}\right)^{1/3} f^{-7/6}
\label{htilde}
\end{eqnarray}
with $D$ being the luminosity distance to the source. The detection horizon distance is the maximum distance at which a GW source can be detected with SNR $\rho$,
\begin{equation}
D_{hor} = \sqrt{\frac{5}{6}}\frac{(G M_{ch})^{5/6}}{c^{3/2}\pi^{2/3}}\frac{1}{\rho} \sqrt{Z(i)\int_0^{f_{isco}}\frac{1}{S_{n}(f)}f^{-7/3}df}
\label{Dhor}
\end{equation}
Substituting $\rho=8$ and $S_{n}(f)$ from equation (\ref{ASD}) gives
\begin{equation}
D_{hor} = \frac{1.214\ {\rm Mpc}}{8}\left(\frac{M_{ch}}{M_{\odot}}\right)^{5/6}\sqrt{Z(i)\int_0^{f_{ISCO}} \frac{f^{-7/3}}{\overline{[S_n}(f)]^{2}}df}
\end{equation}
where $\overline{S_n}(f) = 1.986\times10^{-4} f^{-0.436} + f^{-3.058} + 2.942\times10^{-8} f$. 
As most of the radiation power is concentrated around the binary rotation axis ($i=0$), the GW detectors preferentially select more face-on binaries with smaller $D_{hor}$ \citep{Nissanke10}. The expression for $D_{hor}$ is derived assuming that the GW source is at a small redshift $z$.

\begin{figure*}
    \begin{subfigure}[tp]{0.49\linewidth}
    \centering
    \includegraphics[height=0.7\linewidth,width=0.9\linewidth]{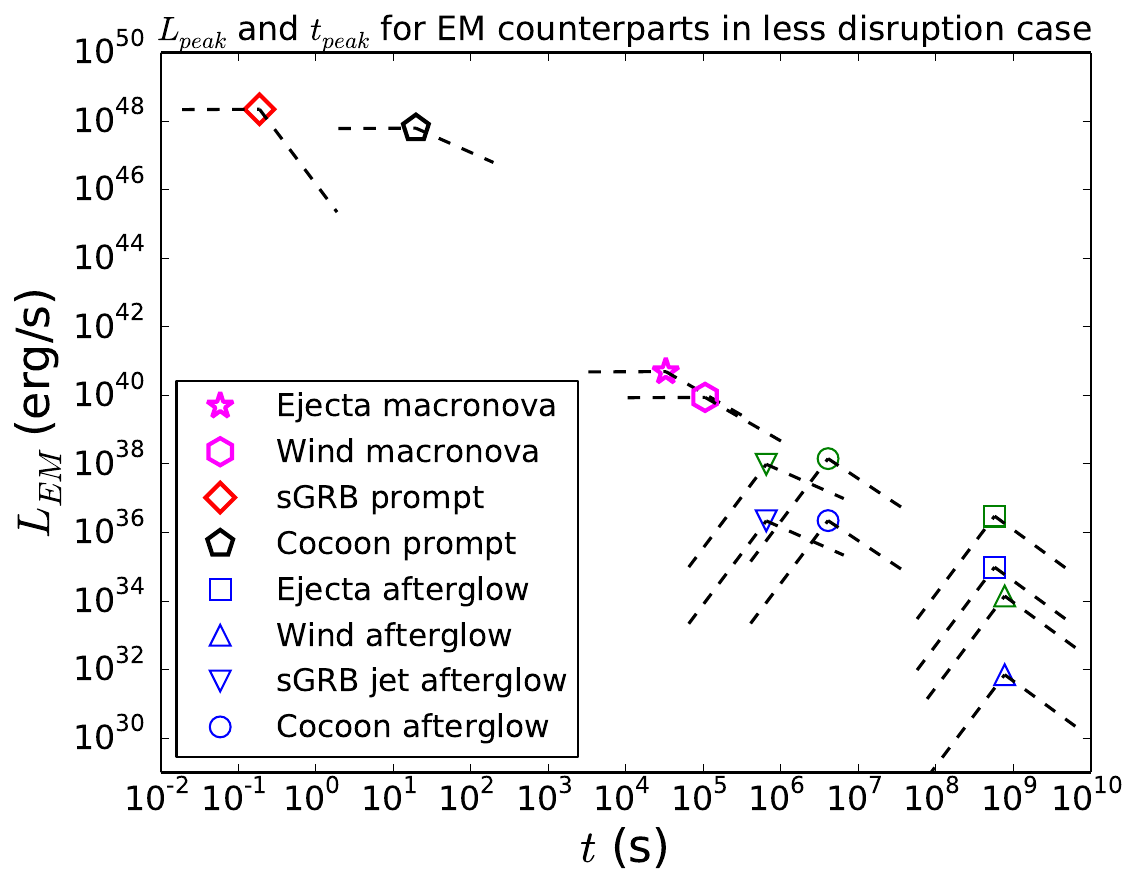} 
  \end{subfigure}
  \begin{subfigure}[tp]{0.49\linewidth}
    \centering
    \includegraphics[height=0.7\linewidth,width=0.9\linewidth]{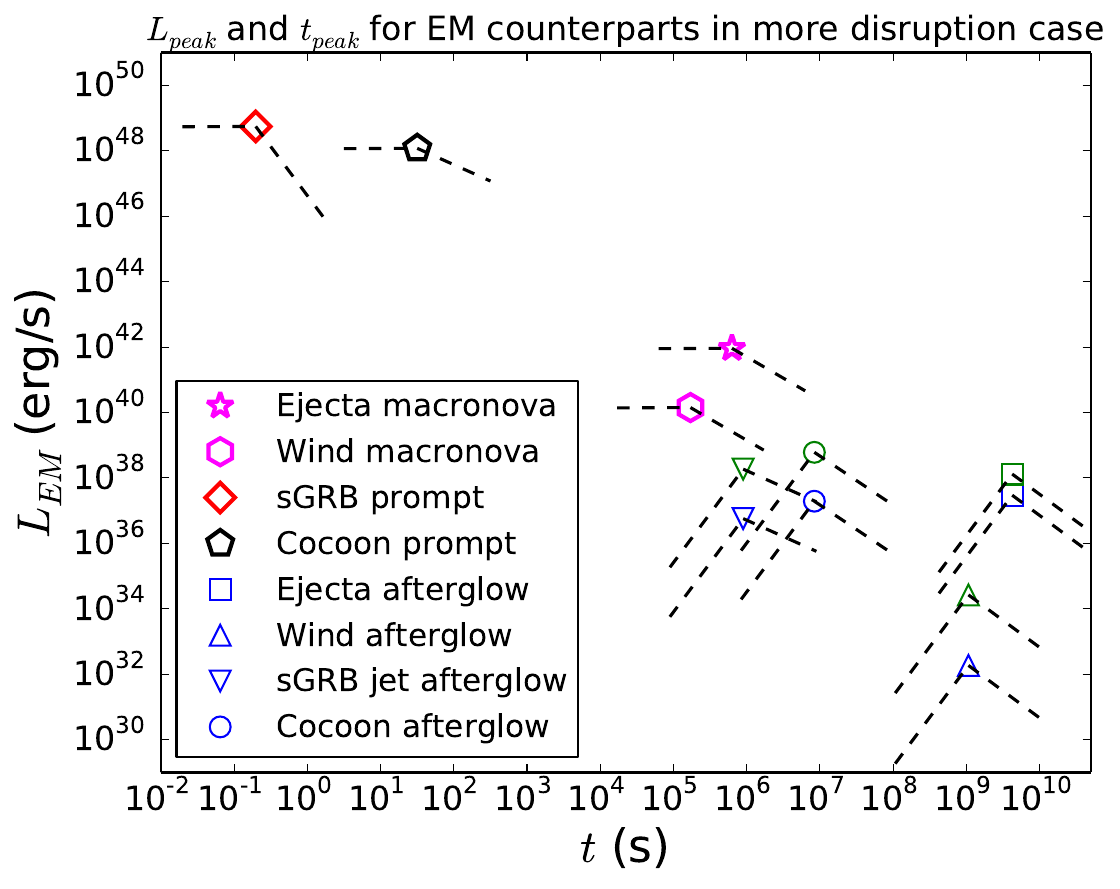}  
  \end{subfigure} \\
  \caption{\emph{Effect of BHNS binary parameters on the EM counterpart $L_{peak}$ and $t_{peak}$ values:}
Simulation results for two BHNS binaries with fixed $M_{BH}$, $a_{BH}$ and NS EoS distributions.
  	{\it Left panel:} Less NS disruption for $R_{NS} = 11.1\ {\rm km}$ (APR4 NS EoS), $M_{BH} = 8.2\ M_{\odot}$ and $a_{BH} = 0.8$,	
	{\it Right panel:} More NS disruption for $R_{NS} = 13.6\ {\rm km}$ (H4 NS EoS), $M_{BH} = 7.4\ M_{\odot}$ and $a_{BH} = 0.97$. 
	Blue and green symbols in both panels denote radio and optical afterglow components, respectively.
	The black dashed lines show the power-law fits for the canonical temporal evolution of the luminosity for each counterpart and the symbols represent their peak quantities.
	The representative bolometric band sensitivities are as listed in Table \ref{Table2}: gamma/X-ray $\nu_{obs} = 10^{20}\ {\rm Hz}$, optical/IR $\nu_{obs} = 5\times10^{14}\ {\rm Hz}$ and radio $\nu_{obs} = 10^{9}\ {\rm Hz}$.
 }
  \label{fig4} 
\end{figure*}

The detection rate $\dot{N}_{GW}$ for a binary system is given by the product of the binary coalescence rate $R_{V}$ and the detector's horizon volume $V_{hor}$, $\dot{N}_{GW} = R_{V}\times V_{hor} = R_{V}\times(4/3)\pi D_{hor}^{3}$. While the upper limit for BNS coalescence rates is determined by the luminosity distribution of the observed binary pulsars in the Milky Way \citep{Kalo04}, it is much more difficult to estimate the BHNS/BBH coalescence rates due to the paucity of observations. The BHNS coalescence rates are based on relatively poorly constrained population synthesis models. The plausible BHNS coalescence rates range from $R_{V,low} \sim 5.80\times10^{-4}\ {\rm Mpc^{-3}\ Myr^{-1}}$ to $R_{V,high} \sim 1.16\ {\rm Mpc^{-3}\ Myr^{-1}}$ with a most probable estimate of $R_{V,mp} \sim 3.48\times10^{-2}\ {\rm Mpc^{-3}\ Myr^{-1}}$ \citep{Shaug08,Abadie10}. For large distances $D_{hor} \gtrsim 30\ {\rm Mpc}$, the GW detection rate from BHNS sources can then be written as \citep{Abadie10}
\begin{align}
\dot{N}_{GW}(R_{V}) &= 
\left(0.0348\ {\rm Myr^{-1}}\right) \frac{4}{3}\pi D_{hor}^{3}R_{V,0} (2.26)^{-3}\nonumber \\
&= \left(0.0126\ {\rm year^{-1}}\right)R_{V,0}  D_{hor,2}^{3} 
\label{NGWRv}
\end{align}
where, $R_{V,0} = R_{V}/R_{V,mp}$, $D_{hor,2} = D_{hor}/(100\: {\rm Mpc})$ and the $(2.26)^{-3}$ factor is included to account for the average over all source orientations and sky locations \citep{FC93}. For typical BHNS binaries containing $1.4\ M_{\odot}$ NS and $7.8\ M_{\odot}$ BH with source orientation averaged $D_{hor} \sim 650\ {\rm Mpc}$, the most likely GW detection rate with aLIGO design sensitivity is $\dot{N}_{GW}(R_{V,mp}) \sim 3.46\ {\rm year^{-1}}$. Equation (\ref{NGWRv}) is obtained for a Milky Way type galaxy and by assuming that $R_V$ is independent of the redshift for distances $D \lesssim D_{hor} \sim 650\ {\rm Mpc}$.
Depending on the value of $R_{V}$, $\dot{N}_{GW}$ can vary by almost three orders of magnitude from $\dot{N}_{GW,low} \sim 5.77\times10^{-2}\ {\rm year^{-1}}$ to $\dot{N}_{GW,high} \sim 1.16\times10^{2}\ {\rm year^{-1}}$.

\section{EM signals associated with BHNS mergers}
In Section 2, we estimated the bound disk mass $M_{disk}$ and the dynamical ejecta mass (velocity) $M_{ej,dyn}\ (v_{ej,dyn})$ for BHNS coalescence events with a given set of binary parameters. While most of the disrupted NS mass remaining outside BH at longer timescales is contained in the remnant accretion disk, a significant fraction is present in the ejecta surrounding the remnant BH. The different ejecta components can be distinguished based on their masses and KEs: an \emph{ultra-relativistic} sGRB jet, \emph{mildly relativistic} cocoon, and \emph{sub-relativistic} dynamical ejecta and magnetic/viscous/neutrino-driven winds. These ejecta components are also directly associated with the EM counterparts of the BHNS merger events. Figure \ref{fig0} shows the approximate geometrical distribution of the ejecta components and their associated early- and late-time EM emission in various observing bands. In the following, we briefly discuss the major possible EM counterparts accompanying GW signals from BHNS mergers.

\subsection{Macronova}
Macronova are thermal transients powered by the radioactive decay of unstable r-process elements via beta-decay and fission \citep{Metzger10,KR10}. This radiation is emitted by the adiabatically expanding ejecta on timescales of $\sim {\rm few\ days-month}$ and is typically peaked in the optical/IR bands \citep{Metzger10b,Roberts11}. The photons escape the bulk and contribute significantly to the EM luminosity only when the density decreases sufficiently such that the photon diffusion timescale matches the expansion timescale. 
Both the dynamical ejecta associated with the tidal tails and the magnetic/viscous/neutrino-driven winds from the remnant accretion disk contribute to the macronova light curve \citep{FM13,Kasen15,Fer17}.

The dynamical ejecta from BHNS mergers is expected to be considerably anisotropic from numerical relativity simulations and is characterised by the opening angle in the equatorial (meridional) plane $\phi_{ej,dyn}$ ($\theta_{ej,dyn}$). The dynamical ejecta sweeps out about half of the equatorial plane with $\phi_{ej,dyn} \approx \pi$ and is mainly concentrated around the equatorial plane with $\theta_{ej,dyn} \approx 10^{\circ}-20^{\circ}$ \citep{Kyu15}.
\citet{Kyu13} approximated the anisotropic ejecta geometry using an axisymmetric cylinder with radial/z-direction velocity $v_{||}$/$v_{\perp}$. The peak time for the emission is determined by photons escaping the ejecta along the z-direction due to shorter distance
\begin{align*}
t_{peak} &= (4\ {\rm day})\ \kappa_{1}^{1/2} M_{dyn,0.03}^{1/2} v_{dyn,0.3}^{-1/2} \theta_{dyn,0} ^{1/2}\phi_{dyn,0}^{-1/2}
\numberthis 
\label{tpeakMac}
\end{align*}
where $\kappa = \kappa_{1}\times 10\ {\rm cm^{2}\ g^{-1}}$ is the opacity with $M_{dyn,0.03} = M_{ej,dyn}/0.03 M_{\odot}$, $v_{dyn,0.3} = v_{ej,dyn}/0.3 c$, $\theta_{dyn,0} = \theta_{ej,dyn}/0.2$ and $\phi_{dyn,0} = \phi_{ej,dyn}/\pi$ being short-hand notations for ejecta parameters. The peak luminosity is estimated using the radioactive heating rate, 
\begin{align*}
&L_{peak} \approx \frac{f M_{ej,dyn} c^2}{t_{peak}} \nonumber \\
&= (1.6\times10^{41}\ {\rm erg/s})\ f_{-6} \kappa_1^{-1/2} M_{dyn,0.03}^{1/2} v_{dyn,0.3}^{1/2} \theta_{dyn,0}^{-1/2} \phi_{dyn,0}^{1/2} \numberthis
\label{LpeakMac}
\end{align*}
where $f = f_{-6}\times10^{-6}$ is the fractional ejecta energy radiated in time $t_{peak}$.

\begin{figure*}
    \begin{subfigure}[tp]{0.49\linewidth}
    \centering
    \includegraphics[height=0.7\linewidth,width=0.9\linewidth]{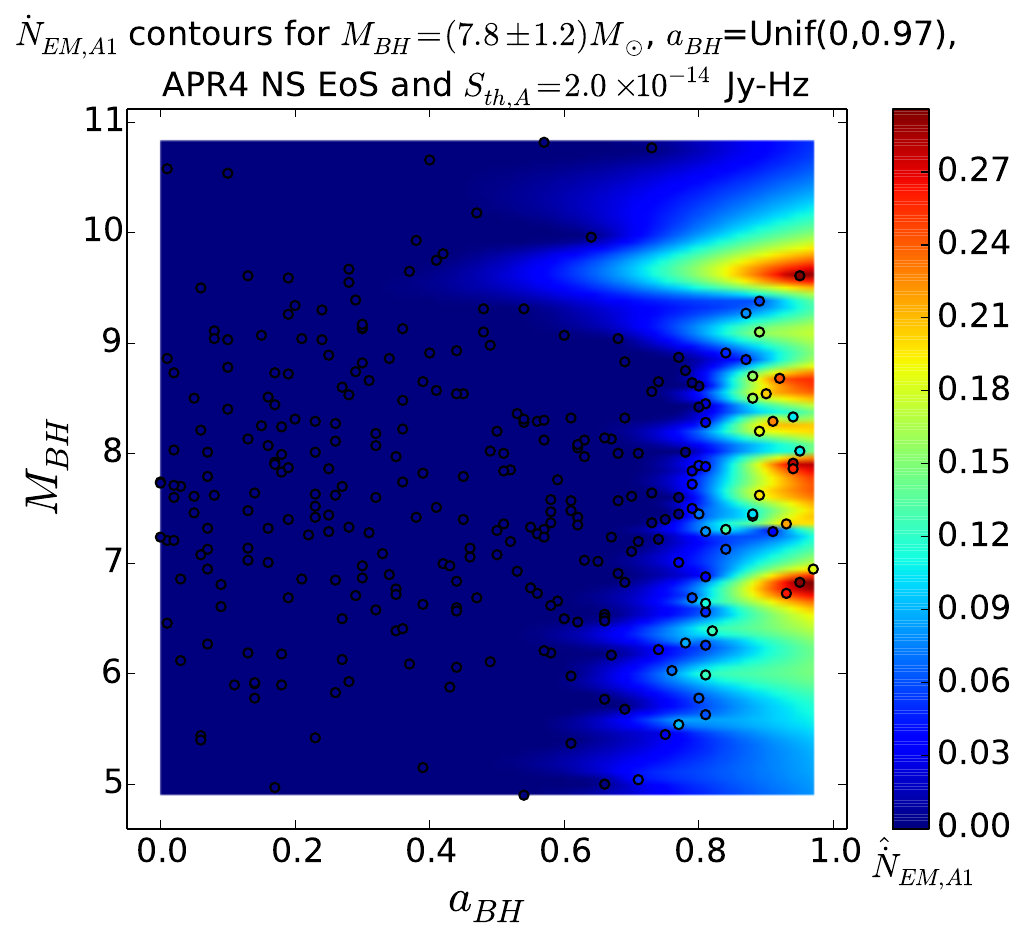} 
  \end{subfigure}
  \begin{subfigure}[tp]{0.49\linewidth}
    \centering
    \includegraphics[height=0.7\linewidth,width=0.9\linewidth]{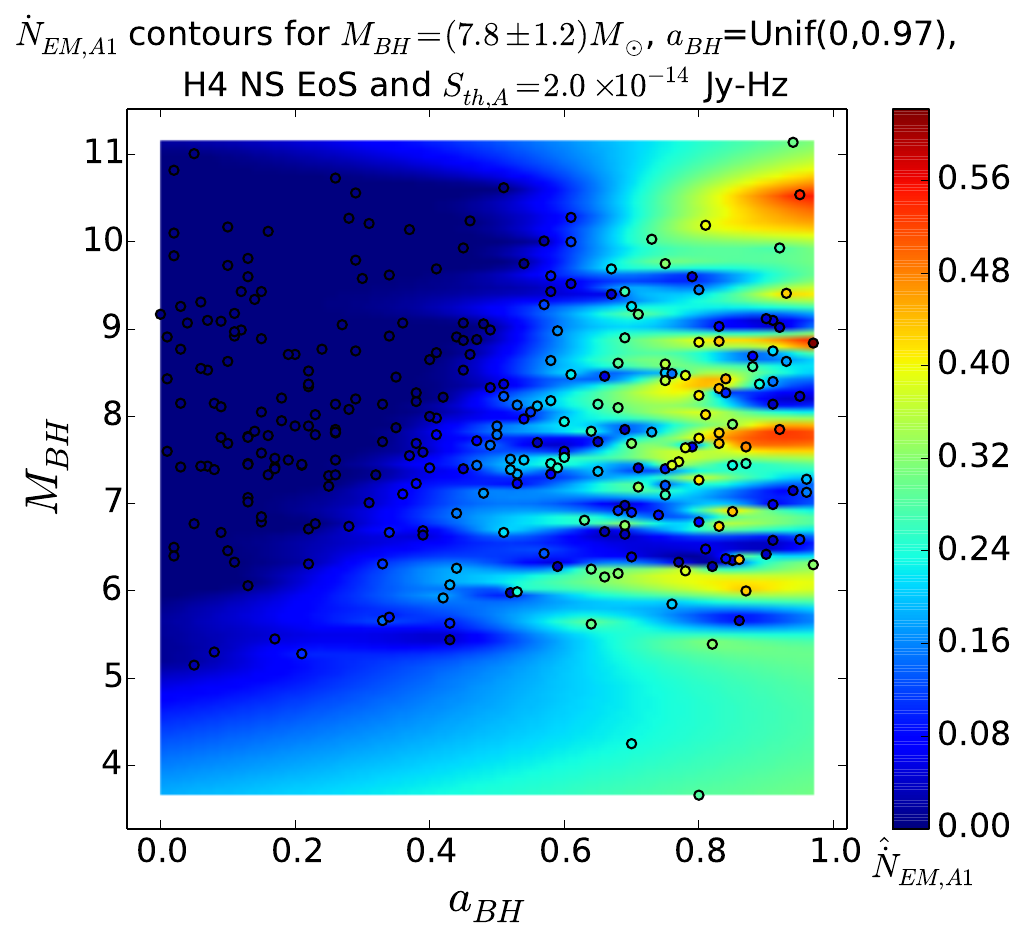}  
  \end{subfigure}\\ 
  \begin{subfigure}[tp]{0.49\linewidth}
    \centering
    \includegraphics[height=0.7\linewidth,width=0.9\linewidth]{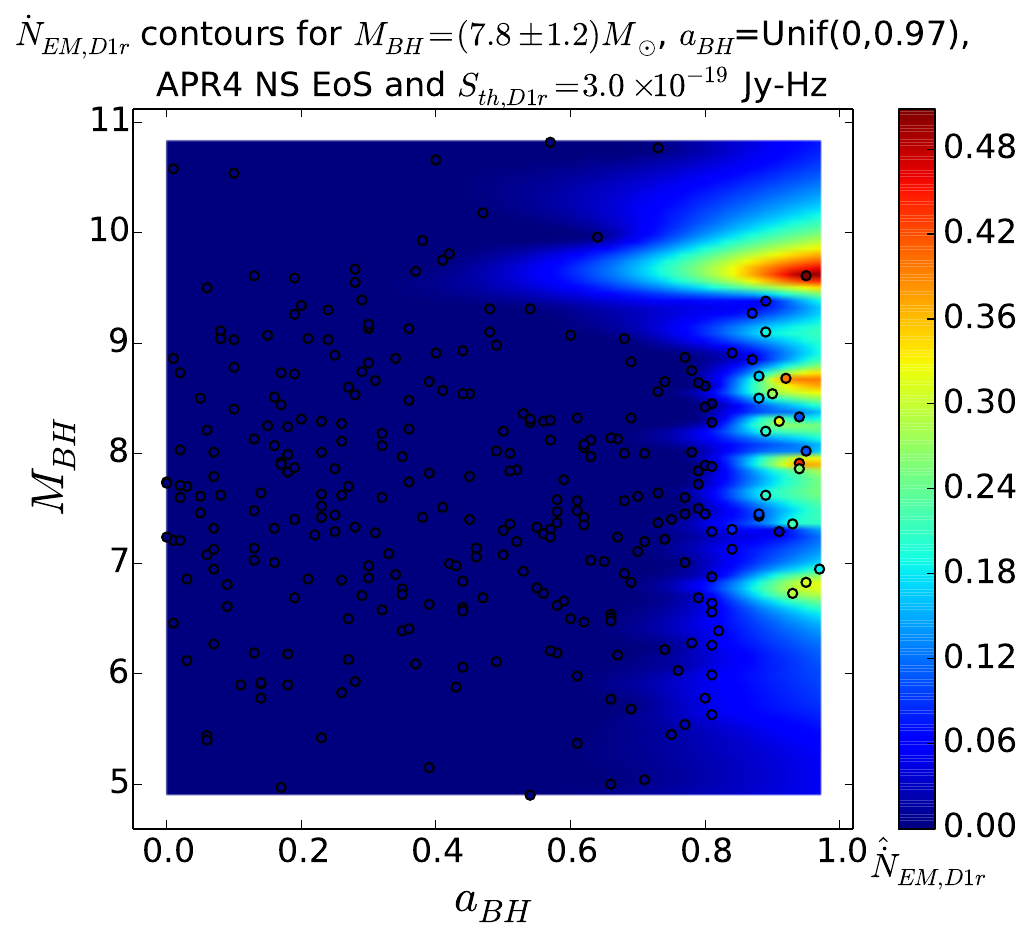} 
  \end{subfigure}
  \begin{subfigure}[tp]{0.49\linewidth}
    \centering
    \includegraphics[height=0.7\linewidth,width=0.9\linewidth]{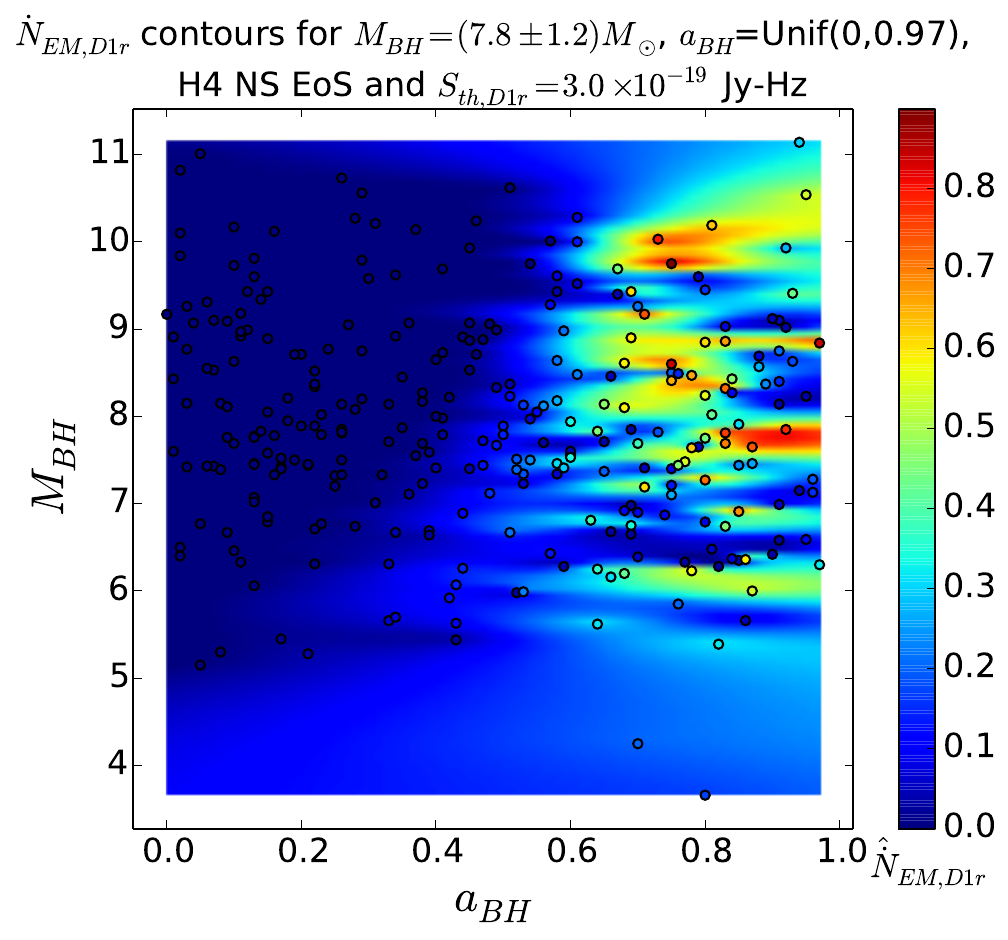}  
  \end{subfigure} 
  \caption{\emph{Effect of binary parameters on the $\dot{N}_{EM}$ values for early-time macronova and late-time radio afterglow emission from anisotropic sub-relativistic dynamical ejecta:} 
Contour plots using simulation results for 300 BHNS binaries with $M_{BH} = (7.8\pm1.2)\ M_{\odot}$, $a_{BH} = {\rm Uniform(0,0.97)}$ and APR4/H4 NS EoS. For the $i$-th EM counterpart, $\hat{\dot{N}}_{EM,i}$ denotes $\dot{N}_{EM,i}/{\rm max}(\dot{N}_{GW})$.
  	{\it Top-left panel:} $\dot{N}_{A1}$ contour plot for APR4 NS EoS,	
	{\it Top-right panel:} $\dot{N}_{A1}$ contour plot for H4 NS EoS,
	{\it Bottom-left panel:} $\dot{N}_{D1r}$ contour plot for APR4 NS EoS,
	{\it Bottom-right panel:} $\dot{N}_{D1r}$ contour plot for H4 NS EoS.
 }
  \label{fig6} 
\end{figure*}

\begin{figure*}
    \begin{subfigure}[tp]{0.49\linewidth}
    \centering
    \includegraphics[height=0.7\linewidth,width=0.9\linewidth]{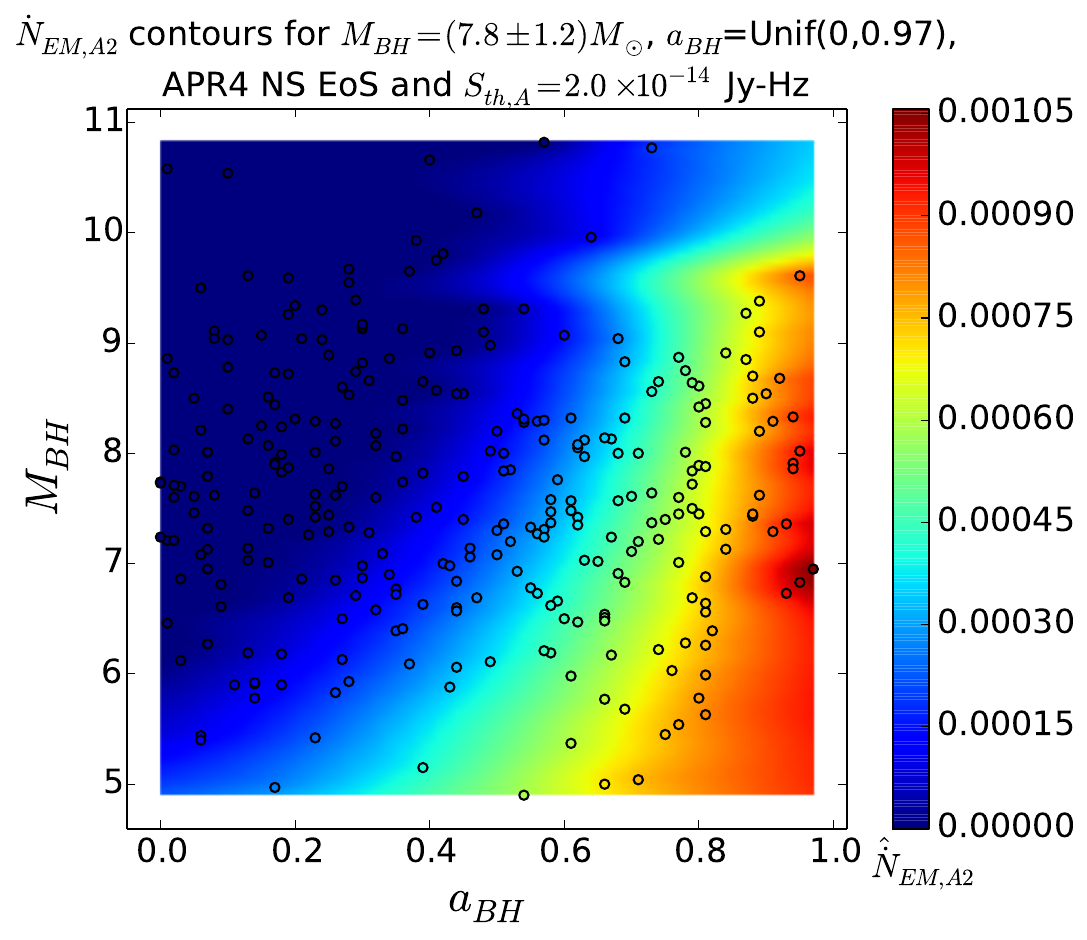} 
  \end{subfigure}
  \begin{subfigure}[tp]{0.49\linewidth}
    \centering
    \includegraphics[height=0.7\linewidth,width=0.9\linewidth]{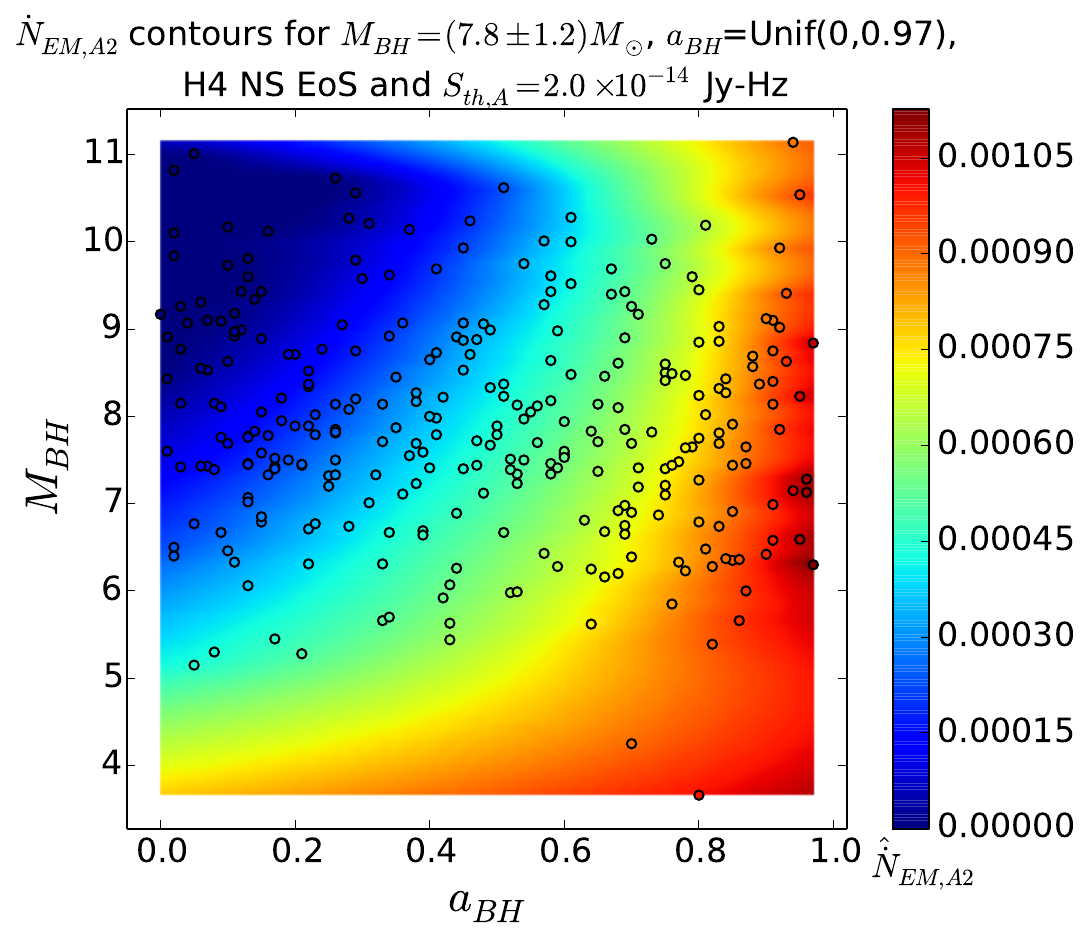} 
  \end{subfigure}\\ 
  \begin{subfigure}[tp]{0.49\linewidth}
    \centering
    \includegraphics[height=0.7\linewidth,width=0.9\linewidth]{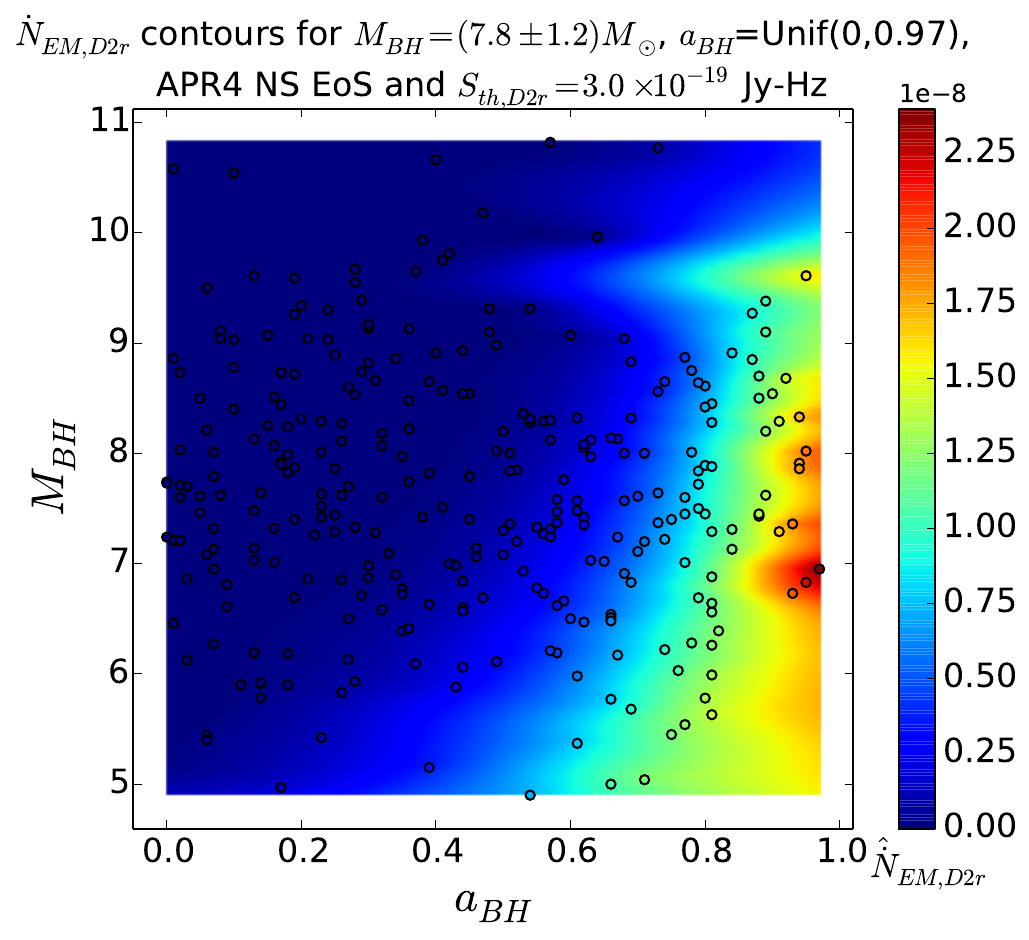} 
  \end{subfigure}
  \begin{subfigure}[tp]{0.49\linewidth}
    \centering
    \includegraphics[height=0.7\linewidth,width=0.9\linewidth]{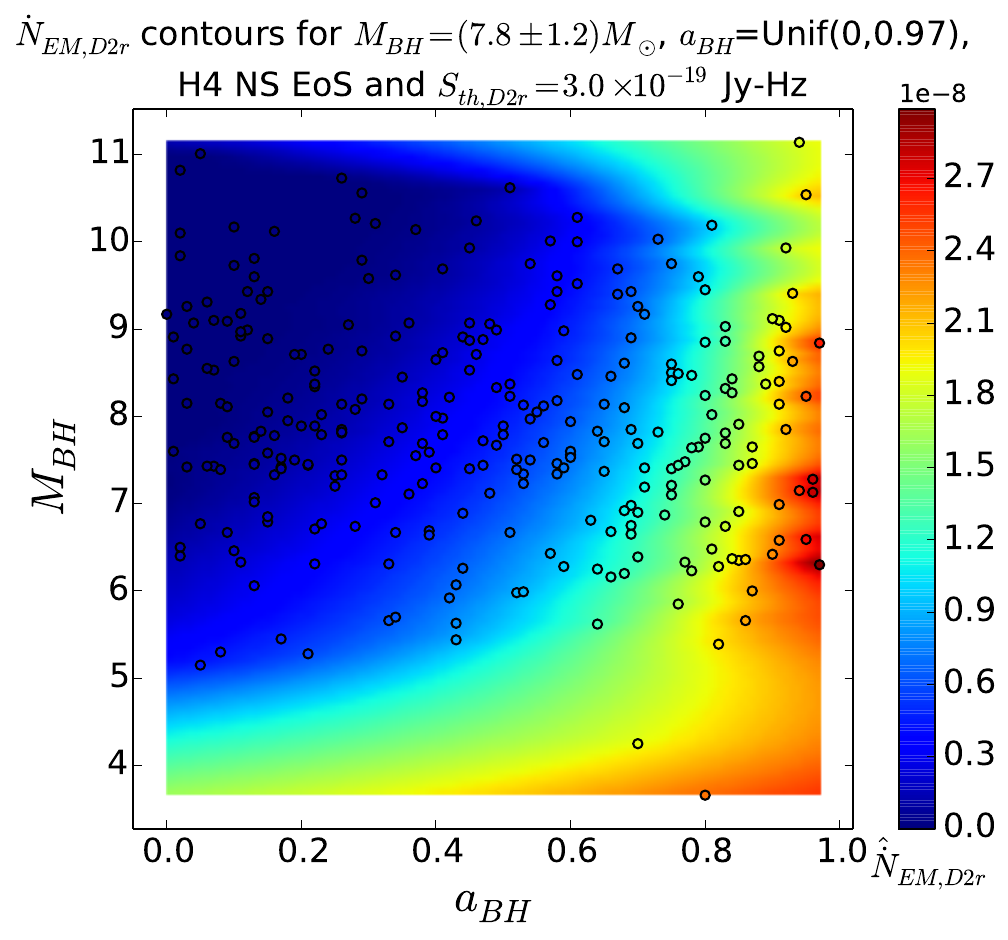} 
   \end{subfigure}
  \caption{\emph{Effect of binary parameters on the $\dot{N}_{EM}$ values for early-time macronova and late-time radio afterglow emission from isotropic sub-relativistic wind:} 
Contour plots using simulation results for 300 BHNS binaries with $M_{BH} = (7.8\pm1.2)\ M_{\odot}$, $a_{BH} = {\rm Uniform(0,0.97)}$ and APR4/H4 NS EoS. 
  	{\it Top-left panel:} $\dot{N}_{A2}$ contour plot for APR4 NS EoS,	
	{\it Top-right panel:} $\dot{N}_{A2}$ contour plot for H4 NS EoS,
	{\it Bottom-left panel:} $\dot{N}_{D2r}$ contour plot for APR4 NS EoS,
	{\it Bottom-right panel:} $\dot{N}_{D2r}$ contour plot for H4 NS EoS.
 }
  \label{fig9} 
\end{figure*}

As opposed to the dynamical ejecta, the wind from the accretion disk is spherically symmetric with a mass $M_{wind} = \eta_{wind} M_{disk}$, where $\eta_{wind}$ is the fraction of the disk mass that gets unbound and contributes to the macronova emission. As the wind is sub-relativistic with velocity $v_{wind} \approx 0.07c$ \citep{HP15}, the peak time and luminosity for the emission are 
\begin{align*}
&t_{peak} = (0.3\ {\rm day})\ \kappa_{1}^{1/2} \eta_{wind,-3}^{1/2} M_{disk,-2}^{1/2} v_{wind,0.07}^{-1/2} \numberthis \label{tpeakMacwind} \\
&L_{peak} = (7.1\times10^{38}\ {\rm erg/s})\ f_{-6} \kappa_1^{-1/2} \eta_{wind,-3}^{1/2} M_{disk,-2}^{1/2} v_{wind,0.07}^{1/2} \numberthis \label{LpeakMacwind}
\end{align*}
where $\eta_{wind,-3} = \eta_{wind}/10^{-3}$, $M_{disk,-2} = M_{disk}/0.01\ M_{\odot}$ and $v_{wind,0.07} = v_{wind}/0.07c$. 
The temporal evolution of the macronova optical/IR luminosity is expected to be $\propto (t/t_{peak})^{-1.3}$ that is the same as the energy injection rate from the radioactive decay of the unstable r-process elements \citep{Metzger10,Tanaka13}.
While the low-opacity blue macronova light curve evolving rapidly over the timescale of $\sim$2 days for GW170817 is expected to result from the isotropic wind component, the high-opacity red macronova light curve with longer timescale $\sim$10 days most likely arises due to the anisotropic dynamical ejecta component \citep{Abbott17b,Hinderer18}.

Due to the anisotropic distribution of dynamical ejecta for BHNS mergers, the macronova emission peaks with larger luminosity $L_{peak}$ and at smaller $t_{peak}$ compared to the isotropic dynamical ejecta ($\theta_{ej,dyn} \approx \pi/2$ and $\phi_{ej,dyn} \approx 2\pi$) from BNS mergers for a similar $M_{ej,dyn}$. However, as $M_{ej,dyn}$ for BNS mergers is expected to be larger by a factor of $\sim 10$, the macronova emission $L_{peak}$ from BNS mergers will be about $\sim 1.5$ times larger even after including the geometrical factors.
The heating efficiency $f$ depends on the electron fraction in the ejecta that is highly uncertain \citep{Wan14}. In this study, we consider an effective value of $f \approx 3\times10^{-6}$ \citep{Metzger10}. Similarly, the opacity $\kappa$ is also very uncertain due to our limited knowledge of the bound-bound transition line features of r-process elements \citep{Kasen13} and we will hereby assume a fiducial value of $\kappa=10\ {\rm cm^{2} g^{-1}}$.

\subsection{sGRB prompt}
One of the most widely studied EM counterparts of BHNS mergers are short duration ($< 2$ s), highly-relativistic collimated outflows, otherwise known as sGRBs. These energetic jets are primarily powered by the gravitational and/or rotational energy of the central compact remnant via accretion. The burst duration for emission in gamma/X-ray bands from sGRB jets is determined by the characteristic viscous timescale for accretion of remnant disk/torus onto the BH and is given by \citep{Fer15}
\begin{eqnarray}
t_{peak} \approx t_{visc} = \frac{R^2}{\nu_a} \approx (0.19\ {\rm s})\ \alpha_{0.03}^{-1} R_2^{3/2} M_{BH,0}^{-1/2} \left(\frac{H}{R}\right)^{2}
\label{tvisc}
\end{eqnarray}
where $R = R_2 \times 100\ {\rm km}$ is the radial extent of the torus in which most of its mass and angular momentum is concentrated, $\nu_a = \alpha c_s H$ is the disk viscosity due to turbulence, $\alpha = \alpha_{0.03}\times0.03$ is the disk viscosity parameter, $c_s$ is the sound speed in the medium, $M_{BH} = M_{BH,0}\times8M_{\odot}$ 
and $H \approx c_s (GM_{BH}/R^3)^{1/2} \sim R$ is the vertical scale-height of the disk. The isotropic energy output of sGRB jets is
\begin{equation}
E_{jet,iso} = \epsilon_{jet} M_{disk} c^2 = (10^{51}\ {\rm erg})\ \epsilon_{jet,-2} M_{disk,-1} \nonumber
\end{equation}
where, $\epsilon_{jet} = \epsilon_{jet,-2}\times10^{-2}$ is the jet energy conversion efficiency factor and $M_{disk,-1} = M_{disk}/0.1\ M_{\odot}$. The jet energy is closely associated with the remnant torus mass $M_{disk}$ which itself is determined by the BHNS binary properties such as $M_{BH}$, $a_{BH}$ and NS EoS (see Section 2). As the jet is highly-relativistic, the observed sGRB energy is obtained by including the relativistic beaming factor: $E_{jet} = E_{jet,iso} \times (1/2)(1 - {\rm cos} \theta_{jet}) = (1/4) \theta_{jet}^2 E_{jet,iso}$, where $\theta_{jet}$ is the jet-opening angle. The peak luminosity for sGRB prompt emission is then given by
\begin{eqnarray}
L_{peak} \approx \frac{E_{jet}}{t_{visc}} = (1.32\times10^{51}\ {\rm erg/s})\nonumber \\ \times \epsilon_{jet,-2} M_{disk,-1} \alpha_{0.03} R_2^{-3/2} M_{BH,0}^{1/2} \left(\frac{H}{R}\right)^{-2} \theta_{jet}^{2}
\label{LpeakC}
\end{eqnarray}
where, all the jet and disk parameters except $\theta_{jet}$ are normalised to their typical values. While the sGRB prompt gamma/X-ray luminosity is roughly constant for $t \lesssim {\rm min}$, it is expected to reduce considerably as $\propto (t/{\rm min})^{-3}$ at later times and until $t \sim 10^{2}\ {\rm s}$ \citep{KZ15}.

\subsection{Cocoon prompt}
\label{Cocoon_prompt}
The observed increase in the afterglow luminosity $\propto t^{0.8}$ for GW170817 up to $t \sim 150$ days after the merger and the subsequent rapid decline can be robustly explained for a compact source with a structured jet including cocoon \citep{Lazzati18,Ghirlanda18}. The VLBI observations of GW170817 also suggest that the early-time radio emission over the first few months was dominated by the mildly-relativistic cocoon and the late-time emission around $t \sim 150$ days was primarily due to the narrowly-collimated relativistic jet with an opening angle $\theta_{jet} \lesssim 5^{\circ}$ seen from a viewing angle $\theta_{v} \sim 14-28^{\circ}$ \citep{Abbott17b,Mooley18b}. 
The interaction of the lighter outgoing ultra-relativistic sGRB jet with the denser material ejected previously (magnetic/viscous/neutrino-driven wind from the disk and/or dynamical ejecta) along the rotation axis of the remnant BH generates a hot mildly-relativistic cocoon surrounding the jet. The cocoon expands approximately spherically after breaking out from the ejecta surface and produces prompt emission in gamma/X-ray bands with energy $E_{c} \sim E_{jet,iso}$.

The cocoon energy is written in terms of the jet luminosity $L_{jet}$ and the time $t_{p}$ that the jet needs to propagate across the denser ejecta as $E_{c} = L_{jet}t_{p}$. As the jet head traverses a distance $R_a$ with velocity $\beta_{jh}$ in time $t_p$, $E_{c} = L_{jet}R_a/(c\beta_{jh})$. The jet head velocity $\beta_{jh}$ is estimated by assuming that the lateral width of the cocoon can be determined by balancing the pressure of the cocoon at equilibrium with the ram pressure of the jet \citep{Matzner03}, $\beta_{jh} \approx (L_{jet}\pi^2/\rho_a t_p^2 \Omega_{jet}^2 c^5)^{1/5}$. Here, $\rho_a$ is the density of the ambient ejecta and $\Omega_{jet}$ is the solid angle occupied by the jet. Substituting $\beta_{jh}$ in $E_{c} = L_{jet}R_a/(c\beta_{jh})$ further gives $E_{c} = (L_{jet}R_a^5 \rho \Omega_{jet}^{2} \pi^{-2})^{1/3} \sim 4\times10^{48}\ {\rm erg}$ for typical parameters: $L_{jet} \sim 2\times10^{50}\ {\rm erg/s}$, $R_a \sim 10^8\ {\rm cm}$, $\rho_a \sim 10^7\ {\rm g\ cm^{-3}}$ and $\theta_{jet} \sim 16^{\circ}$ \citep{Lazzati17}. 

After breaking out from the surrounding ejecta, the cocoon fireball accelerates outwards adiabatically until it attains a saturation Lorentz factor $\Gamma_c$ at radial distance $R_{sat}$ and finally releases the advected radiation at the photosphere with radius \citep{MR00}
\begin{eqnarray}
R_{c} = \left(\frac{E_{c} \sigma_T}{8\pi m_p \Gamma_c^3 c^2}\right)^{1/2} = (4.2\times10^{11}\ {\rm cm})\ E_{c,49}^{1/2} \Gamma_{c,1}^{-3/2}
\label{Rphc}
\end{eqnarray}
where $E_{c,49} = E_c/10^{49}\ {\rm erg}$ and $\Gamma_{c,1} = \Gamma_c/10$. As the cocoon fireball cools adiabatically, its temperature drops as $T_{c} = T_{in,c} (R_c/R_{sat})^{-2/3} \approx 10\ {\rm keV}$ for typical parameters \citep{Lazzati17}, where $T_{in,c}$ is the initial cocoon temperature. The peak cocoon prompt luminosity can then be written assuming a blackbody spectrum and including the cocoon relativistic beaming factor
\begin{eqnarray}
L_{peak} = (1.0\times10^{49}\ {\rm erg/s}) E_{c,49}^{2/3} \Gamma_{c,1}^{5/3} R_{a,8}^{-1/3} \theta_c^2
\label{Lcprompt}
\end{eqnarray}
where $R_{a,8} = R_a/10^8\ {\rm cm}$ and $\theta_c$ is the cocoon-opening angle. For the typical jet and cocoon parameters, the prompt emission lasts for the angular timescale $t_{ang}$ given by
\begin{eqnarray}
t_{peak} \sim t_{ang} \approx (0.14\ {\rm s}) R_{c,0} \Gamma_{c,1}^{-2} = (0.14\ {\rm s}) E_{c,49}^{1/2} \Gamma_{c,1}^{-7/2}
\label{tcprompt}
\end{eqnarray}
where $R_{c,0} = R_c/4.2\times10^{11}\ {\rm cm}$. The gamma/X-ray luminosity for the prompt emission from the mildly-relativistic cocoon is expected to be approximately constant until $t \sim {\rm few\ min}$ and then fall off at later times as $\propto (t/{\rm min})^{-1}$ \citep{Lazzati17}.

As the relativistic jet propagates through the ejecta, the surrounding material forms a cocoon that accelerates outwards adiabatically and with mildly-relativistic velocities. For a jet with given opening angle, the condition for a cocoon to develop but not overcome the jet is \citep{Matzner03}
\begin{eqnarray}
\left(\frac{\theta_{jet}}{90^{\circ}}\right)^{4} < \tilde{L} < \theta_{jet}^{-4}
\label{cjcond}
\end{eqnarray}
where $\tilde{L} = \beta_{jh}^{2} \approx 5.31\times10^{-3} \theta_{jet}^{-8/5}$, for $t_p \sim R_a/c \approx 0.01\ {\rm s}$ and $\Omega_{jet} = 4\pi (1-{\rm cos}\theta_{jet}) = 2\pi \theta_{jet}^{2}$. Substituting $\tilde{L}$ in equation (\ref{cjcond}) further gives $\theta_{jet} < 31.05^{\circ}$ and thereby two possible outcomes: (1) a jet with wide opening angle $\theta_{jet} \gtrsim 30^{\circ}$ that cannot overcome the cocoon fireball and gets choked, (2) a narrow opening angle jet with $\theta_{jet} \lesssim 30^{\circ}$ that can penetrate the surrounding ejecta to look like an on-axis sGRB. While the cocoon geometry is still somewhat uncertain, the cocoon-opening angle is expected to be directly correlated with the jet-opening angle. We discuss the effect of jet- and cocoon-opening angles on their corresponding EM counterpart $L_{peak}$ and $t_{peak}$ along with the detection rate in the next section.

Although BHNS mergers do not produce polar-shocked material unlike BNS mergers and the dynamical ejecta is mostly concentrated around the equatorial plane, the presence of strong magnetic field/viscous flow/neutrino-driven disk winds and sufficient jet-launching time delay $t_{jet,d}$ can result in the accumulation of the required ejecta mass around the polar regions.
The cocoon mass can be written as $M_{cocoon} = \eta_{dyn}M_{dyn} + \eta_{disk}M_{disk}$, where $\eta_{dyn/disk}$ is the contribution to the cocoon mass from the dynamical ejecta/disk wind component. 
The relative contributions from the dynamical ejecta and the wind components depends on $t_{jet,d}$, with the ratio $\eta_{disk}/\eta_{dyn} \propto t_{jet,d}$ as the sub-relativistic winds progressively drive more unbound disk mass towards the existing ejecta material along the BH rotation axis.

\begin{figure*}
    \begin{subfigure}[tp]{0.49\linewidth}
    \centering
    \includegraphics[height=0.7\linewidth,width=0.9\linewidth]{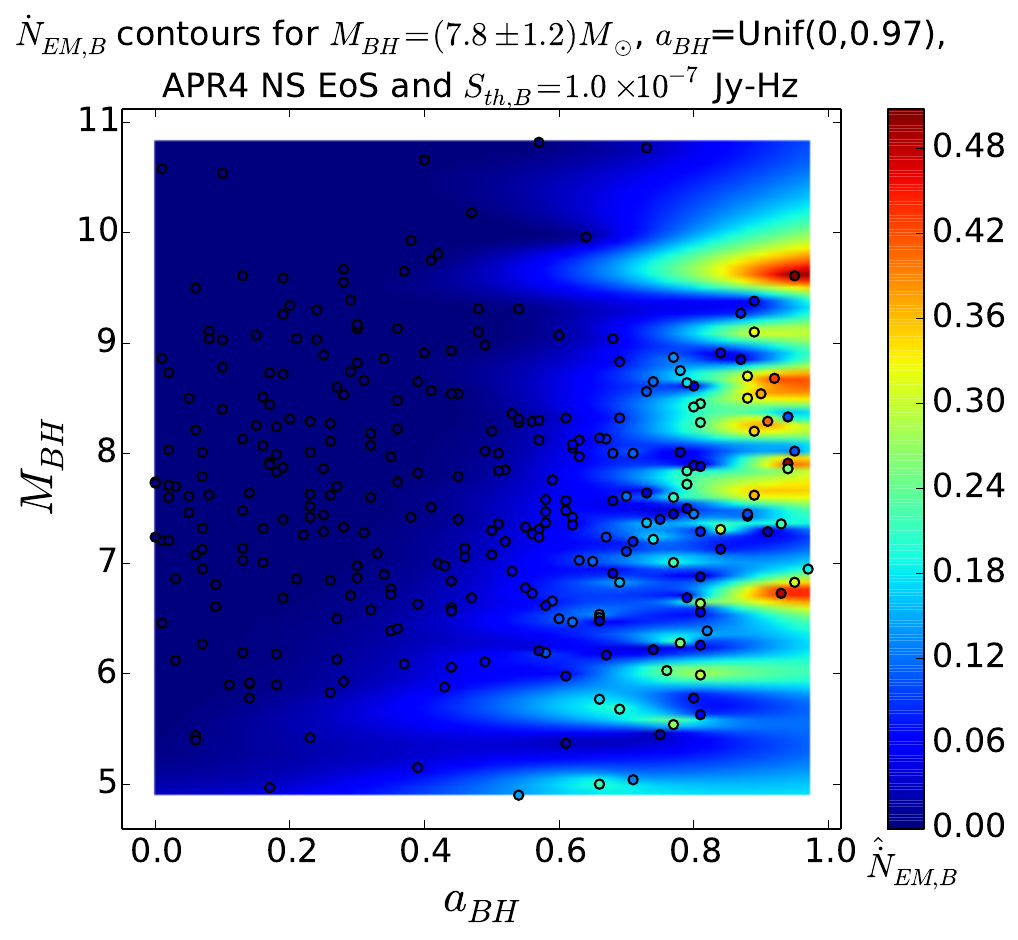} 
  \end{subfigure}
  \begin{subfigure}[tp]{0.49\linewidth}
    \centering
    \includegraphics[height=0.7\linewidth,width=0.9\linewidth]{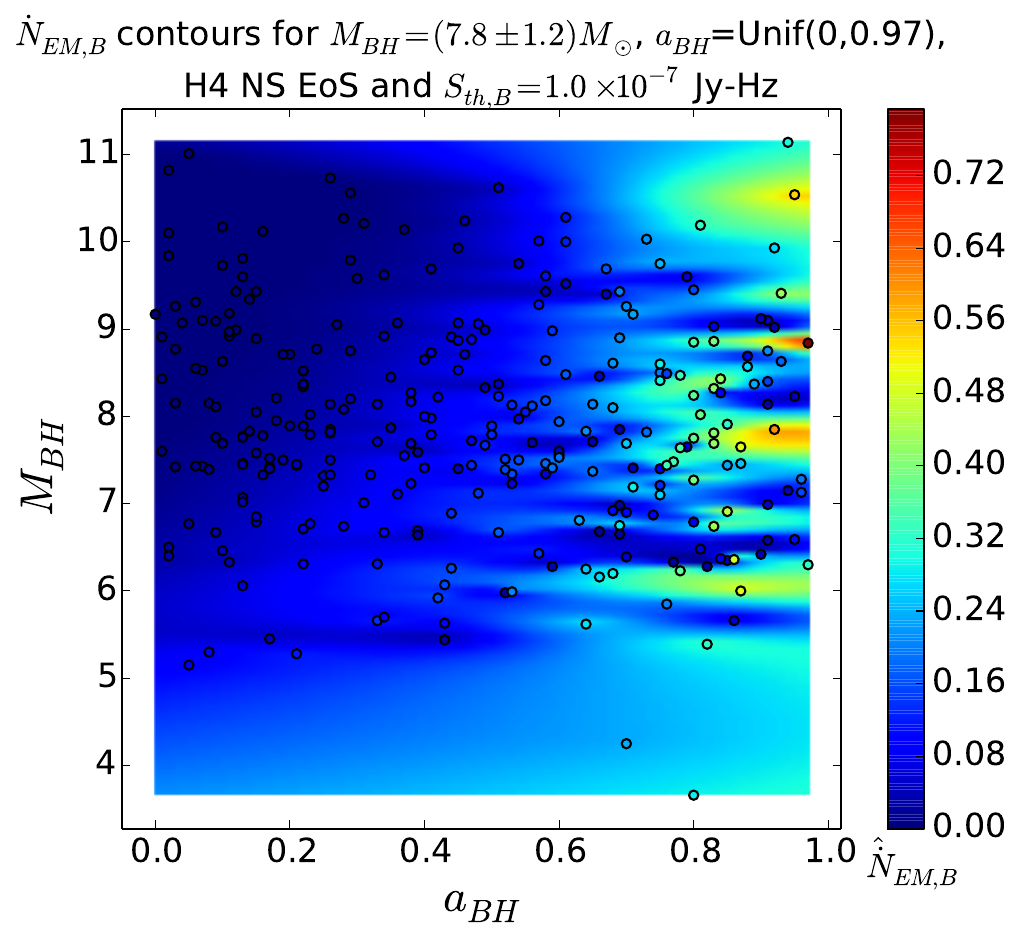}  
  \end{subfigure}\\ 
  \begin{subfigure}[tp]{0.49\linewidth}
    \centering
    \includegraphics[height=0.7\linewidth,width=0.9\linewidth]{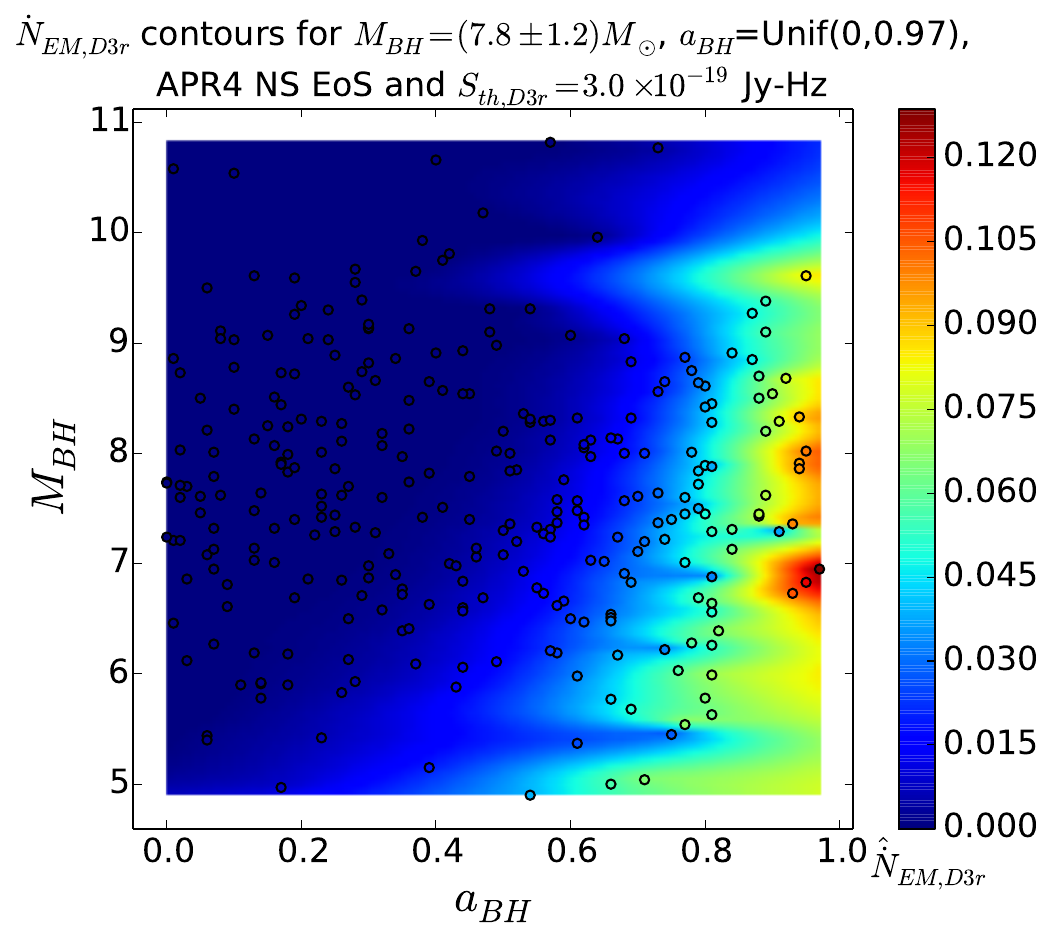} 
  \end{subfigure}
  \begin{subfigure}[tp]{0.49\linewidth}
    \centering
    \includegraphics[height=0.7\linewidth,width=0.9\linewidth]{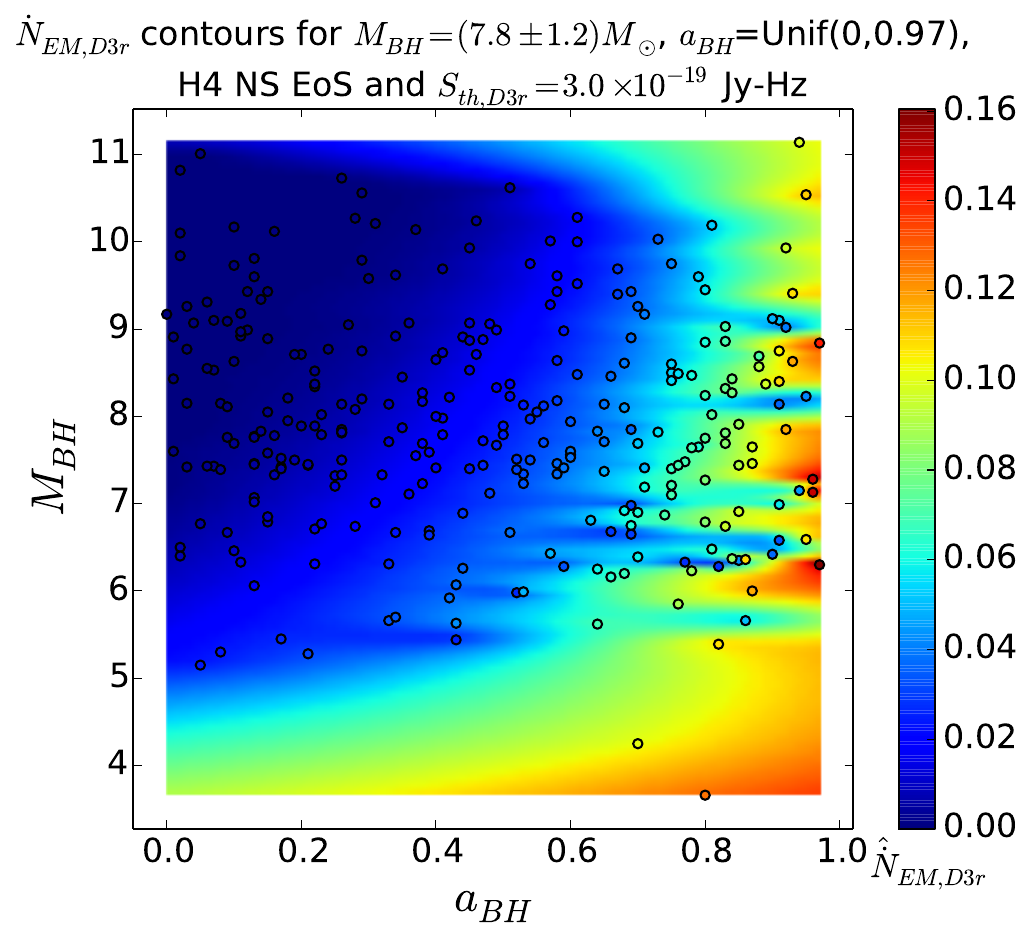}  
  \end{subfigure} 
  \caption{\emph{Effect of binary parameters on the $\dot{N}_{EM}$ values for early-time prompt and late-time radio afterglow emission from anisotropic ultra-relativistic sGRB jet:} 
Contour plots using simulation results for 300 BHNS binaries with $M_{BH} = (7.8\pm1.2)\ M_{\odot}$, $a_{BH} = {\rm Uniform(0,0.97)}$ and APR4/H4 NS EoS. 
  	{\it Top-left panel:} $\dot{N}_{B}$ contour plot for APR4 NS EoS,	
	{\it Top-right panel:} $\dot{N}_{B}$ contour plot for H4 NS EoS,
	{\it Bottom-left panel:} $\dot{N}_{D3r}$ contour plot for APR4 NS EoS,
	{\it Bottom-right panel:} $\dot{N}_{D3r}$ contour plot for H4 NS EoS
 }
  \label{fig7} 
\end{figure*}

\begin{figure*}
    \begin{subfigure}[tp]{0.49\linewidth}
    \centering
    \includegraphics[height=0.7\linewidth,width=0.9\linewidth]{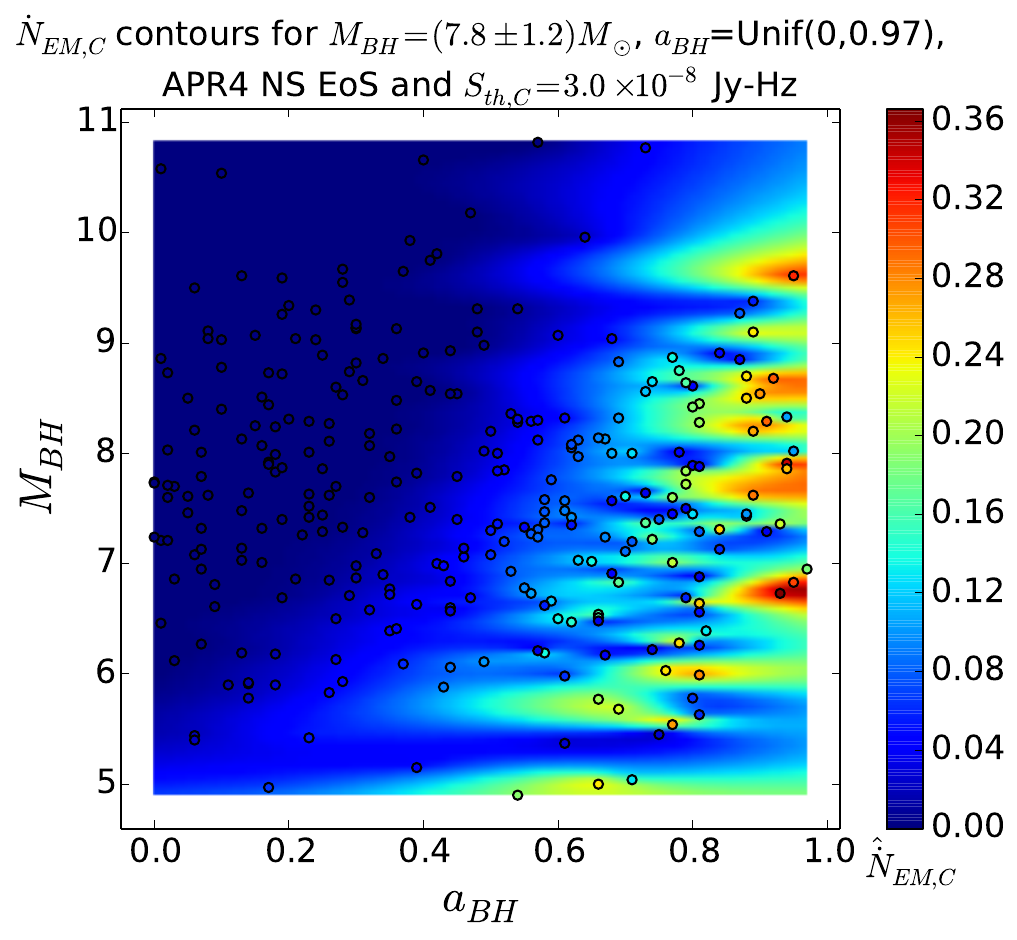} 
  \end{subfigure}
  \begin{subfigure}[tp]{0.49\linewidth}
    \centering
    \includegraphics[height=0.7\linewidth,width=0.9\linewidth]{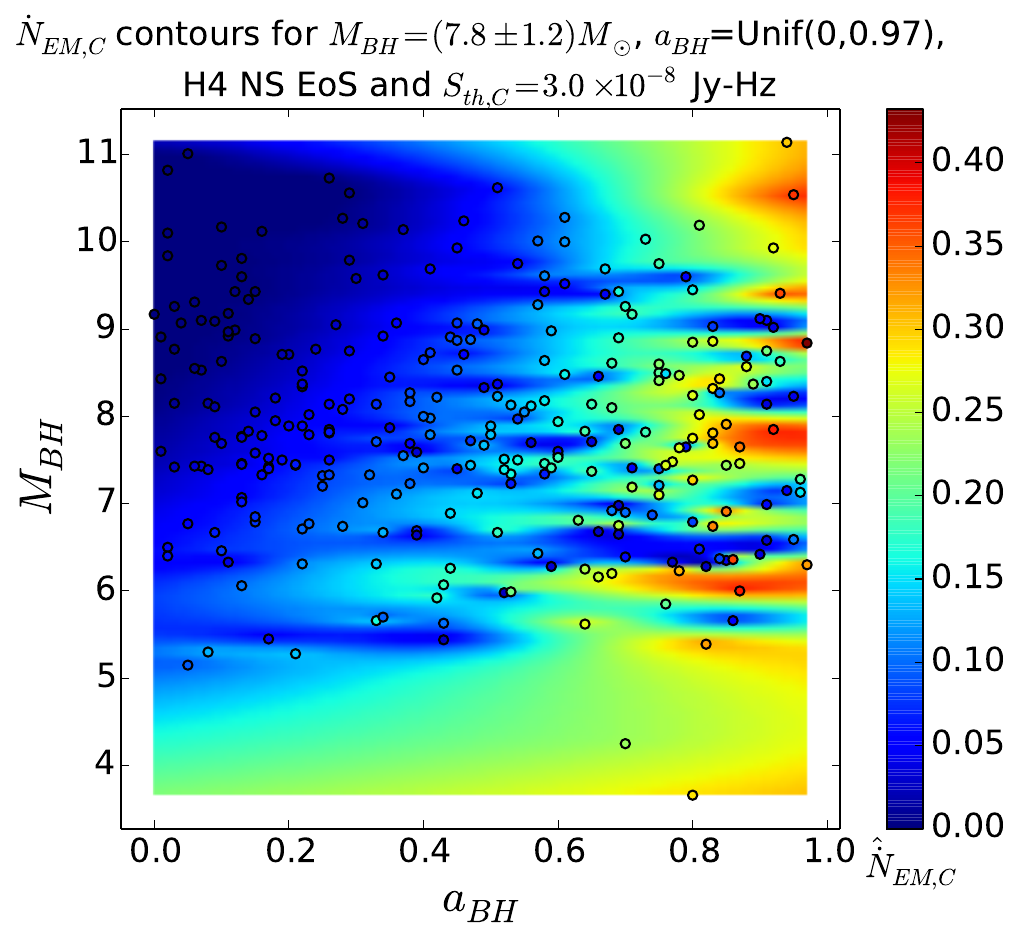}  
  \end{subfigure}\\ 
  \begin{subfigure}[tp]{0.49\linewidth}
    \centering
    \includegraphics[height=0.7\linewidth,width=0.9\linewidth]{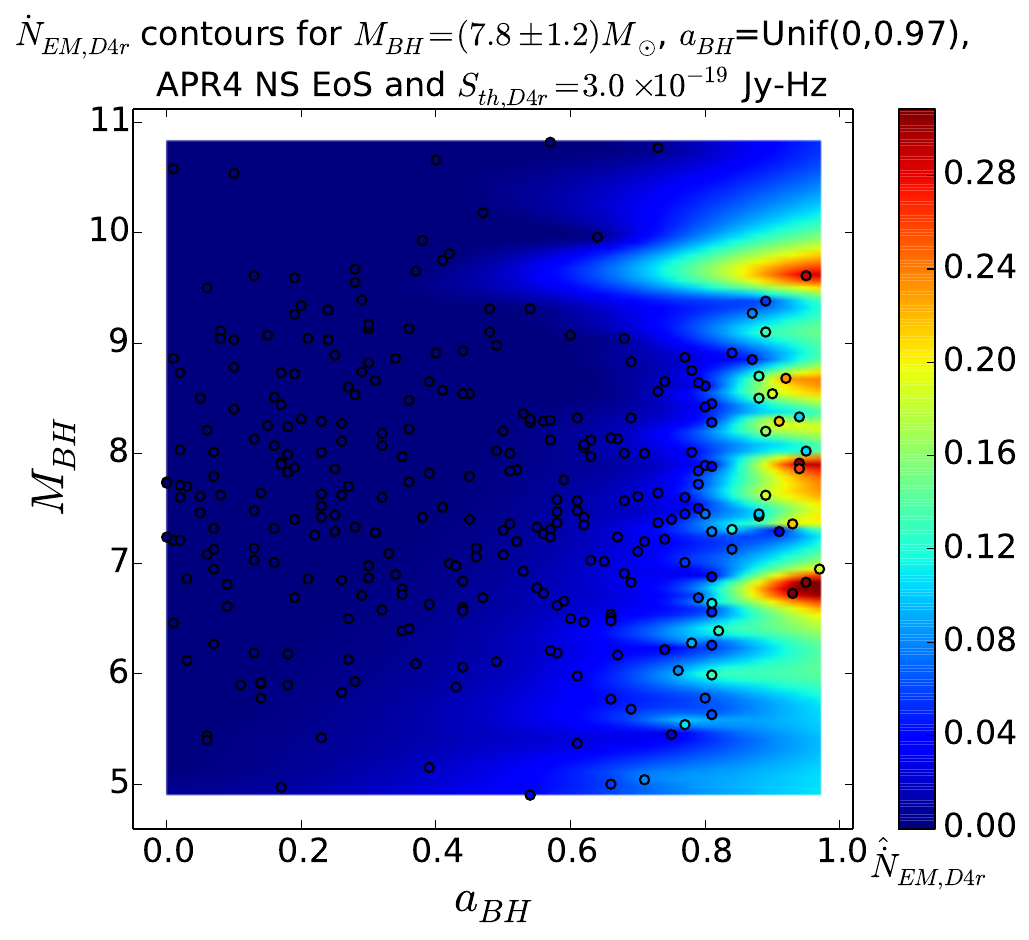} 
  \end{subfigure}
  \begin{subfigure}[tp]{0.49\linewidth}
    \centering
    \includegraphics[height=0.7\linewidth,width=0.9\linewidth]{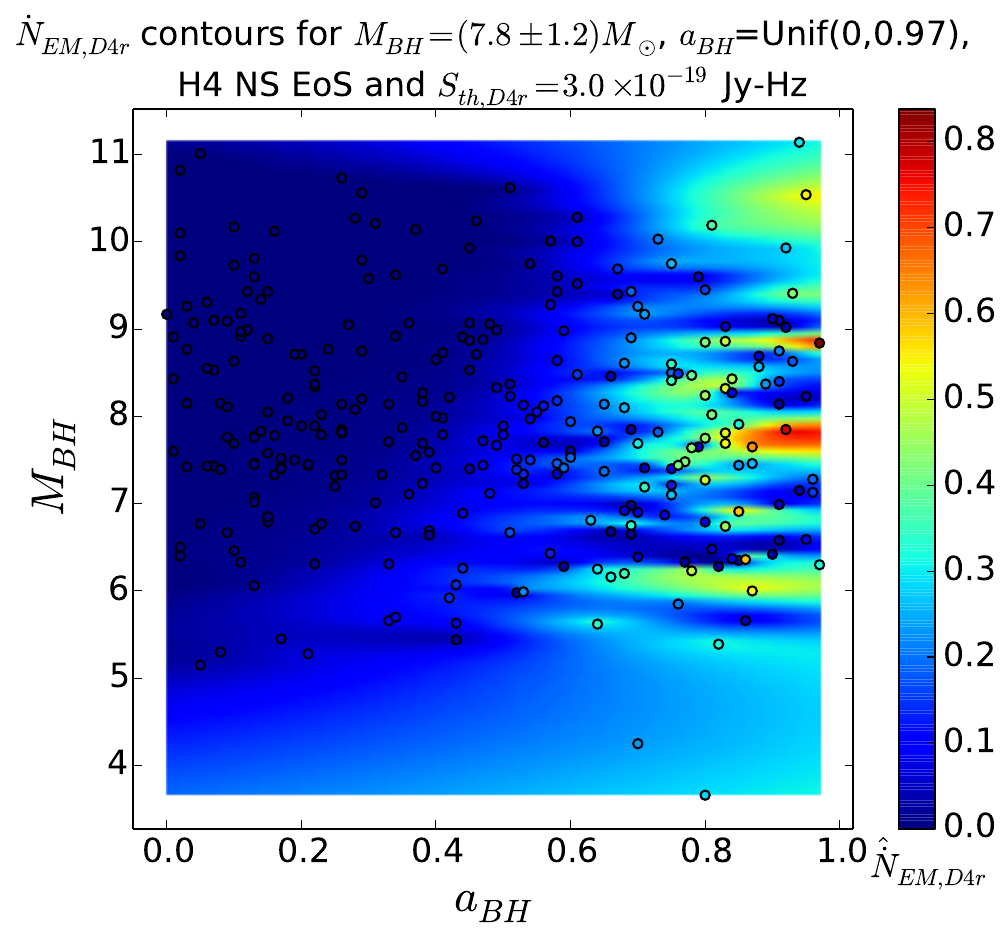}  
  \end{subfigure}\\ 
  \caption{\emph{Effect of binary parameters on the $\dot{N}_{EM}$ values for early-time prompt and late-time radio afterglow emission from anisotropic mildly-relativistic cocoon:} 
Contour plots using simulation results for 300 BHNS binaries with $M_{BH} = (7.8\pm1.2)\ M_{\odot}$, $a_{BH} = {\rm Uniform(0,0.97)}$ and APR4/H4 NS EoS. 
  	{\it Top-left panel:} $\dot{N}_{C}$ contour plot for APR4 NS EoS,	
	{\it Top-right panel:} $\dot{N}_{C}$ contour plot for H4 NS EoS,
	{\it Bottom-left panel:} $\dot{N}_{D4r}$ contour plot for APR4 NS EoS,
	{\it Bottom-right panel:} $\dot{N}_{D4r}$ contour plot for H4 NS EoS
 }
  \label{fig8} 
\end{figure*}

\subsection{Synchrotron afterglow}
As the energetic ejecta launched from BHNS mergers expands outwards, it interacts with the surrounding ISM and its KE gets converted into the internal energy of a propagating shock. A significant portion of this energy is transferred to the electrons that are collected from the ambient ISM. Synchrotron afterglow radiation is then emitted by these accelerated non-thermal electrons in the presence of $B$ on timescales of $\sim {\rm few\ weeks - year}$, and is detectable in radio and (possibly) optical bands. While all ejecta components (sGRB jet, dynamical ejecta, magnetic/viscous/neutrino-driven winds and cocoon) contribute to this emission, different components have different $L_{peak}$ and $t_{peak}$ due to their distinct masses, KEs and geometrical distributions \citep{HP15}. It should be noted that the \emph{ultra-relativistic} jet, \emph{mildly-relativistic} cocoon and \emph{sub-relativistic} dynamical ejecta are spherically asymmetric while the \emph{sub-relativistic} wind ejecta component is spherically symmetric. Here we assume that the dynamical ejecta/jet velocity $v_{ej,dyn}/v_{jet}$ is practically unchanged until $t \sim t_{peak}$ resulting in fixed ejecta/jet geometry. 

Most of the synchrotron emission is radiated once the ejecta decelerates considerably by accumulating a mass comparable to its own at a radius $R_{dec,s} = (3M_{ej}/4\pi m_p n)^{1/3} = (3.2\ {\rm pc}) M_{ej,0.03}^{1/3} n_{-2}^{-1/3}$. Here $M_{ej} = M_{ej,0.03}\times0.03 M_{\odot}$ is the spherically symmetric equivalent of ejecta mass, $m_p$ is the proton mass and $n = n_{-2} \times {\rm 0.01\ cm^{-3}}$ is the constant density of the ambient ISM. The value of $n$ can vary by orders of magnitude based on the location of the BHNS coalescence in the host galaxy and here we assume an average density $n \approx 0.01\ {\rm cm^{-3}}$ that is similar to the typical sGRB median densities and slightly larger than the GW170817 expected circumburst density $\sim 10^{-4} - 5\times10^{-3}\ {\rm cm^{-3}}$ \citep{Fong15,Hoto16,Mooley18b}. While the ejecta expands at a constant velocity $v_{ej}$ upto $R=R_{dec,s}$, it decelerates at larger radius according to the Sedov-Taylor solution, $v_{ej}(R/R_{dec,s})^{-3/2}$. The deceleration time for spherical ejecta is then given by
\begin{eqnarray}
t_{dec,s} = \frac{R_{dec,s}}{\Gamma_{ej}^2 v_{ej}} = (32.5\ {\rm year}) M_{ej,0.03}^{1/3}n_{-2}^{-1/3}v_{ej,0.3}^{-1}\Gamma_{ej,0}^{-2}
\label{tdecs}
\end{eqnarray}
where $\Gamma_{ej} = \Gamma_{ej,0}\times1.0$ is the Lorentz factor of the ejecta component.
For the anisotropic ejecta components (jet, cocoon and dynamical ejecta), the deceleration radius includes angular factors, $R_{dec} = R_{dec,s}(\pi/2\theta_{ej})^{1/3}(2\pi/\phi_{ej})^{1/3} = (7.9\ {\rm pc}) M_{ej,0.03}^{1/3} n_{-2}^{-1/3}\theta_{ej,0}^{-1/3}\phi_{ej,0}^{-1/3}$ and the deceleration time is \citep{Kyu13}
\begin{equation}
t_{dec} = \frac{R_{dec}}{\Gamma_{ej}^2 v_{||}} = (83.6\ {\rm year}) M_{ej,0.03}^{1/3}n_{-2}^{-1/3}v_{ej,0.3}^{-1}\theta_{ej,0}^{-1/3}\phi_{ej,0}^{-1/3}\Gamma_{ej,0}^{-2}
\label{tdec}
\end{equation}
where we have used the relation $v_{||} \approx v_{ej}$. The peak emission time $t_{peak} \sim t_{dec}$ for synchrotron afterglow varies within a significant range based on the properties of the ejecta components: $\sim$ few days--week for the \emph{ultra-relativistic} jet, $\sim$ few weeks--month for \emph{mildly relativistic} cocoon, $\sim$ few years--decade for \emph{sub-relativistic} dynamical ejecta and winds. 

A considerable fraction $\epsilon_{e}/\epsilon_{B}$ of the post-shock internal energy is transferred to the non-thermal electrons/magnetic fields. For typical blast waves with $\nu_{min} < \nu_a < \nu_c$ (see Appendix \ref{syn_char_freq} for definitions), the synchrotron spectrum at a given time is
\begin{equation}
F_{\nu} = F_{\nu,min} 
\left\{
\begin{array}{ll}
\left(\frac{\nu_a}{\nu_{min}}\right)^{-(p+4)/2}\left(\frac{\nu}{\nu_{min}}\right)^{2},& \ \nu < \nu_{min} \\
\left(\frac{\nu_a}{\nu_{min}}\right)^{-(p-1)/2}\left(\frac{\nu}{\nu_{a}}\right)^{5/2},& \ \nu_{min} \leq \nu < \nu_{a}\\
\left(\frac{\nu}{\nu_{min}}\right)^{-(p-1)/2},& \ \nu_{a} \leq \nu < \nu_{c}\\
\left(\frac{\nu_c}{\nu_{min}}\right)^{-(p-1)/2}\left(\frac{\nu}{\nu_{c}}\right)^{-p/2},& \ \nu \geq \nu_{c}
\end{array}
\right.  
\label{synspec}
\end{equation}
where $p \approx 2.5$ is the power-law index for the energy distribution of the accelerated electrons.


For most sensitive radio (optical) observations, $\nu_{a} \leq \nu < \nu_{c}$ ($\nu \geq \nu_{c}$), and the peak specific flux/luminosity $F_{\nu,peak}$/$L_{peak}$ are computed using the respective $F_{\nu}$ expressions from equation (\ref{synspec}). Flux $F_{\nu}$ across the whole spectrum increases for $t < t_{dec}$ and peaks at $t=t_{dec}$ for typical radio/optical observing frequencies $\nu_{obsr/o} > \nu_{a}(t_{dec}),\ \nu_{min}(t_{dec})$. The peak of the observed radio/optical specific flux $F_{\nu_{obsr/o}}$ at frequency $\nu_{obsr/o}$
\begin{eqnarray}
F_{\nu_{obsr}} = (1.1\ {\rm mJy}) \epsilon_{e,-1}^{p-1} \epsilon_{B,-1}^{(p+1)/4} n_{-2}^{(p+1)/4} M_{ej,0.03} \nonumber \\ \times \ v_{ej,0.3}^{(5p-3)/2}D_2^{-2}\left(\frac{\nu_{obsr}}{1\ {\rm GHz}}\right)^{-(p-1)/2} \label{Fnur} \\
F_{\nu_{obso}} = (0.2\ {\rm Jy}) \epsilon_{e,-1}^{p-1} \epsilon_{B,-1}^{(p-2)/4} n_{-2}^{(3p-2)/12} M_{ej,0.03}^{2/3} \nonumber \\ \times \ v_{ej,0.3}^{(5p-4)/2} \theta_{ej,0}^{1/3}\phi_{ej,0}^{1/3} D_2^{-2}\left(\frac{\nu_{obso}}{1\ {\rm GHz}}\right)^{-p/2}
\label{Fnuo}
\end{eqnarray}
where $D = D_2 \times{\rm 100\ Mpc}$ is the distance to the binary source ignoring any cosmological effects. We have normalized $\epsilon_e \approx \epsilon_B \sim 0.1$ to their typical values with $\epsilon_{e,-1} = \epsilon_{e}/0.1$ and $\epsilon_{B,-1} = \epsilon_{B}/0.1$.
The peak radio/optical luminosity $L_{peakr/o}$ is then estimated from the peak specific flux $F_{\nu_{obsr/o}}$ as $L_{peakr/o} = 4\pi D^{2} (\nu_{obsr/o} F_{\nu_{obsr/o}})$. 

As $\nu_{a}(t_{dec}),\ \nu_{min}(t_{dec}) < \nu_{obsr/o}$, the synchrotron radio/optical afterglow luminosity for the \emph{sub-relativistic} dynamical ejecta and wind increases as $\propto (t/t_{peak})^{3}$ until $t \approx t_{peak}$ and subsequently falls off as $\propto (t/t_{peak})^{-(15p-21)/10} \propto t^{-1.65}$ at later times \citep{PNR13}. For the \emph{ultra-relativistic} jet, the afterglow luminosity increases as $\propto (t/t_{peak})^{3}$ before $t = t_{peak}$ and then drops as $\propto (t/t_{peak})^{-1}$ for $t \gtrsim t_{peak}$ \citep{KZ15}, whereas for the \emph{mildly-relativistic} cocoon, the afterglow luminosity increases as $\propto (t/t_{peak})^{3}$ for $t \lesssim t_{peak}$ and drops beyond $t \sim t_{peak}$ as $\propto (t/t_{peak})^{-1.5}$ \citep{RK15,Lyman18}.
The synchrotron radio afterglow peak luminosity from the dynamical ejecta for BHNS mergers will be about an order of magnitude smaller compared to BNS mergers due to the smaller dynamical ejecta mass for the former (see equation \ref{Fnur}). 

\begin{table*}
\begin{center}
\caption{\small EM telescope sensitivities in different observing bands} 
\label{Table2}
\begin{tabular}{| c | c | c | c | c | c | c | c |}
\hline
\hline
\centering
\emph{Observing band} & \emph{Survey} & \emph{Frequency range} & \emph{Nominal $\nu_{obs}$} & \emph{Sensitivity} & \emph{Cadence} & \emph{Reference}\\
& & \emph{(Hz)} & \emph{(Hz)} & \emph{(Jy-Hz)} & \emph{(days)} & 
\\ \hline \hline
Gamma-ray & Fermi GBM & $(1.93-9.67)\times10^{21}$ & $4.84\times10^{21}$ & $10^{-6}$ & 1 & \citet{Mee09}\\ \hline
X-ray & Swift BAT & $(0.36-3.63)\times10^{19}$ & $1.21\times10^{19}$ & $10^{-8}$ & 2 & \citet{Cus10}\\ \hline
Optical & LSST & $(4.28-5.44)\times10^{14}$ & $4.84\times10^{14}$ & $2.81\times10^{-15}$ & 3 & \citet{Ive08}\\ \hline
& PTF & $(4.28-5.44)\times10^{14}$ & $4.84\times10^{14}$ & $6.78\times10^{-14}$ & 5 & \citet{Law09} \\ \hline
& ZTF & $(4.28-5.44)\times10^{14}$ & $4.84\times10^{14}$ & $2.76\times10^{-14}$ & 1 & \citet{Cao16} \\ \hline
Radio & Apertif & $(1.00-1.75)\times10^9$ & $1.32\times10^9$ & $1.32\times10^{-21}$ & 1 & \citet{OVC10} \\ \hline
& ASKAP & $(0.70 - 1.80)\times10^9$ & $1.12\times10^9$ & $1.12\times10^{-18}$ & 1 & \citet{JFG09} \\ \hline
& LOFAR & $(0.10 - 2.00)\times10^8$ & $4.50\times10^7$ & $4.50\times10^{-19}$ & 1 & \citet{Fen06} \\ \hline
\hline
\end{tabular}
\end{center}
\end{table*}

\section{EM follow-up of GW triggers}
In the previous section, we discussed the EM signals associated with GW emission from BHNS mergers. Here we estimate the detection rates corresponding to each of those EM counterparts for low-latency follow-up of GW triggers. In addition to enhancing the confidence on the astrophysical origin of the GW signal, the detection of EM counterparts aids precise source localization, redshift estimation and further provides information about the jet and the ejecta physics. The EM localization for a GW event via follow-up also makes it considerably easier to search for the transient in a relatively smaller sky area as compared to blind surveys across large sky areas.

\subsection{Event rates for EM counterparts}
\label{EM_prompt_radio}
For a given EM counterpart with peak luminosity $L_{peak}$, the maximum distance at which it can be detected using a telescope/survey with bolometric flux detection threshold $S_{th}$ in a given observing band is $D_{max} = \sqrt{L_{peak}/4\pi S_{th}}$. The $S_{th}$ values for some of the current gamma/X-ray, optical and radio surveys that can be used to observe the EM signals accompanying BHNS GW triggers are listed in Table \ref{Table2}. The detection rate for the EM follow-up of BHNS GW events can then be written similarly to equation (\ref{NGWRv}) as
\begin{align}
\dot{N}_{EM} &= \left(0.0348\ {\rm year^{-1}}\right) \times (2.26)^{-3} \int_0^{D_{EM}} R_{V,0}\ . \ 4\pi D_2^2 dD_2 \nonumber \\
&= \left(0.0126\ {\rm year^{-1}}\right) R_{V,0} D_{EM,2}^{3}
\label{NEM}
\end{align}
where $D_{EM} = {\rm min}(D_{hor},D_{max})$ and $D_2 = D/(100\ {\rm Mpc})$. For obtaining equation (\ref{NEM}), we have assumed that the BHNS coalescence rate $R_{V}$ is broadly independent of the distance which is expected for relatively small distances, $D_{EM} \lesssim D_{hor} \sim 650\ {\rm Mpc}$. 

We use the following subscripts to denote the EM counterparts: $i = A1/2$ for macronova from dynamical ejecta/wind, $B/C$ for sGRB/cocoon prompt emission and $D1/2/3/4$ for dynamical ejecta/wind/sGRB jet/cocoon afterglow.
We denote the radio/optical afterglow emission with the subscript $r/o$, and estimate the masses of all the ejecta components in terms of $M_{disk}$ and $M_{ej,dyn}$ (see Section 2). 
In Table \ref{Table3}, we list $t_{peak}(M_{ej},v_{ej})$ and $L_{peak}(M_{ej},v_{ej})$ used to evaluate $D_{max}(M_{ej},v_{ej},S_{th})$ and $\dot{N}_{EM}$ (from equation \ref{NEM}) for each EM counterpart.

\begin{table*}
\begin{center}
\caption{\small The EM follow-up detection parameters are listed for each component. We use the definitions: $\nu_{obsr,GHz} = \nu_{obsr}/(1\ {\rm GHz})$, $M_{dyn,-3} = M_{ej,dyn}/10^{-3}M_{\odot}$, $v_{dyn,0.2} = v_{ej,dyn}/0.2c$, $\epsilon_{jet,-3} = \epsilon_{jet}/10^{-3}$, $\eta_{jet,-4} = \eta_{jet}/10^{-4}$, $\Gamma_{jet,1} = \Gamma_{jet}/10$, $\eta_{dyn,-2} = \eta_{dyn}/10^{-2}$ and $\eta_{disk,-3} = \eta_{disk}/10^{-3}$. The optical/IR, gamma/X-ray and radio band sensitivities used to compute EM detection rates are $S_{OIR} = 2\times10^{-14}\ {\rm Jy\ Hz}$, $S_{GX} = 10^{-7}\ {\rm Jy\ Hz}$ and $S_{R} = 3\times10^{-17}\ {\rm Jy\ Hz}$, respectively. The EM follow-up rate for a given component is $\dot{N}_{EM}/\dot{N}_{GW} = {\rm min}(1,D_{max}^{3}/D_{hor}^{3})$ for telescope sensitivity $S_{th}$.} 
\label{Table3}
\begin{tabular}{| c | c | c | c | c | c |}
\hline
\hline
\centering
\emph{EM component} & $t_{peak}$ & $L_{peak}$ & $D_{max}$ & Relevant equations
\\ \hline \hline
[A1]: Ejecta & $(0.84\ {\rm day})\times$ & $(7.66\times10^{40}\ {\rm erg/s})$ & $(178.94\ {\rm Mpc})M_{dyn,-3}^{1/4}\times$ & \ref{tpeakMac}, \ref{LpeakMac}\\ 
macronova & $M_{dyn,-3}^{1/2}v_{dyn,0.2}^{-1/2}$ & $\times M_{dyn,-3}^{1/2}v_{dyn,0.2}^{1/2}$ & $v_{dyn,0.2}^{1/4}(S_{th}/S_{OIR})^{-1/2}$ & \\ \hline
[A2]: Wind & $(0.3\ {\rm day})\eta_{wind,-3}^{1/2}\times$ & $(2.13\times10^{39}\ {\rm erg/s})\eta_{wind,-3}^{1/2}$ & $(29.84\ {\rm Mpc})\eta_{wind,-3}^{1/4}M_{disk,-2}^{1/4}$ & \ref{tpeakMacwind}, \ref{LpeakMacwind}\\ 
macronova & $M_{disk,-2}^{1/2}v_{wind,0.07}^{-1/2}$ & $\times M_{disk,-2}^{1/2}v_{wind,0.07}^{1/2}$ & $\times v_{wind,0.07}^{1/4}(S_{th}/S_{OIR})^{-1/2}$ & \\ \hline
[B]: sGRB & $\sim 0.19\ {\rm s}$ & $(1.32\times10^{47}\ {\rm erg/s})\times$ & $(105.05\ {\rm Mpc})\epsilon_{jet,-3}^{1/2}\times$ & \ref{tvisc}, \ref{LpeakC}\\ 
prompt & & $\epsilon_{jet,-3}M_{disk,-2}\Gamma_{jet,1}^{-2}$ & $M_{disk,-2}^{1/2}\Gamma_{jet,1}^{-1}(S_{th}/S_{GX})^{-1/2}$ & \\ \hline
[C]: Cocoon & $(4.81\ {\rm s})\eta_{jet,-4}^{1/2}\times$ & $(9.56\times10^{46}\ {\rm erg/s})\times$ & $(89.40\ {\rm Mpc})\eta_{jet,-4}^{1/3}\times$ & \ref{Lcprompt}, \ref{tcprompt} \\ 
prompt & $M_{disk,-2}^{1/2}\Gamma_{jet,1}^{-7/2}$ & $\eta_{jet,-4}^{2/3} M_{disk,-2}^{2/3}\Gamma_{jet,1}^{-1/3}$ & $M_{disk,-2}^{1/3}\Gamma_{jet,1}^{-1/6}(S_{th}/S_{GX})^{-1/2}$ & \\ \hline \hline
[D1]: Ejecta & $(38.99\ {\rm year})\times$ & $(6.31\times10^{34}\ {\rm erg/s})M_{dyn,-3}$ & $(4.19\ {\rm Mpc})M_{dyn,-3}^{1/2}\times$ & \ref{tdec}, \ref{Fnur} \\ 
afterglow & $M_{dyn,-3}^{1/3}v_{dyn,0.2}^{-1}$ & $\times v_{dyn,0.2}^{(5p-3)/2}\nu_{obsr,GHz}^{-(p-3)/2}$ & $v_{dyn,0.2}^{(5p-3)/4}(S_{th}/S_{R})^{-1/2}$ & \\ \hline
[D2]: Wind & $(9.61\ {\rm year})\eta_{wind,-3}^{1/3}$ & $(4.30\times10^{30}\ {\rm erg/s})\eta_{wind,-3}$ & $(0.035\ {\rm Mpc})\eta_{wind,-3}^{1/2}M_{disk,-2}^{1/2}$ & \ref{tdecs}, \ref{Fnur} \\ 
afterglow & $\times M_{disk,-2}^{1/3}v_{wind,0.07}^{-1}$ & $\times M_{disk,-2}v_{wind,0.07}^{(5p-3)/2}\nu_{obsr,GHz}^{-(p-3)/2}$ & $\times v_{wind,0.07}^{(5p-3)/4}(S_{th}/S_{R})^{-1/2}$ & \\ \hline
[D3]: Jet & $(70.69\ {\rm hour})\eta_{jet,-4}^{1/3}$ & $(1.32\times10^{35}\ {\rm erg/s})\eta_{jet,-4}$ & $(6.06\ {\rm Mpc})\eta_{jet,-4}^{1/2}$ & \ref{tdec}, \ref{Fnur} \\ 
afterglow & $\times M_{disk,-2}^{1/3}\Gamma_{jet,1}^{-5/3}$ & $\times M_{disk,-2}\nu_{obsr,GHz}^{-(p-3)/2}$ & $\times M_{disk,-2}^{1/2}(S_{th}/S_{R})^{-1/2}$ & \\ \hline
[D4]: Cocoon & $(39.50\ {\rm day})\Gamma_{jet,1}^{-5/3}$ & $(1.32\times10^{36}\ {\rm erg/s})\times$ & $(19.18\ {\rm Mpc})(S_{th}/S_{R})^{-1/2}$ & \ref{tdec}, \ref{Fnur} \\
afterglow & $[\eta_{dyn,-2}M_{dyn,-3} +$ & $[\eta_{dyn,-2}M_{dyn,-3} +$ & $[\eta_{dyn,-2}M_{dyn,-3} +$ & \\
& $\eta_{disk,-3}M_{disk,-2}]^{1/3}$ & $\eta_{disk,-3}M_{disk,-2}]\nu_{obsr,GHz}^{-(p-3)/2}$ & $\eta_{disk,-3}M_{disk,-2}]^{1/2}$ & \\
\hline
\hline
\end{tabular}
\end{center}
\end{table*}

\begin{itemize}
\item \emph{Early-time emission:} The early-time EM emission consists of macronova from dynamical ejecta/wind, sGRB jet prompt and cocoon prompt. 

\begin{itemize}
\item \emph{Macronova optical/IR follow-up:} 
The quasi-thermal emission from the decay of unstable r-process elements in the wind/dynamical ejecta peaks in the optical/IR bands. We compute $\dot{N}_{EM,A1/2}$ for typical optical/IR sensitivities $S_{th,OIR} \sim 10^{-15}-10^{-13}\ {\rm Jy\ Hz}$ (see Table \ref{Table2}) and ejecta parameters: $\kappa \sim 10\ {\rm cm^2\ g^{-1}}$, $\theta_{ej,dyn} \sim 0.1745\ (10^{\circ})$, $\phi_{ej,dyn} \sim \pi$ and $f \sim 3\times10^{-6}$ (see Table \ref{Table3}).\\

\item \emph{sGRB and cocoon prompt gamma/X-ray follow-up:} \\
The short-duration sGRB prompt and cocoon prompt emission peak in the gamma/X-ray bands. We estimate the sGRB/cocoon prompt follow-up rate $\dot{N}_{EM,B/C}$ for typical gamma/X-ray sensitivities $S_{th,GX} \sim 10^{-8}-10^{-6}\ {\rm Jy\ Hz}$ (see Table \ref{Table2}). We use characteristic accretion disk parameters: $\alpha \sim 0.03$, $R \sim 100\ {\rm km}$, $M_{BH} \sim 8\ M_{\odot}$ and $H \sim R \approx 100\ {\rm km}$ with a jet-opening angle $\theta_{jet} = 1/\Gamma_{jet} \approx 0.1$. The hot mildly-relativistic cocoon surrounding the jet has energy $E_c \sim \eta_c (\Gamma_{jet}/\Gamma_{coc}) (\eta_{jet}M_{disk}c^2) \approx 3 \eta_{jet}M_{disk}c^2 = (5.4\times10^{48}\ {\rm erg}) \eta_{jet,-4}M_{disk,-2}$ and opening angle $\theta_{coc} = 1/\Gamma_{coc} = 3\theta_{jet}$, where $\eta_c \sim 1$ is the fraction of the jet energy that is transferred to the cocoon. We consider $\Gamma_{coc} = (1/3)\Gamma_{jet}$ and $R_a = 10^8\ {\rm cm}$ as the typical cocoon parameters.  \\
\end{itemize}

\item \emph{Late-time emission:} The late-time emission consists of synchrotron radio and optical afterglows from the interaction of the dynamical ejecta, wind, jet and cocoon components with the ambient ISM. The $t_{peak}$ and $L_{peak}$ values for the afterglow emission from a particular ejecta component critically depends on its mass and velocity along with the geometry (see Table \ref{Table1}). Here we assume $n=0.01\ {\rm cm^{-3}}$, $\epsilon_{e} = \epsilon_{B} = 0.1$, $p=2.5$ and $\nu_{obsr} = 1\ {\rm GHz}$ to estimate the radio emission detection rates for each ejecta component (see Appendix \ref{EM_opt} for optical emission detection rates). The typical radio band sensitivity varies within $S_{th,R} \sim 10^{-21} - 10^{-18}\ {\rm Jy\ Hz}$ (see Table \ref{Table2}).

We consider $\theta_{ej,dyn} \sim 0.1745$ and $\phi_{ej,dyn} \sim \pi$ for the anisotropic \emph{sub-relativistic} dynamical ejecta. A significant fraction $\eta_{wind}$ of the disk mass can also become unbound due to the spherically symmetric \emph{sub-relativistic} magnetic field/viscous flow/neutrino-driven winds and contribute to the afterglow emission. The wind afterglow follow-up detection rate in radio band is significantly smaller compared to the other ejecta components for typical ejecta parameters and telescope sensitivities, thereby making it very challenging to detect. 

The mass of the \emph{ultra-relativistic} jet can be written as $M_{jet} = \eta_{jet} M_{disk}$, where $\eta_{jet}$ is the fraction of the remnant accretion disk mass contributing to the jet kinetic energy. As the jet propagates outwards with highly relativistic velocities $v_{jet} \approx c$, the emission is significantly anisotropic with $\theta_{jet} = 1/\Gamma_{jet}$ and $\phi_{jet} = 2\pi$. We assume $\Gamma_{jet} \sim 10$ to be the typical jet Lorentz factor to compute the EM follow-up rate. We consider $\eta_{disk}/\eta_{dyn} \sim 0.1$ for the \emph{mildly relativistic} anisotropic cocoon generated by the interaction of the outgoing ultra-relativistic jet with the previously ejected material (see Section \ref{Cocoon_prompt}). The cocoon Lorentz factor $\Gamma_{coc}$ is expected to be directly correlated with $\Gamma_{jet}$ and here we assume a typical value of $\Gamma_{coc} \sim (1/3) \Gamma_{jet}$. For $\Gamma_{jet} \sim 10-30$ (see Table \ref{Table1}), $\Gamma_{jet} \sim 3-10$ with a cocoon velocity $v_{coc} \approx c$. The emission from the relativistic cocoon is anisotropic with $\theta_{coc} = 1/\Gamma_{coc} = 3\theta_{jet}$ and $\phi_{coc} = 2\pi$.

\end{itemize}

\subsection{Effect of binary parameters on the EM follow-up rates}
\label{EM_rates}
Here we study the effect of the BHNS binary parameters ($M_{BH}$, $a_{BH}$ and NS EoS) on the $L_{peak}$ and $t_{peak}$ values of the EM signals accompanying the GW trigger. 
We determine the range of binary parameters for which NS tidal disruption is favoured and the accompanying EM signals are detectable with the current telescope sensitivities.
Figure \ref{fig4} shows the effect of binary parameters on the $L_{peak}$ and $t_{peak}$ values of the EM counterparts for two different cases: (a) \emph{less NS disruption} for APR4 $R_{NS} = 11.1\ {\rm km}$, $M_{BH} = 8.2\ M_{\odot}$ and $a_{BH} = 0.8$, and (b) \emph{more NS disruption} for H4 $R_{NS} = 13.6\ {\rm km}$, $M_{BH} = 7.4\ M_{\odot}$ and $a_{BH} = 0.97$. 
For less (more) NS disruption, $M_{ej,tot}/M_{tot} = 1.736\%\ (6.155\%)$ with $M_{ej,dyn}/M_{tot} = 0.003\%\ (1.190\%)$, $M_{disk}/M_{tot} = 1.733\%\ (4.965\%)$ and $v_{ej,dyn} = 0.284c\ (0.275c)$.
We normalize the values of the ejecta parameters and the observing band sensitivities as in Section \ref{EM_prompt_radio} for early time emission and radio afterglow components, and Appendix \ref{EM_opt} for optical afterglow components. 

The early time sGRB prompt emission lasts for $t_{peak,B} = (1.88-1.98)\times10^{-1}\ {\rm s} \lesssim {\rm sec}$ with a peak luminosity $L_{peak,B} = (2.21-5.52)\times10^{48}\ {\rm erg/s}$ in gamma/X-ray. 
The cocoon prompt emission peaks around $t_{peak,C} = (1.96-3.17)\times10^{1}\ {\rm s} \sim {\rm sec-min}$ with a peak luminosity $L_{peak,C} = (0.62-1.18)\times10^{48}\ {\rm erg/s}$ in gamma/X-ray. 
The macronova emission from dynamical ejecta/wind peaks at a later time with $t_{peak,A1/A2} = (0.38-7.27)/(1.22-1.97)\ {\rm day} \sim {\rm hours-week}$ and a peak optical/IR luminosity $L_{peak,A1/A2} = (0.49-9.16)\times10^{41}/(0.87-1.41)\times10^{40}\ {\rm erg/s}$. 
Although the macronova emission peaks with significantly smaller luminosities compared to both sGRB and cocoon prompt, it has comparable detection rates in the optical/IR bands and is a relatively easy follow-up target due to its peak times of $\sim {\rm hours-week}$. The late time afterglow emission from the ultra-relativistic jet peaks at $t_{peak,D3} = (0.75-1.04)\times10^{1}\ {\rm days}$ with radio/optical luminosity $L_{peak,D3r/o} = (2.18-5.73)\times10^{36}/(0.98-1.86)\times10^{38}\ {\rm erg/s}$. 
The mildly-relativistic cocoon afterglow peaks at longer timescales with $t_{peak,D4} = (4.71-9.71)\times10^{1}\ {\rm days} \sim {\rm month}$ with a radio/optical luminosity $L_{peak,D4r/o} = (0.22-1.95)\times10^{37}/(1.43-6.07)\times10^{38}\ {\rm erg/s}$. 
The afterglow from sub-relativistic dynamical ejecta peaks around $t_{peak,D1} = (0.18-1.34)\times10^{2}\ {\rm years} \sim {\rm decade}$ with luminosity in radio/optical band $L_{peak,D1r/o} = (0.01-2.97)\times10^{37}/(0.03-1.31)\times10^{38}\ {\rm erg/s}$. 
The late time afterglow from the sub-relativistic wind component peaks with similar timescales of $t_{peak,D2} = (2.45-3.38)\times10^{1}\ {\rm years} \sim {\rm decade}$ and radio/optical luminosity $L_{peak,D2r/o} = (0.71-1.87)\times10^{32}/(1.40-2.67)\times10^{34}\ {\rm erg/s}$.  

We propagate the error in $M_{ej,tot}$ and $M_{ej,dyn}$ obtained for both less and more NS disruption cases to estimate the error in $L_{peak}$ and $t_{peak}$ for each EM counterpart. 
We find that the gamma/X-ray $\Delta L_{peak,B} = (1.90-5.52)\times10^{47}\ {\rm erg/s}$ for sGRB prompt emission whereas the error in peak time and gamma/X-ray luminosity for cocoon prompt emission are $\Delta t_{peak,C} = (0.84-1.58)\ {\rm s}$ and $\Delta L_{peak,C} = (3.54-7.91)\times10^{46}\ {\rm erg/s}$, respectively. 
While the error in the peak time for macronova emission from dynamical ejecta/wind is $\Delta t_{peak,A1/A2} = (0.40-3.24)/(0.05-0.10)\ {\rm day}$, the corresponding error in the optical/IR peak luminosity is $\Delta L_{peak,A1/A2} = (0.51-4.22)\times10^{41}/(3.75-7.05)\times10^{38}\ {\rm erg/s}$. 
The jet afterglow peak time and radio/optical peak luminosity error are relatively small with $\Delta t_{peak,D3} = (0.22-0.34)\ {\rm day}$ and $\Delta L_{peak,D3r/o} = (1.88-5.73)\times10^{35}/(0.84-1.86)\times10^{37}\ {\rm erg/s}$, while the corresponding quantities for the late-time cocoon emission are marginally higher with $\Delta t_{peak,D4} = (1.84-3.30)\ {\rm days}$ and $\Delta L_{peak,D4r/o} = (0.26-1.99)\times10^{36}/(1.67-6.19)\times10^{37}\ {\rm erg/s}$. 
The magnitude of the error in the ejecta masses implies that the error in dynamical ejecta afterglow peak time and radio/optical luminosity are significantly large with $\Delta t_{peak,D1} = (0.05-1.04)\times10^{2}\ {\rm years}$ and $\Delta L_{peak,D1r/o} = (1.66-3.27)\times10^{36}/(1.44-5.15)\times10^{37}\ {\rm erg/s}$. Similarly, the error in the peak quantities for the wind afterglow emission are $\Delta t_{peak,D2} = (0.71-1.12)\ {\rm year}$ and $\Delta L_{peak,D2r/o} = (0.61-1.87)\times10^{31}/(1.20-2.67)\times10^{33}\ {\rm erg/s}$.
 
We further evaluate the EM follow-up rates $\dot{N}_{EM}$ from the $L_{peak}$ values of different counterparts and for given bolometric sensitivities $S_{th}$. Figures \ref{fig6}, \ref{fig9}, \ref{fig7} and \ref{fig8} show the effect of BHNS binary parameters $M_{BH}$, $a_{BH}$ and NS EoS on the EM follow-up rate for the early-time and late-time EM emissions from the \emph{anisotropic sub-relativistic} dynamical ejecta, the \emph{isotropic sub-relativistic} winds, the \emph{anisotropic ultra-relativistic} sGRB jet and the \emph{anisotropic mildly-relativistic} cocoon, respectively. The Monte Carlo simulation results are shown for 300 BHNS binaries with gaussian $M_{BH} = (7.8\pm1.2)\ M_{\odot}$, $a_{BH} = {\rm Uniform}(0,0.97)$ and APR4/H4 NS EoS. 

Figure \ref{fig6} shows the contour plots for the EM follow-up rates of early-time macronova and late-time radio afterglow emission from the dynamical ejecta for APR4 and H4 NS EoS. 
For the soft APR4 NS EoS, the ejecta macronova emission has an optical/IR follow-up rate of $0.10 \leq \dot{N}_{A1}/\dot{N}_{GW} \leq 0.30$ for $0.75 \leq a_{BH} \leq 1.0$ and $6.0 \leq M_{BH}/M_{\odot} \leq 10.0$. However, the optical/IR follow-up rate is larger by a factor of $\sim$2 for the stiffer H4 EoS, with $0.22 \leq \dot{N}_{A1}/\dot{N}_{GW} \leq 0.60$ for $0.7 \leq a_{BH} \leq 1.0$ and $5.0 \leq M_{BH}/M_{\odot} \leq 11.0$.
The radio afterglow emission from the dynamical ejecta is considerable for $0.75 \leq a_{BH} \leq 1.0$ ($0.6 \leq a_{BH} \leq 1.0$) and $6.5 \leq M_{BH}/M_{\odot} \leq 10.0$ ($6.0 \leq M_{BH}/M_{\odot} \leq 11.0$) with $0.18 \leq \dot{N}_{D1r}/\dot{N}_{GW} \leq 0.50$ ($0.3 \leq \dot{N}_{D1r}/\dot{N}_{GW} \leq 0.9$) in the case of APR4 (H4) EoS. 
The radio afterglow follow-up rates are larger by a factor of $\sim$1.5-2 for the stiffer H4 EoS compared to the softer APR4 NS EoS due to more ejecta from NS disruption. Figure \ref{fig9} shows the contour plots for the follow-up rates of early-time wind macronova and late-time wind radio afterglow for APR4 and H4 NS EoS. The wind macronova has a significantly smaller optical/IR follow-up rate compared to the ejecta macronova with $4.0\times10^{-4} \leq \dot{N}_{A2}/\dot{N}_{GW} \leq 10^{-3}$ for $0.5 \leq a_{BH} \leq 0.8$ and $5.0 \leq M_{BH}/M_{\odot} \leq 10.0$ in the region where $15.0a_{BH} - M_{BH} \geq 2.5$ and for APR4 NS EoS. The optical/IR follow-up rate is similar for H4 NS EoS with $4.0\times10^{-4} \leq \dot{N}_{A2}/\dot{N}_{GW} \leq 1.1\times10^{-3}$ for $0.2 \leq a_{BH} \leq 0.7$ and $5.0 \leq M_{BH}/M_{\odot} \leq 10.5$ in the region where $M_{BH} - 12.0a_{BH} \leq 2.0$.
The radio afterglow emission from the wind ejecta component has a negligibly small follow-up detection rate: for APR4 NS EoS, $0.85\times10^{-8} \leq \dot{N}_{D2r}/\dot{N}_{GW} \leq 2.35\times10^{-8}$ when $0.7 \leq a_{BH} \leq 0.9$ and $5.0 \leq M_{BH}/M_{\odot} \leq 10.0$ whereas for H4 NS EoS, $1.0\times10^{-8} \leq \dot{N}_{D2r}/\dot{N}_{GW} \leq 3.0\times10^{-8}$ when $0.55 \leq a_{BH} \leq 0.85$ and $4.5 \leq M_{BH}/M_{\odot} \leq 10.5$ in the region where $22.0a_{BH} - M_{BH} \geq 8.7$. Due to the significantly small $L_{peak,D2r}$ and consequently $\dot{N}_{D2r}$, any subsequent radio follow-up of GW triggers using wind afterglow emission is extremely difficult with the current telescope sensitivities (see Table {\ref{Table2}). 

The contour plots for the EM follow-up rates of early-time prompt and late-time radio afterglow emission from the sGRB jet for APR4 and H4 EoS are shown in figure \ref{fig7}. 
The jet prompt emission has a follow-up rate of $0.18 \leq \dot{N}_{B}/\dot{N}_{GW} \leq 0.50$ for $0.8 \leq a_{BH} \leq 1.0$ and $6.0 \leq M_{BH}/M_{\odot} \leq 10.0$ in case of APR4 EoS. The corresponding rate is larger by a factor of $\sim 1.5$ for H4 NS EoS with $0.28 \leq \dot{N}_{B}/\dot{N}_{GW} \leq 0.78$ for $0.8 \leq a_{BH} \leq 1.0$ and $6.0 \leq M_{BH}/M_{\odot} \leq 11.0$. 
For APR4 EoS, the jet radio afterglow has follow-up rate $0.045 \leq \dot{N}_{D3r}/\dot{N}_{GW} \leq 0.125$ for $0.7 \leq a_{BH} \leq 1.0$ and $5.0 \leq M_{BH}/M_{\odot} \leq 10.0$. However, the radio afterglow follow-up rate for H4 EoS is slightly larger with $0.06 \leq \dot{N}_{D3r}/\dot{N}_{GW} \leq 0.16$ for $0.7 \leq a_{BH} \leq 1.0$ and $4.5 \leq M_{BH}/M_{\odot} \leq 10.5$. 
Figure \ref{fig8} shows the EM follow-up rate contour plots for the early-time cocoon prompt and the late-time cocoon radio afterglow for APR4 and H4 NS EoS. 
For the softer APR4 EoS, the cocoon prompt emission has rate $0.14 \leq \dot{N}_{C}/\dot{N}_{GW} \leq 0.36$ for $0.7 \leq a_{BH} \leq 1.0$ and $6.0 \leq M_{BH}/M_{\odot} \leq 10.0$. The gamma/X-ray follow-up rate for H4 EoS is similar with $0.15 \leq \dot{N}_{C}/\dot{N}_{GW} \leq 0.42$ for $0.6 \leq a_{BH} \leq 1.0$ and $4.5 \leq M_{BH}/M_{\odot} \leq 11.0$.  
The cocoon afterglow has a rate of $0.1 \leq \dot{N}_{D4r}/\dot{N}_{GW} \leq 0.3$ for $0.75 \leq a_{BH} \leq 1.0$, $6.0 \leq M_{BH}/M_{\odot} \leq 10.0$ and APR4 EoS. The follow-up rate for H4 EoS is larger by a factor of $\sim 3$ with $0.3 \leq \dot{N}_{D4r}/\dot{N}_{GW} \leq 0.82$ for $0.7 \leq a_{BH} \leq 1.0$ and $6.0 \leq M_{BH}/M_{\odot} \leq 10.5$.

\section{Discussions and Conclusions}
\label{disc_conc}
In this paper, we studied the effect of BHNS binary parameters on the properties of the ejected material as well as their associated EM emission. The tidal disruption of the NS and the outcome of the subsequent BHNS merger are primarily determined by $M_{BH}$, $a_{BH}$ and the EoS of the in-falling NS. While the majority of the tidally disrupted NS mass gets directly accreted onto the BH on short viscous timescales, a considerable fraction of the ejecta with mass $M_{disk}$ forms a remnant disk around the BH as the remaining ejecta material with mass $M_{ej,dyn}$ becomes gravitationally unbound from the compact remnant. We consider $M_{BH}$ and $a_{BH}$ distributions motivated from transient LMXB observations in order to estimate the ejecta masses $M_{disk}$ and $M_{ej,dyn}$. 
We find that $M_{disk}$ and $M_{ej,dyn}$ are practically unaffected by the specific $M_{BH}$ distribution considered (see Appendix \ref{dist_sims}), provided that the mean BH masses are similar whereas both the ejecta masses increase with the increase in BH spin magnitude. The ejecta masses are expectedly larger for the stiffer H4 NS EoS due to larger disruption.   

We obtain the aLIGO design strain sensitivity as a function of detection frequency to estimate the detection horizon volume and thereby the event detection rates for a BHNS binary with a given orientation using the coalescence rates from population synthesis models. Although the GW event detection rates can vary by about three orders of magnitude around the mean value depending on the poorly constrained BHNS coalescence rate, the mean rate for a fixed binary configuration
is practically independent of the choice of $M_{BH}$ and $a_{BH}$ distributions and is unaffected by the NS EoS. In addition to being promising sources of GWs detectable by aLIGO, BHNS binaries are expected to have significant EM emission providing further information about the properties of the ejecta material as well as the post-merger remnant. 

We discuss the possible EM counterparts for these BHNS mergers and estimate their peak emission parameters. The EM emission is primarily due to four ejecta components: the \emph{ultra-relativistic} sGRB jet, the \emph{mildly-relativistic} cocoon, the \emph{sub-relativistic} dynamical ejecta and the \emph{sub-relativistic} magnetic/viscous/neutrino-driven winds. The early-time emission consists of the jet prompt, cocoon prompt and macronova whereas the late-time emission comprises of synchrotron afterglow from the jet, cocoon, dynamical ejecta and the wind. In contrast to the isotropic emission from the wind, the emissions from the jet, cocoon and dynamical ejecta components are expected to be fairly anisotropic. 

The low latency EM follow-up observations of GW triggers helps us in constraining the EoS of neutron degenerate matter and jet+ejecta properties post-merger in addition to improving the source localisation. The EM follow-up volume in all observing bands is restricted to the GW detection horizon volume dictated by the aLIGO sensitivity and makes it easier for the transient search in a smaller sky area compared to the blind surveys. 
We perform Monte Carlo simulations for 300 BHNS binaries with gaussian $M_{BH} = (7.8\pm1.2)\ M_{\odot}$ and $a_{BH} = {\rm Uniform}(0,0.97)$ distributions and APR4/H4 NS EoS in order to estimate the follow-up rates of the EM counterparts in terms of the binary parameters. 
The EM follow-up rates for the stiffer H4 NS EoS are larger compared to the corresponding rates for APR EoS by a factor of $\sim 1.5-2$ and $\sim 1.5-3$ for early-time and late-time emissions, respectively. Almost all EM counterparts are detectable for typically small $M_{BH}$ and large $a_{BH}$ leading to significant NS disruption and subsequent stronger EM emission from the ejecta material. The wind macronova and afterglow detectability are especially challenging with the current radio telescope sensitivities due to the small isotropic luminosity of this sub-relativistic component. 

Below we summarize the main results of this work:

(i) The mass ejected in a BHNS merger event is primarily determined by the binary parameters: mass ratio $q = M_{BH}/M_{NS}$, magnitude of $a_{BH}$, and $R_{NS}$ based on the NS EoS. 
We find that both $M_{ej,tot}$ and $M_{ej,dyn}$ increase as NS disruption increases with a decrease in $M_{BH}$ (regardless of the choice of the distribution, provided the mean mass is similar), increase in $a_{BH}$ and stiffer NS EoS (see Figures \ref{fig1} and \ref{fig2}). 
Due to the smaller dynamical ejecta masses and merger rates for typical BHNS mergers compared to BNS mergers, the BHNS mergers are estimated to synthesise only about $0.1\%$ of the total r-process elements found in the universe.
The \emph{ultra-relativistic} sGRB jet, the \emph{mildly-relativistic} cocoon, the \emph{sub-relativistic} dynamical ejecta and the \emph{sub-relativistic} magnetic/viscous/neutrino-driven winds are the different BHNS merger ejecta components, each associated with some early- and/or late-time EM emission. 
Figure \ref{fig0} shows a schematic sketch of all the ejecta components along with their characteristic emission and Table \ref{Table1} lists their properties. 

(ii) For the typical ejecta parameters and range of BHNS binary parameters allowing NS disruption (see Figure \ref{fig4}), we estimate the EM peak luminosities and emission timescales for both early- and late-time emissions.\\ 
The early-time emissions are in chronological order:

1. sGRB jet prompt emission in gamma/X-ray ($\nu_{obs} \sim 10^{20}\ {\rm Hz}$) with $t_{peak,B} \lesssim {\rm sec}$ and $L_{peak,B} \sim 10^{48-49}\ {\rm erg/s}$ until $t \lesssim {\rm min}$ and $\propto (t/{\rm min})^{-3}$ afterwards,

2. cocoon prompt emission in gamma/X-ray with $t_{peak,C} \sim {\rm sec-min}$ and $L_{peak,C} \sim 10^{47-48}\ {\rm erg/s}$ until $t \approx t_{peak,C}$ and $\propto (t/t_{peak,C})^{-1}$ at later times, 

3. ejecta macronova emission in optical/IR ($\nu_{obs} \sim 5\times10^{5}\ {\rm GHz}$) with $t_{peak,A1} \sim {\rm day-week}$ and $L_{peak,A1} \sim 10^{40-42}\ {\rm erg/s}$,

4. wind macronova emission in optical/IR with $t_{peak,A2} \sim {\rm hour-day}$ and $L_{peak,A2} \sim 10^{39-40}\ {\rm erg/s}$. \\ 
The macronova luminosity for both dynamical ejecta and magnetic/viscous/neutrino-driven winds are roughly constant until $t \approx t_{peak}$ and fall off as $\propto (t/t_{peak})^{-1.3}$ at later times. \\
The corresponding order for the late-time radio synchrotron afterglow emissions with $\nu_{obs} \sim 1\ {\rm GHz}$ is:

1. jet afterglow with $t_{peak,D3} \sim {\rm day-week}$ and $L_{peak,D3} \sim 10^{36-38}\ {\rm erg/s}$. $L(t) \propto (t/t_{peak,D3})^{3}$ for $t \lesssim t_{peak,D3}$ whereas $L(t) \propto (t/t_{peak,D3})^{-1}$ for $t > t_{peak,D3}$,

2. cocoon afterglow with $t_{peak,D4} \sim {\rm week-month}$ and $L_{peak,D4} \sim 10^{36-39}\ {\rm erg/s}$. The cocoon luminosity steadily increases as $\propto (t/t_{peak,D4})^{3}$ for $t \lesssim t_{peak,D4}$ and drops as $\propto (t/t_{peak,D4})^{-1.5}$ at later times,

3. dynamical ejecta afterglow with $t_{peak,D1} \sim {\rm year-decade}$ and $L_{peak,D1} \sim 10^{35-38}\ {\rm erg/s}$,

4. wind afterglow with $t_{peak,D2} \sim {\rm year-decade}$ and $L_{peak,D2} \sim 10^{32-34}\ {\rm erg/s}$.\\ 
The afterglow luminosity for the sub-relativistic dynamical ejecta and wind components are expected to increase as $t^3$ upto $t \approx t_{peak}$ and then reduce as $t^{-1.65}$ at later times. 

(iii) We estimate the follow-up detection rates of all early- and late-time EM counterparts for representative bolometric sensitivities: optical/IR $S_{th} \sim 2\times10^{-14}\ {\rm Jy\ Hz}$, gamma/X-ray $S_{th} \sim 10^{-7}\ {\rm Jy\ Hz}$ and radio $S_{th} \sim 3\times10^{-19}\ {\rm Jy\ Hz}$ (see Figures \ref{fig6}-\ref{fig8}, \ref{fig10} and \ref{fig11}). We find that the early-time sGRB prompt gamma/X-ray, cocoon prompt gamma/X-ray, ejecta macronova optical/IR and wind macronova optical/IR follow-up rates are $\dot{N}_{B} \sim (0.2-0.8)\ \dot{N}_{GW}$, $\dot{N}_{C} \sim (0.1-0.4)\ \dot{N}_{GW}$, $\dot{N}_{A1} \sim (0.1-0.6)\ \dot{N}_{GW}$ and $\dot{N}_{A2} \sim (0.4-1.1)\times10^{-3}\ \dot{N}_{GW}$, respectively, for BH mass range $6.0\ M_{\odot} \lesssim M_{BH} \lesssim 10.0\ M_{\odot}$ and spin range $0.7 \lesssim a_{BH} \lesssim 1.0$. The late-time synchrotron afterglow radio follow-up rates for the jet, cocoon, dynamical ejecta and wind ejecta components are $\dot{N}_{D3r} \sim (0.04-0.16)\ \dot{N}_{GW}$, $\dot{N}_{D4r} \sim (0.10-0.82)\ \dot{N}_{GW}$, $\dot{N}_{D1r} \sim (0.18-0.90)\ \dot{N}_{GW}$ and $\dot{N}_{D2r} \sim (0.85-3.00)\times10^{-8}\ \dot{N}_{GW}$, respectively, for $6.0\ M_{\odot} \lesssim M_{BH} \lesssim 10.0\ M_{\odot}$ and $0.7 \lesssim a_{BH} \lesssim 1.0$. 

(iv) The low-latency EM follow-up rate $\dot{N}_{EM} \propto D_{EM}^{3}$ (from equation \ref{NEM}) for each counterpart. Based on the follow-up rates for $6\ M_{\odot} \lesssim M_{BH} \lesssim 10\ M_{\odot}$ and $0.7  \lesssim a_{BH} \lesssim 1.0$, the relative ease of follow-up of BHNS EM counterparts for typical ejecta parameters and telescope band sensitivities is in the order: ejecta afterglow (radio) $>$ cocoon afterglow (radio) $\gtrsim$ jet prompt (gamma/X-ray) $>$ ejecta macronova (optical/IR) $>$ cocoon prompt (gamma/X-ray) $>$ jet afterglow (radio) $>>$ wind macronova (optical/IR) $>>$ wind afterglow (radio).

\begin{appendix}

\section{Effect of binary parameters on ejecta masses}
Here we discuss the effect of the BHNS binary parameters ($M_{BH}$, $a_{BH}$ and NS EoS) on the remnant disk and the dynamical ejecta masses.
Figure \ref{fig1} shows the effect of $M_{BH}$ distribution on $M_{disk}$ and $M_{ej,dyn}$ for $a_{BH}=0.97$ and APR4/H4 NS EoS. 
For APR4 EoS, we find that $M_{disk} \sim 2.26\times10^{-1}-4.15\times10^{-1}\ M_{\odot}$ with similar mean mass $\sim 3.5\times10^{-1}\ M_{\odot}$ for all $M_{BH}$ distributions. The dispersion in $M_{disk}$ is larger in case of power-law distribution ($3.2\times10^{-2}\ M_{\odot}$) as compared to exponential ($2.8\times10^{-2}\ M_{\odot}$) and gaussian ($2.6\times10^{-2}\ M_{\odot}$) distributions. 
In case of H4 EoS, $M_{disk} \sim 2.5\times10^{-1}-4.6\times10^{-1}\ M_{\odot}$ with mean mass $\sim 4.3\times10^{-1}\ M_{\odot}$ for all $M_{BH}$ distributions.

\label{dist_sims}
\begin{figure*}
    \begin{subfigure}[tp]{0.49\linewidth}
    \centering
    \includegraphics[height=0.65\linewidth,width=0.85\linewidth]{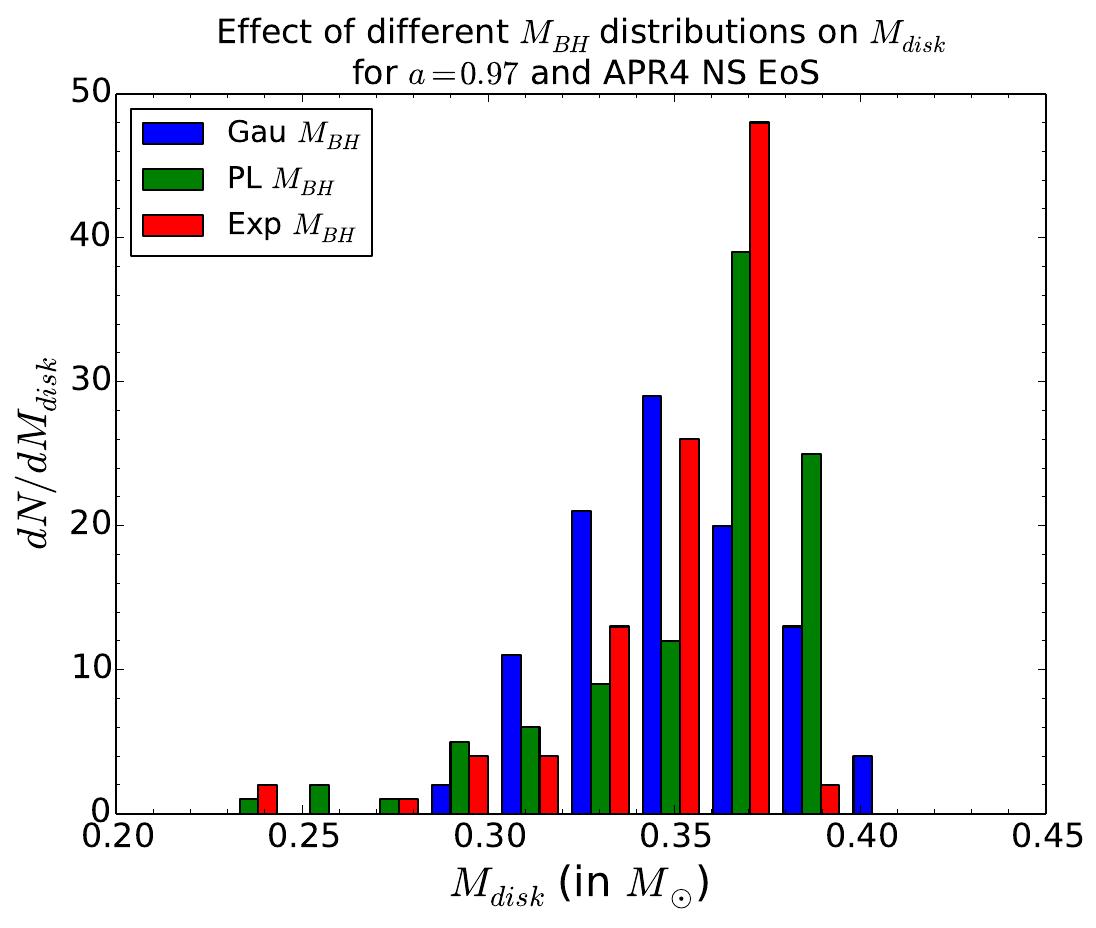} 
  \end{subfigure}
  \begin{subfigure}[tp]{0.49\linewidth}
    \centering
    \includegraphics[height=0.65\linewidth,width=0.85\linewidth]{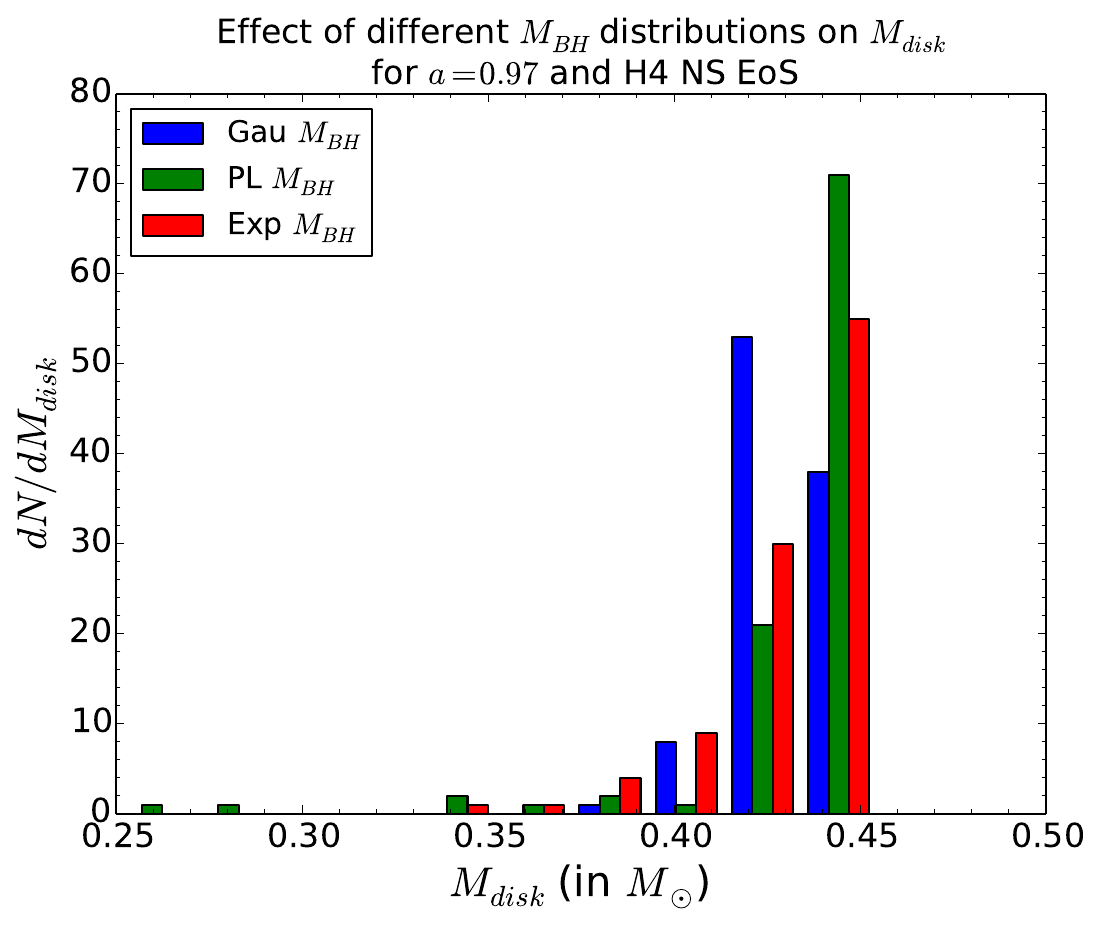}  
  \end{subfigure} \\ 
  \begin{subfigure}[tp]{0.49\linewidth}
    \centering
    \includegraphics[height=0.65\linewidth,width=0.85\linewidth]{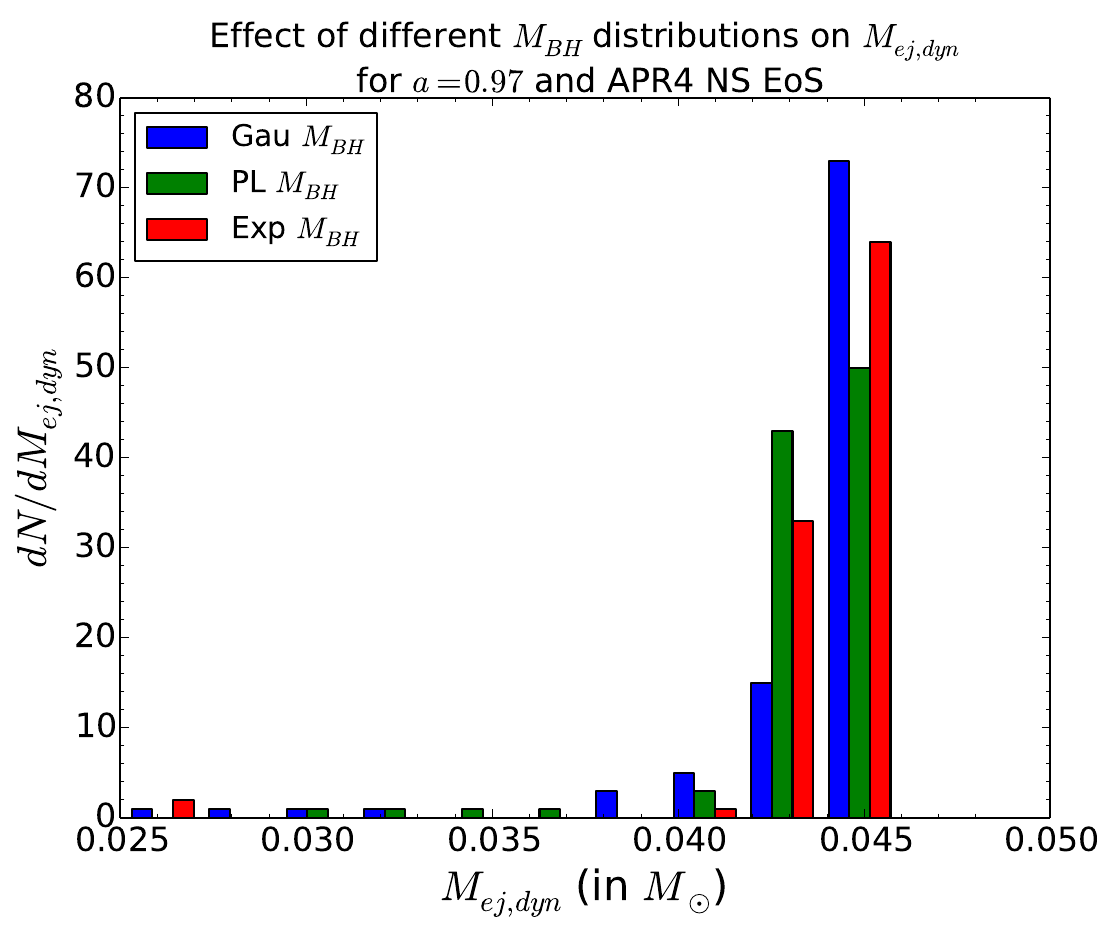} 
  \end{subfigure} 
    \begin{subfigure}[tp]{0.49\linewidth}
    \centering
    \includegraphics[height=0.65\linewidth,width=0.85\linewidth]{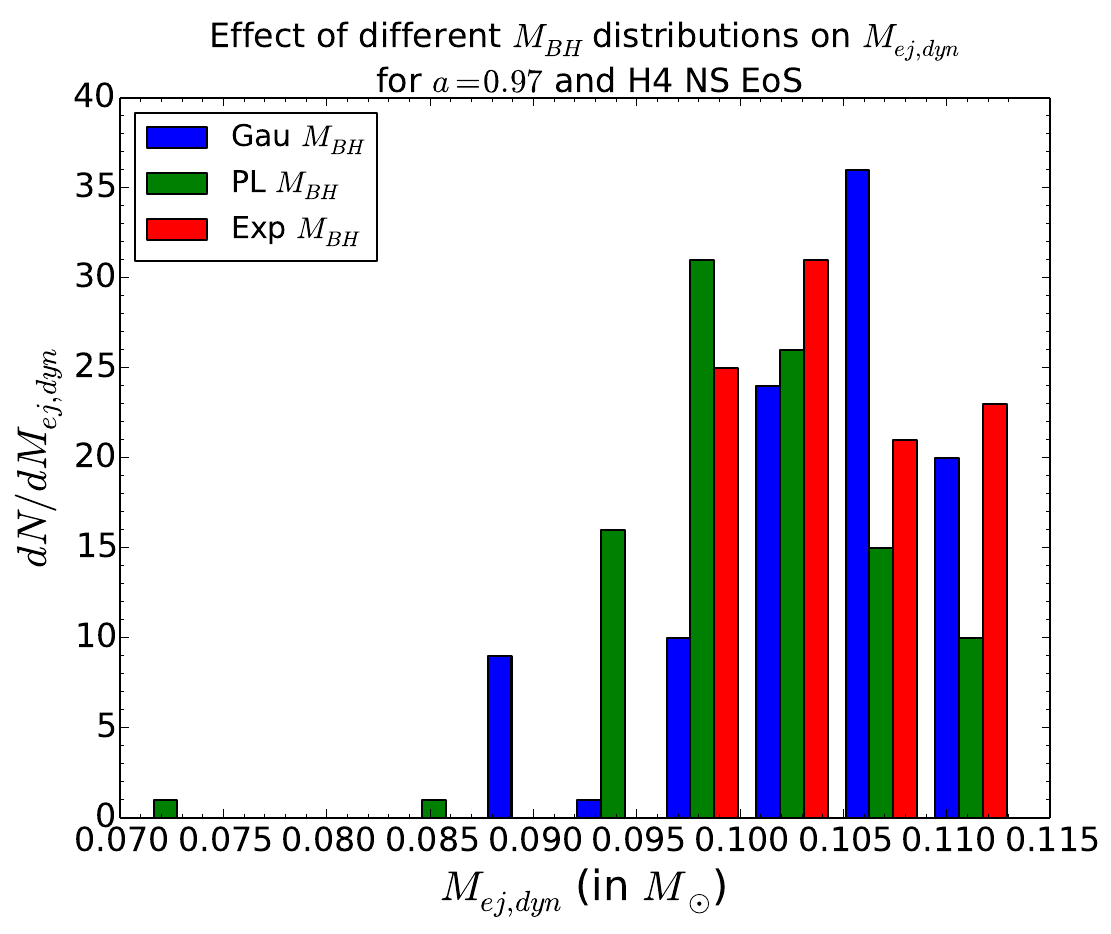} 
  \end{subfigure} \\
  \caption{\emph{Effect of $M_{BH}$ distribution on the bound disk mass $M_{disk}$ and the unbound dynamical ejecta mass $M_{ej,dyn}$:} Simulation results for 100 BHNS binaries with $a_{BH}=0.97$, two different NS EoS (APR4 and H4) and three different $M_{BH}$ distributions (gaussian, power-law and exponential). The vertical axes show the number of data points in each mass bin.
  	{\it Top-left panel:} $M_{disk}$ distribution for APR4 EoS,
	{\it Top-right panel:} $M_{disk}$ distribution for H4 EoS,
	{\it Bottom-left panel:} $M_{ej,dyn}$ distribution for APR4 EoS,
	{\it Bottom-right panel:} $M_{ej,dyn}$ distribution for H4 EoS
 }
  \label{fig1} 
\end{figure*}

The dispersion in $M_{disk}$ for power-law $M_{BH}$ distribution ($3.1\times10^{-2}\ M_{\odot}$) is significantly larger compared to exponential ($1.9\times10^{-2}\ M_{\odot}$) and gaussian ($1.4\times10^{-2}\ M_{\odot}$) distributions.
For APR4 EoS, the dynamical ejecta mass $M_{ej,dyn}$ lies in the range $\sim 2.5\times10^{-2}-4.6\times10^{-2}\ M_{\odot}$ with mean mass $\sim 4.4\times10^{-2}\ M_{\odot}$ for all $M_{BH}$ distributions. The dispersion in $M_{ej,dyn}$ is in the order: gaussian ($3.4\times10^{-3}\ M_{\odot}$) $>$ exponential ($2.9\times10^{-3}\ M_{\odot}$) $>$ power-law ($2.5\times10^{-3}\ M_{\odot}$).
For the stiffer H4 NS EoS, $M_{ej,dyn} \sim 7.0\times10^{-2}-1.1\times10^{-1}\ M_{\odot}$ with equal mean mass $\sim 1.0\times10^{-1}\ M_{\odot}$ for all three distributions. The $M_{ej,dyn}$ dispersion for gaussian ($6.4\times10^{-3}\ M_{\odot}$) is larger compared to power-law ($6.1\times10^{-3}\ M_{\odot}$) and exponential ($5.0\times10^{-3}\ M_{\odot}$) distributions. 
As expected, the mean masses for both $M_{disk}$ and $M_{ej,dyn}$ are smaller for the softer APR4 NS EoS compared to the stiffer H4 NS EoS and are unaffected by the choice of the $M_{BH}$ distribution. The mass dispersion for $M_{disk}$ and $M_{ej,dyn}$ are relatively unaffected by the NS EoS.

\begin{figure*}
    \begin{subfigure}[tp]{0.49\linewidth}
    \centering
    \includegraphics[height=0.65\linewidth,width=0.85\linewidth]{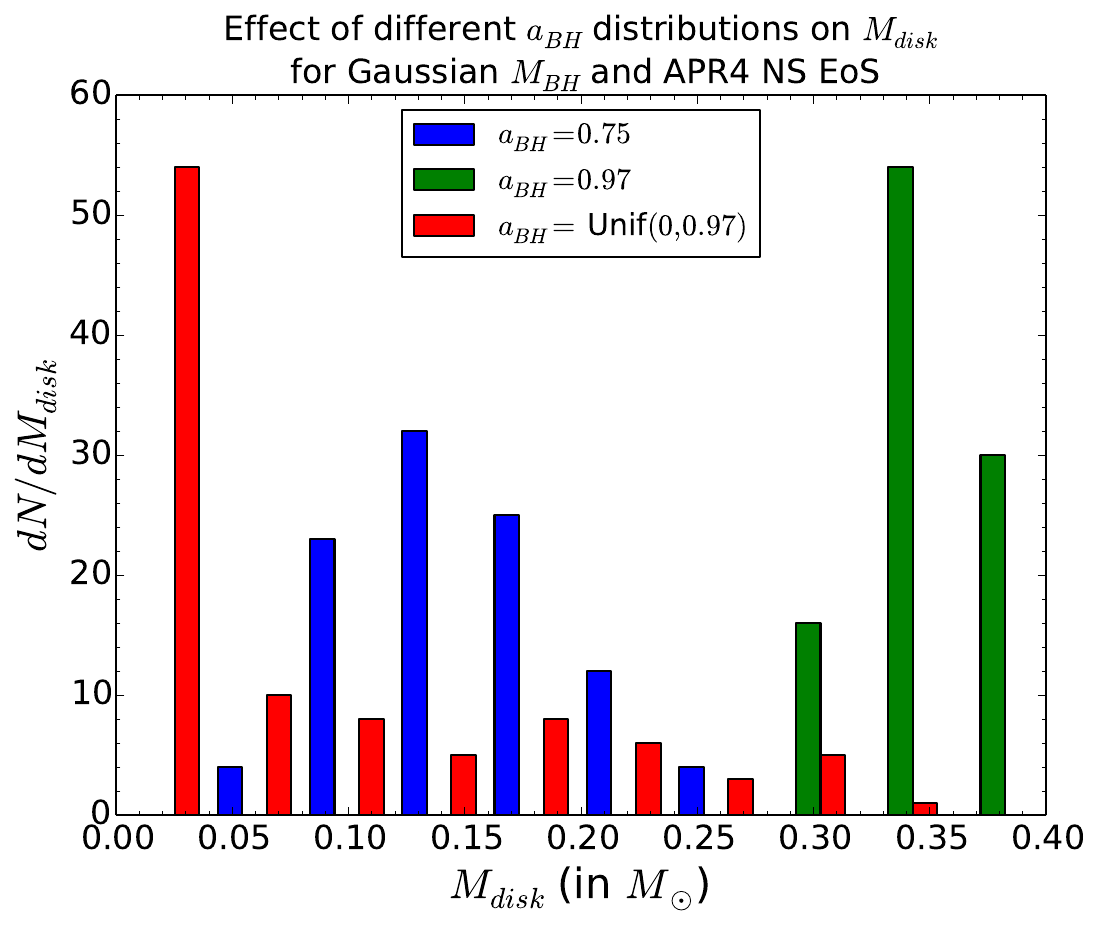} 
  \end{subfigure}
  \begin{subfigure}[tp]{0.49\linewidth}
    \centering
    \includegraphics[height=0.65\linewidth,width=0.85\linewidth]{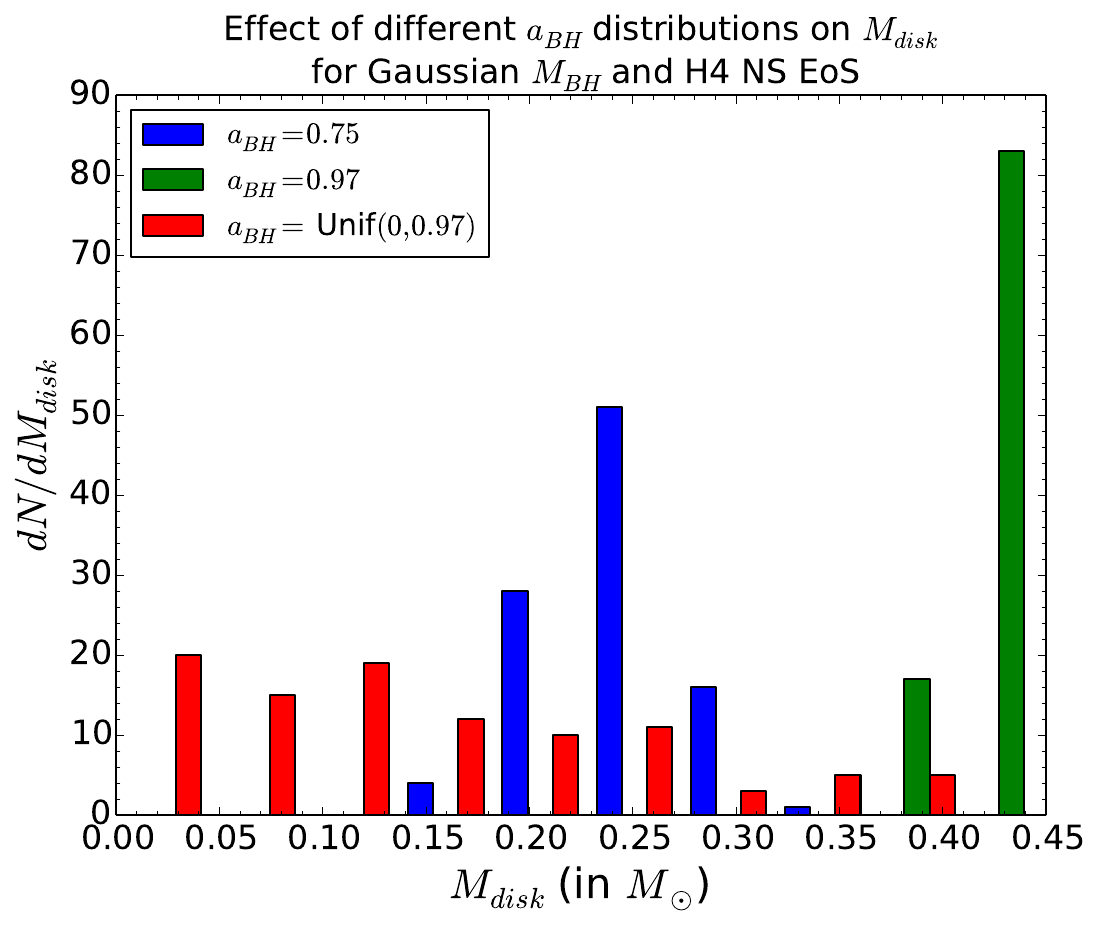}  
  \end{subfigure} \\ 
  \begin{subfigure}[tp]{0.49\linewidth}
    \centering
    \includegraphics[height=0.65\linewidth,width=0.85\linewidth]{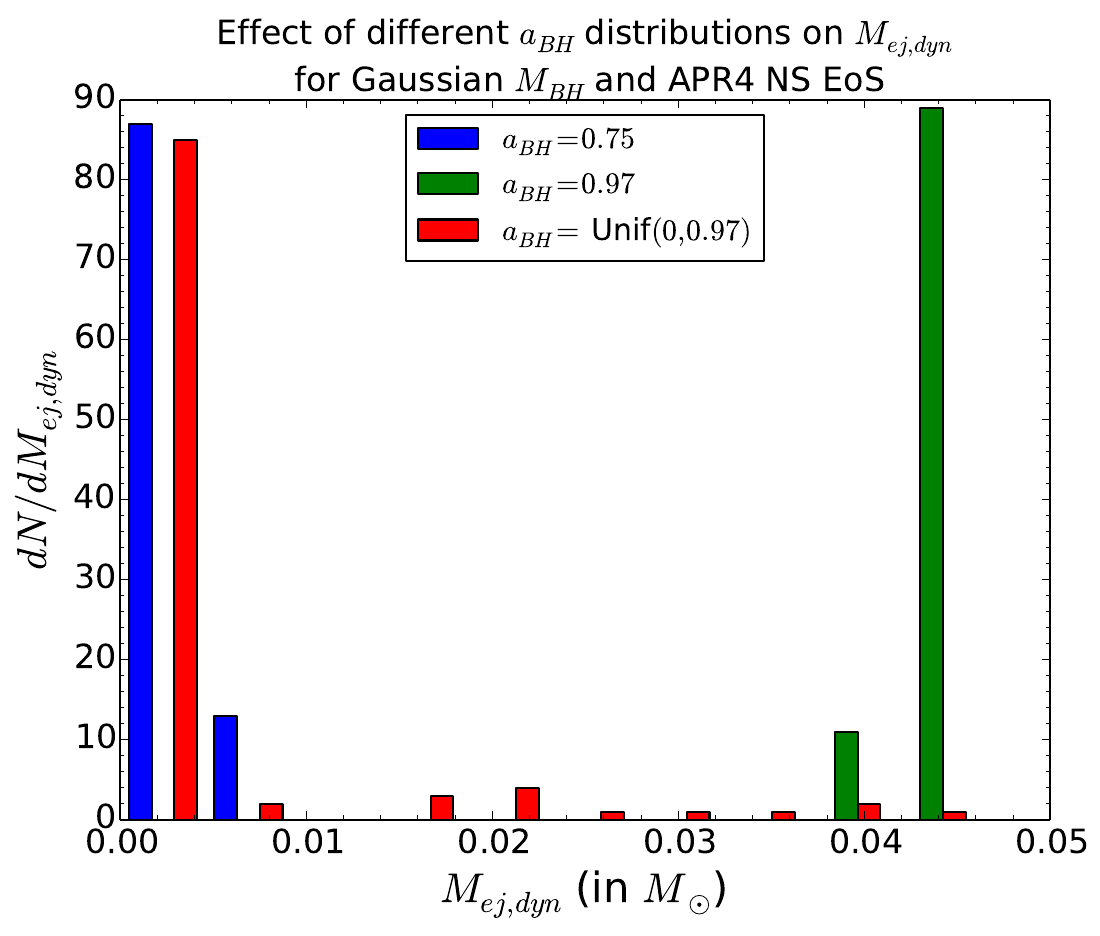} 
  \end{subfigure} 
    \begin{subfigure}[tp]{0.49\linewidth}
    \centering
    \includegraphics[height=0.65\linewidth,width=0.85\linewidth]{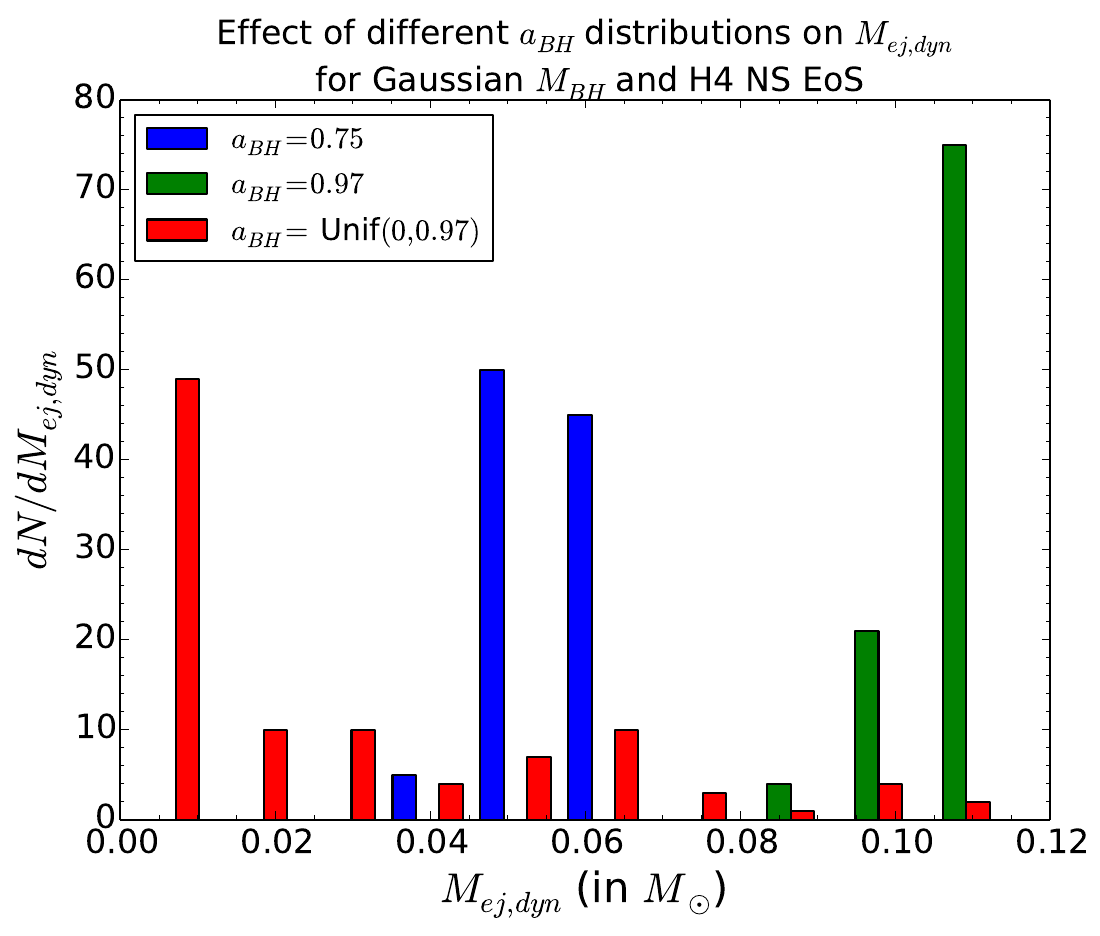} 
  \end{subfigure} \\
  \caption{\emph{Effect of $a_{BH}$ distribution on the bound disk mass $M_{disk}$ and the unbound dynamical ejecta mass $M_{ej,dyn}$:} Simulation results for 100 BHNS binaries with gaussian $M_{BH}$ distribution, two different NS EoS (APR4 and H4) and three different $a_{BH}$=0.75, 0.97 ${\rm and\ Uniform(0,0.97)}$.
  	{\it Top-left panel:} $M_{disk}$ distribution for APR4 EoS,
	{\it Top-right panel:} $M_{disk}$ distribution for H4 EoS,
	{\it Bottom-left panel:} $M_{ej,dyn}$ distribution for APR4 EoS,
	{\it Bottom-right panel:} $M_{ej,dyn}$ distribution for H4 EoS
 }
  \label{fig2} 
\end{figure*}

Figure \ref{fig2} shows the effect of $a_{BH}$ distribution on $M_{disk}$ and $M_{ej,dyn}$ for gaussian $M_{BH}$ and APR4/H4 NS EoS. 
For APR4 NS EoS, the bound disk mass $M_{disk}$ lies within the range $\sim 0 - 4.0\times10^{-1}\ M_{\odot}$ while the mean mass for $a_{BH} = {\rm Unif(0,0.97)}$ ($7.8\times10^{-2}\ M_{\odot}$) is smaller compared to that for $a_{BH} = 0.75$ ($1.5\times10^{-1}\ M_{\odot}$) and $a_{BH} = 0.97$ ($3.5\times10^{-1}\ M_{\odot}$). The $M_{disk}$ dispersion is in the order: $a_{BH} = {\rm Unif(0,0.97)}$ ($9.5\times10^{-2}\ M_{\odot}$) $>$ $a_{BH} = 0.75$ ($4.5\times10^{-2}\ M_{\odot}$) $>$ $a_{BH} = 0.97$ ($2.4\times10^{-2}\ M_{\odot}$). 
For stiffer H4 NS EoS, $M_{disk} \sim 0-4.6\times10^{-1}\ M_{\odot}$ with mean mass in the order: $a_{BH} = 0.97$ ($4.3\times10^{-1}\ M_{\odot}$) $>$ $a_{BH} = 0.75$ ($2.4\times10^{-1}\ M_{\odot}$) $>$ $a_{BH} = {\rm Unif(0,0.97)}$ ($1.5\times10^{-1}\ M_{\odot}$). The dispersion in $M_{disk}$ for $a_{BH} = {\rm Unif(0,0.97)}$ ($1.1\times10^{-1}\ M_{\odot}$) is larger than that of $a_{BH} = 0.75$ ($3.4\times10^{-2}\ M_{\odot}$) and $a_{BH} = 0.97$ ($1.5\times10^{-2}\ M_{\odot}$). 
The $M_{ej,dyn}$ for APR4 EoS lies in the range $\sim 0-4.6\times10^{-2}\ M_{\odot}$ with mean $M_{ej,dyn}$ in the order: $a_{BH} = 0.97$ ($4.4\times10^{-2}\ M_{\odot}$) $>$ $a_{BH} = {\rm Unif}(0,0.97)$ ($3.6\times10^{-3}\ M_{\odot}$) $>$ $a_{BH} = 0.75$ ($1.1\times10^{-3}\ M_{\odot}$). The mass dispersion also depends on the $a_{BH}$ distribution with: $a_{BH} = {\rm Unif}(0,0.97)$ ($9.4\times10^{-3}\ M_{\odot}$) $>$ $a_{BH} = 0.75$ ($2.2\times10^{-3}\ M_{\odot}$) $>$ $a_{BH} = 0.97$ ($1.9\times10^{-3}\ M_{\odot}$). 
For H4 NS EoS, $M_{ej,dyn} \sim 0-1.1\times10^{-1}\ M_{\odot}$ and mean $M_{ej,dyn}$ for $a_{BH} = 0.97$ ($1.0\times10^{-1}\ M_{\odot}$) is larger than that for $a_{BH} = 0.75$ ($5.5\times10^{-2}\ M_{\odot}$) and $a_{BH} = {\rm Unif}(0,0.97)$ ($2.6\times10^{-2}\ M_{\odot}$). The $M_{ej,dyn}$ dispersion is in the order: $a_{BH} = {\rm Unif(0,0.97)}$ ($3.0\times10^{-2}\ M_{\odot}$) $>$ $a_{BH} = 0.97$ ($6.2\times10^{-3}\ M_{\odot}$) $>$ $a_{BH} = 0.75$ ($4.5\times10^{-3}\ M_{\odot}$).
The mean masses for both $M_{disk}$ and $M_{ej,dyn}$ increase with the BH spin whereas the corresponding mass dispersions decrease with $a_{BH}$ magnitude. This is expected as a larger $a_{BH}$ results in more NS disruption and thereby larger $M_{disk}$ and $M_{ej,dyn}$ values. Similar to the varying $M_{BH}$ distribution case, the mean masses for $M_{disk}$ and $M_{ej,dyn}$ are smaller for APR4 EoS compared to H4 EoS while the mass dispersions are unaffected by the choice of NS EoS in general.

\section{Synchrotron emission characteristic frequencies}
\label{syn_char_freq}
The ejecta geometry does not influence either the magnitude of $B$ or the energy distribution of the non-thermal electrons. The distribution of Lorentz factor $\gamma_e$ of the accelerated electrons is expected to be a power-law, $dN_e/d\gamma_e \propto \gamma_e^{-p}$ for $\gamma_e > \gamma_{e,min} = (4.5\times10^{-3}) \epsilon_{e,-1} [(p-2)/(p-1)] (m_p/m_e) v_{ej,0.3}^2$, where $m_e$ is the electron mass and $\gamma_{e,min}$ is the minimum electron Lorentz factor. The field strength $B$ is written using the Rankine-Hugoniot relation, $B = \sqrt{9\pi \epsilon_B n m_p} v_{ej} = (2.0\times10^{-3}\ {\rm G}) \epsilon_{B,-1}^{1/2} n_{-2}^{1/2} v_{ej,0.3}$.  

The radio spectrum for synchrotron emission is characterised by the frequency of electrons $\nu_{min}$ with Lorentz factor $\gamma_{e,min}$ and the self-absorption frequency $\nu_a$. The frequency and specific flux in the absence of self-absorption corresponding to $\gamma_{e,min}$ are \citep{Kyu15}
\begin{eqnarray}
\nu_{min} = (3.6\times10^{5}\ {\rm Hz})\left(\frac{p-2}{p-1}\right)^{2} \epsilon_{e,-1}^{2} \epsilon_{B,-1}^{1/2} n_{-2}^{1/2} v_{ej,0.3}^5 \\
F_{\nu,min} = (2.2\ {\rm Jy}) \epsilon_{B,-1}^{1/2} n_{-2}^{1/2} M_{ej,0.03} v_{ej,0.3} D_2^{-2}
\label{nuFnumin}
\end{eqnarray}
The self-absorption frequency $\nu_a$ is determined by equating the Rayleigh blackbody flux with the unabsorbed flux $F_{\nu,min}$. The cooling frequency 
\begin{equation}
\nu_c = (3.2\times10^{13}\ {\rm Hz}) \epsilon_{B,-1}^{-3/2} n_{-2}^{-5/6} M_{ej,0.03}^{-2/3} v_{ej,0.3}^{-1}  D_2^{-2}\theta_{ej,0}^{2/3}\phi_{ej,0}^{2/3}
\end{equation}
corresponding to the electron Lorentz factor $\gamma_{e,c}$ above which radiative energy losses are significant, is unimportant in the radio band and only affects the optical spectrum.

 \begin{figure*}
 \begin{subfigure}[tp]{0.49\linewidth}
    \centering
    \includegraphics[height=0.7\linewidth,width=0.9\linewidth]{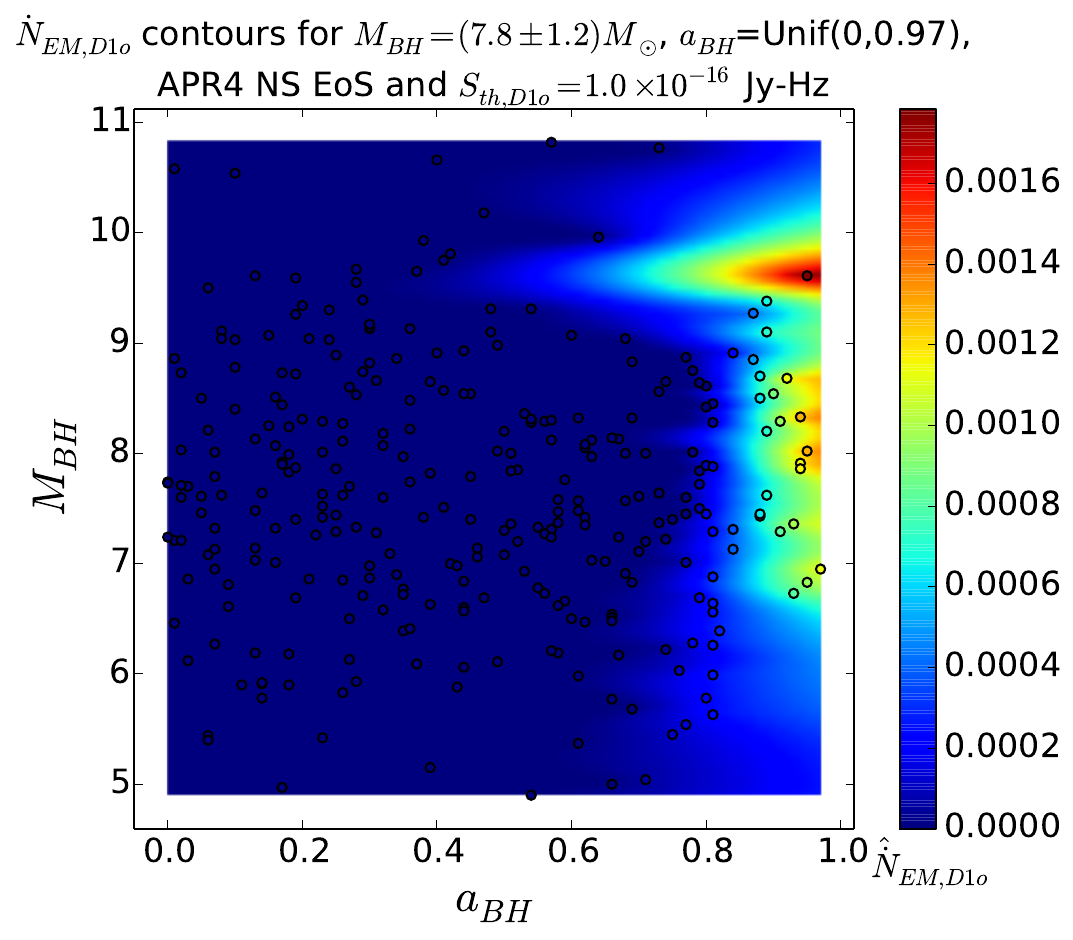}  
  \end{subfigure} 
 \begin{subfigure}[tp]{0.49\linewidth}
    \centering
    \includegraphics[height=0.7\linewidth,width=0.9\linewidth]{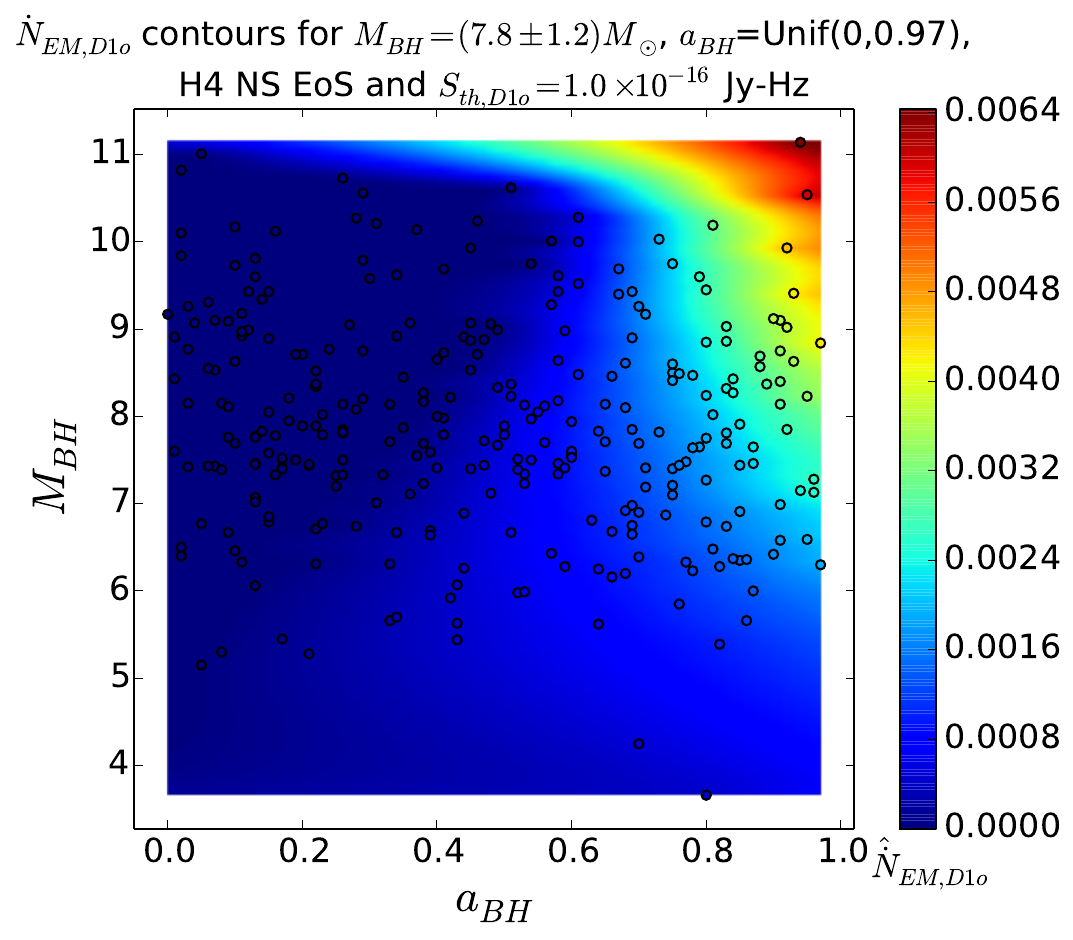}  
  \end{subfigure} \\
 \begin{subfigure}[tp]{0.49\linewidth}
    \centering
    \includegraphics[height=0.7\linewidth,width=0.9\linewidth]{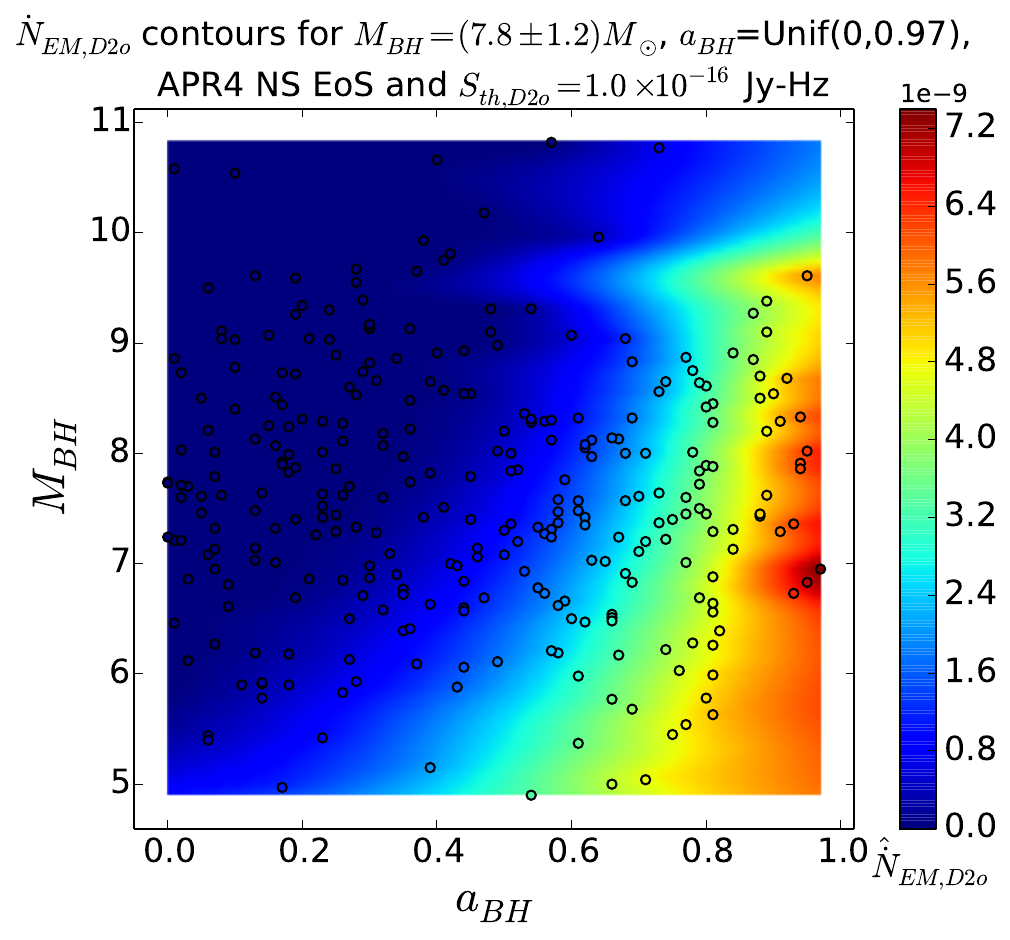}  
  \end{subfigure} 
  \begin{subfigure}[tp]{0.49\linewidth}
    \centering
    \includegraphics[height=0.7\linewidth,width=0.9\linewidth]{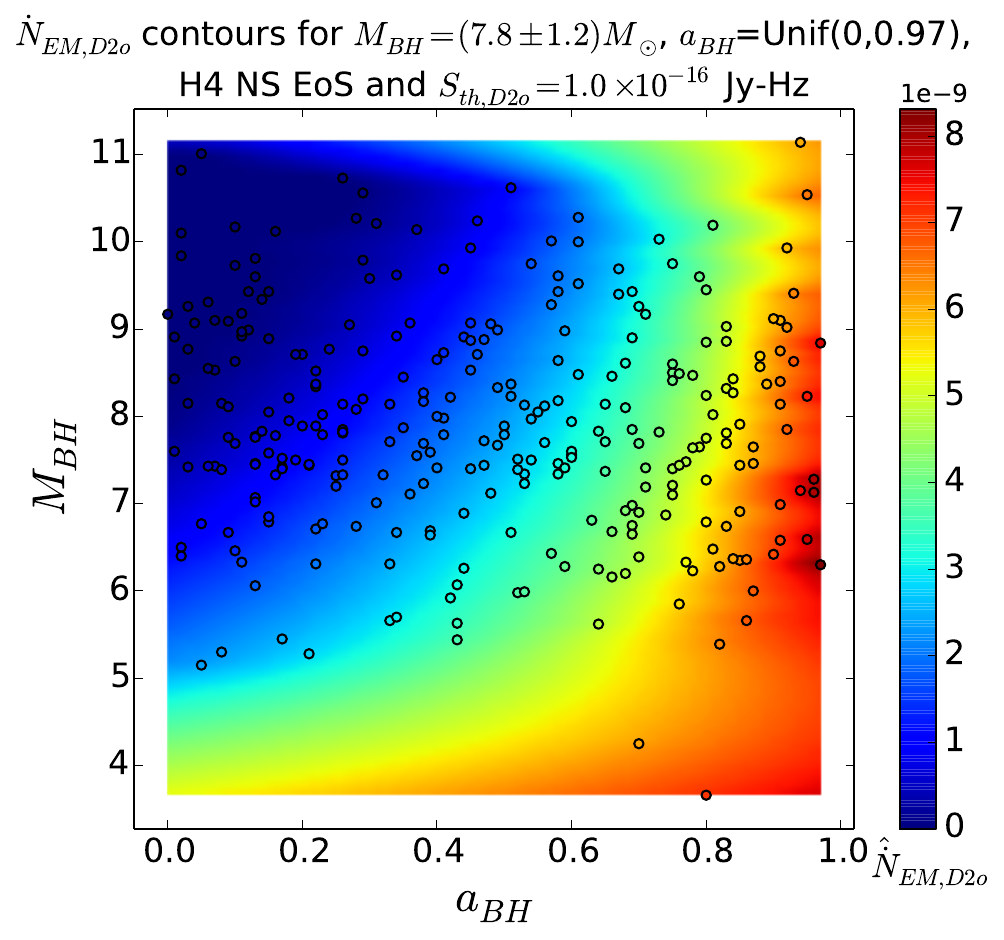}  
  \end{subfigure} \\
  \caption{\emph{Effect of binary parameters on the $\dot{N}_{EM}$ values for late-time optical afterglow emission from anisotropic dynamical ejecta and isotropic wind:} 
Contour plots using simulation results for 300 BHNS binaries with $M_{BH} = (7.8\pm1.2)\ M_{\odot}$, $a_{BH} = {\rm Uniform(0,0.97)}$ and APR4/H4 NS EoS. 
  	{\it Top-left panel:} $\dot{N}_{D1o}$ contour plot for APR4 NS EoS,	
	{\it Top-right panel:} $\dot{N}_{D1o}$ contour plot for H4 NS EoS,
 	{\it Bottom-left panel:} $\dot{N}_{D2o}$ contour plot for APR4 NS EoS,	
	{\it Bottom-right panel:} $\dot{N}_{D2o}$ contour plot for H4 NS EoS
 }
  \label{fig10} 
\end{figure*}

\begin{figure*}
\begin{subfigure}[tp]{0.49\linewidth}
    \centering
    \includegraphics[height=0.7\linewidth,width=0.9\linewidth]{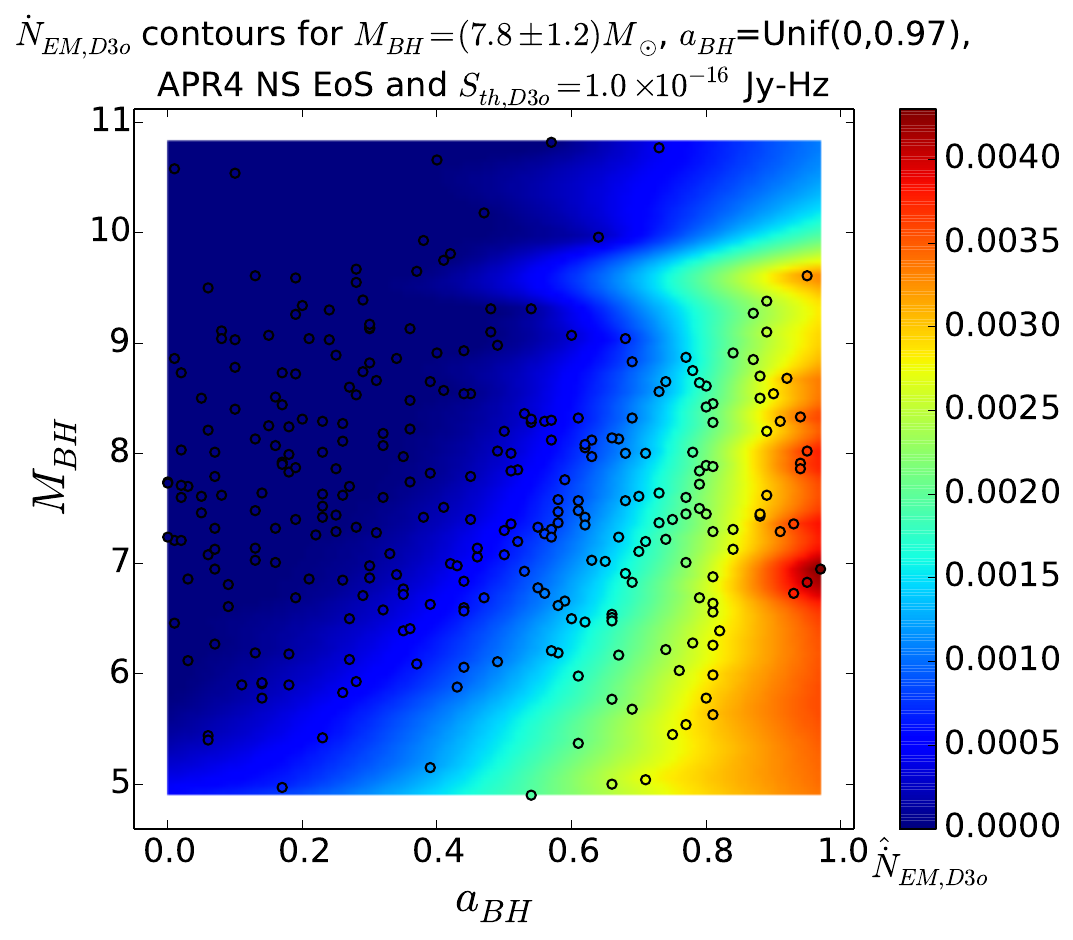}  
  \end{subfigure} 
 \begin{subfigure}[tp]{0.49\linewidth}
    \centering
    \includegraphics[height=0.7\linewidth,width=0.9\linewidth]{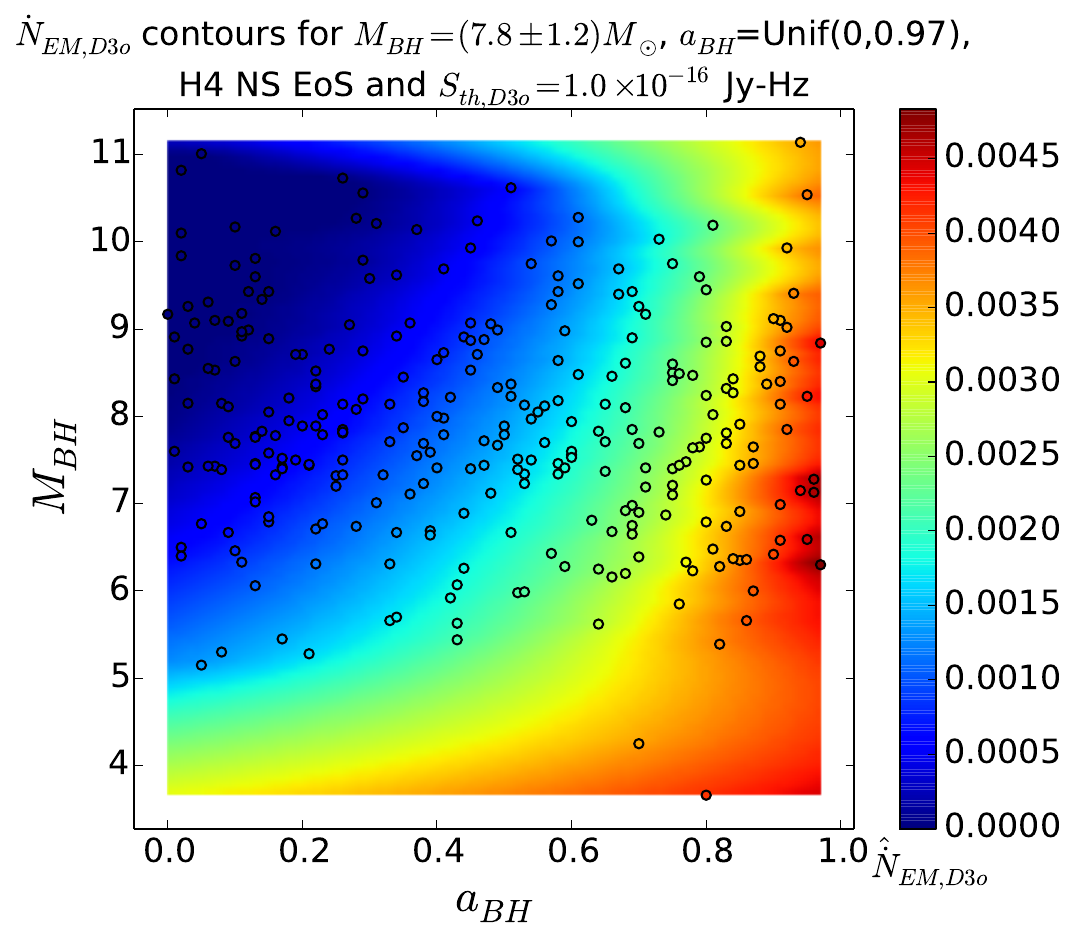}  
  \end{subfigure} \\
  \begin{subfigure}[tp]{0.49\linewidth}
    \centering
    \includegraphics[height=0.7\linewidth,width=0.9\linewidth]{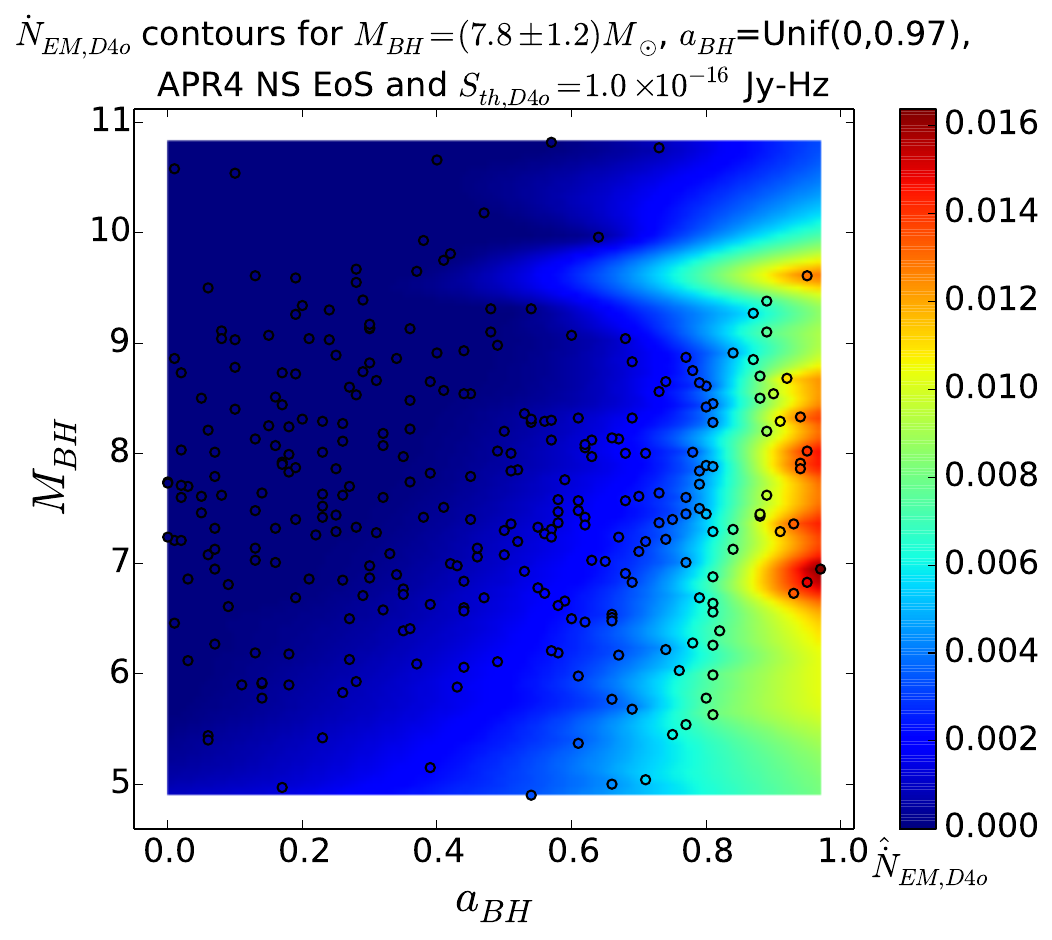}  
  \end{subfigure} 
  \begin{subfigure}[tp]{0.49\linewidth}
    \centering
    \includegraphics[height=0.7\linewidth,width=0.9\linewidth]{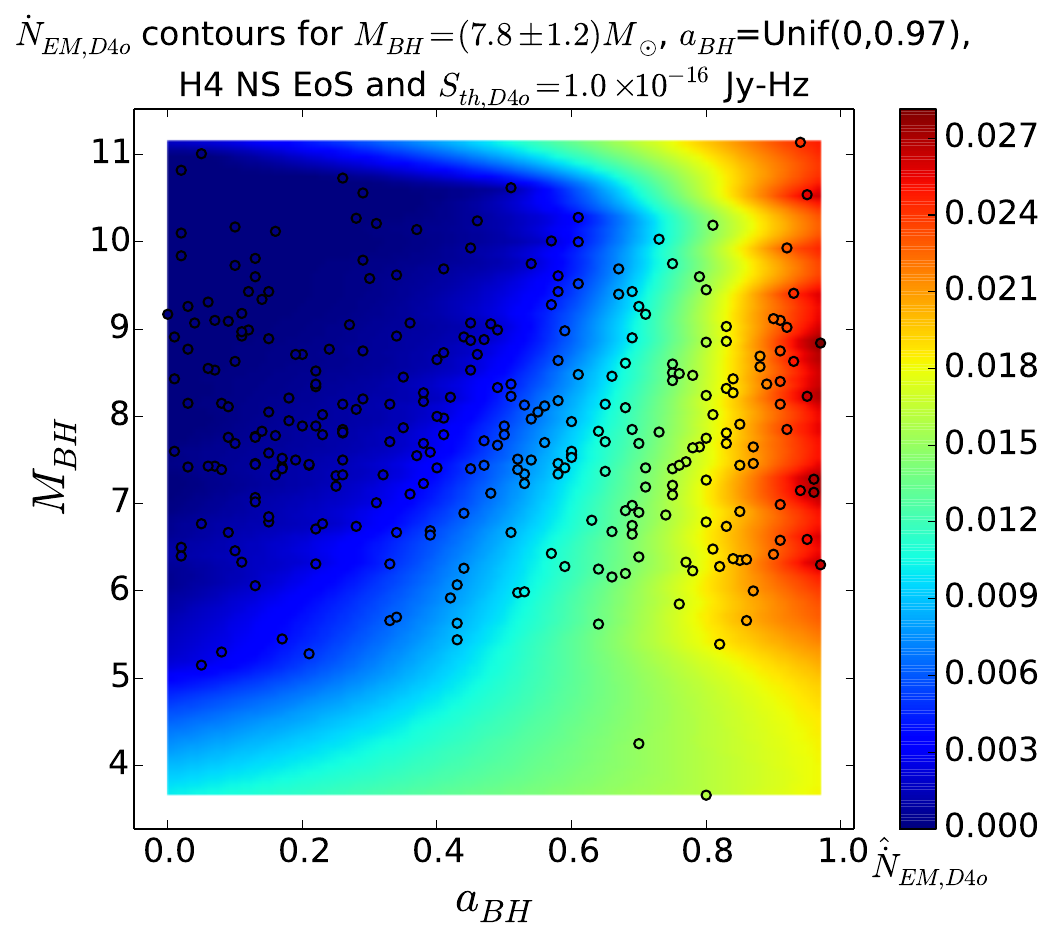}  
    \end{subfigure} \\
 \caption{\emph{Effect of binary parameters on the $\dot{N}_{EM}$ values for late-time optical afterglow emission from anisotropic sGRB jet and anisotropic cocoon:} 
Contour plots using simulation results for 300 BHNS binaries with $M_{BH} = (7.8\pm1.2)\ M_{\odot}$, $a_{BH} = {\rm Uniform(0,0.97)}$ and APR4/H4 NS EoS. 
  	{\it Top-left panel:} $\dot{N}_{D3o}$ contour plot for APR4 NS EoS,	
	{\it Top-right panel:} $\dot{N}_{D3o}$ contour plot for H4 NS EoS,
 	{\it Bottom-left panel:} $\dot{N}_{D4o}$ contour plot for APR4 NS EoS,	
	{\it Bottom-right panel:} $\dot{N}_{D4o}$ contour plot for H4 NS EoS
 }
  \label{fig11} 
\end{figure*}

\section{Optical afterglow emission detection rates}
\label{EM_opt}
Here we estimate the optical afterglow emission follow-up rates for the \emph{ultra-relativistic} jet, the \emph{mildly relativistic} cocoon, the \emph{sub-relativistic} dynamical ejecta and the wind. As in Section \ref{EM_prompt_radio}, we assume $n=0.01\ {\rm cm^{-3}}$, $\epsilon_{e} = \epsilon_{B} = 0.1$ and $p=2.5$ for all ejecta components. We use $\nu_{obso} = \nu_{obso,GHz}\times{\rm 1\ GHz} = 5\times10^{5}\ {\rm GHz}$ as the observing frequency in the optical band. The optical afterglow peak luminosity for the anisotropic \emph{sub-relativistic} dynamical ejecta, isotropic \emph{sub-relativistic} wind, anisotropic \emph{ultra-relativistic} jet and anisotropic \emph{mildly-relativistic} cocoon are
\begin{align*}
L_{peak,D1o} &= (4.09\times10^{37}\ {\rm erg/s}) M_{dyn,-3}^{2/3} \\
&\times v_{dyn,0.2}^{(5p-4)/2} \nu_{obso,GHz}^{-(p-2)/2} \label{Lpeak_B1o_EM} \numberthis \\
L_{peak,D2o} &= (5.75\times10^{34}\ {\rm erg/s}) \eta_{wind,-3}^{2/3}M_{disk,-2}^{2/3} \\
&\times v_{wind,0.07}^{(5p-4)/2} \nu_{obso,GHz}^{-(p-2)/2} \label{Lpeak_B3o_EM} \numberthis \\
L_{peak,D3o} &= (4.01\times10^{38}\ {\rm erg/s}) \eta_{jet,-4}^{2/3}M_{disk,-2}^{2/3} \Gamma_{jet,1}^{-1/3} \\
&\times \nu_{obso,GHz}^{-(p-2)/2} \label{Lpeak_B2o_EM} \numberthis \\
L_{peak,D4o} &= (2.68\times10^{39}\ {\rm erg/s}) [\eta_{dyn,-2}M_{dyn,-3} \\
&+ \eta_{disk,-3}M_{disk,-2}]^{2/3} \Gamma_{jet,1}^{-1/3} \times \nu_{obso,GHz}^{-(p-2)/2} \label{Lpeak_B4o_EM} \numberthis
\end{align*}
with the corresponding follow-up rates for $\nu_{obso}=5\times10^{5}\ {\rm GHz}$
\begin{align*}
\dot{N}_{D1o} &= (1.51\times10^{3}\ {\rm year^{-1}}) M_{dyn,-3} v_{dyn,0.2}^{3(5p-4)/4} \\
&\times (S_{th,D1o}/S_{th,O})^{-3/2} \label{NdotB1o_EM} \numberthis \\
\dot{N}_{D2o} &= (1.86\times10^{-4}\ {\rm year^{-1}}) \eta_{wind,-3} M_{disk,-2} \\
&\times v_{wind,0.07}^{3(5p-4)/4} (S_{th,D2o}/S_{th,O})^{-3/2} \label{NdotB3o_EM} \numberthis \\
\dot{N}_{D3o} &= (1.08\times10^{2}\ {\rm year^{-1}}) \eta_{jet,-4} M_{disk,-2} \Gamma_{jet,1}^{-1/2} \\
&\times (S_{th,D3o}/S_{th,O})^{-3/2} \label{NdotB2o_EM} \numberthis \\
\dot{N}_{D4o} &= (1.88\times10^{3}\ {\rm year^{-1}}) [\eta_{dyn,-2}M_{dyn,-3} \\
&+ \eta_{disk,-3}M_{disk,-2}]\Gamma_{jet,1}^{-1/2}\times (S_{th,D4o}/S_{th,O})^{-3/2} \label{NdotB4o_EM} \numberthis 
\end{align*}
where $S_{th,O} = 3\times10^{-20}\ {\rm Jy\ Hz}$. 

As in Section \ref{EM_rates}, we study the effect of binary parameters on the optical follow-up rates of the afterglow emission from each ejecta component using Monte Carlo simulations for 300 BHNS binaries with gaussian $M_{BH} = (7.8\pm1.2)\ M_{\odot}$, $a_{BH} = {\rm Uniform}(0,0.97)$ and APR4/H4 NS EoS. Figure \ref{fig10} shows the contour plots for the optical follow-up rates of the afterglow emission from the dynamical ejecta and wind for APR4/H4 NS EoS.  
For APR4 EoS, the dynamical ejecta optical afterglow has a follow-up rate $6.0\times10^{-4} \leq \dot{N}_{D1o}/\dot{N}_{GW} \leq 1.8\times10^{-3}$ for $0.75 \leq a_{BH} \leq 1.0$ and $6.5 \leq M_{BH}/M_{\odot} \leq 10.0$. The optical follow-up rate for H4 EoS is larger by a factor of $\sim$4 compared to APR4 EoS due to more ejecta material, with $2.2\times10^{-3} \leq \dot{N}_{D1o}/\dot{N}_{GW} \leq 6.4\times10^{-3}$ for $0.65 \leq a_{BH} \leq 1.0$ and $7.0 \leq M_{BH}/M_{\odot} \leq 11.0$ in the region $6.7a_{BH} + M_{BH} \geq 15.0$. 
The afterglow emission from the wind has a very small follow-up rate $2.6\times10^{-9} \leq \dot{N}_{D2o}/\dot{N}_{GW} \leq 7.2\times10^{-9}$ for $0.55 \leq a_{BH} \leq 0.85$, $5.0 \leq M_{BH}/M_{\odot} \leq 9.5$ and APR4 NS EoS in the region $22.5a_{BH} - M_{BH} \geq 8.5$. For the stiffer H4 EoS, the corresponding rate is also considerably small with $2.8\times10^{-9} \leq \dot{N}_{D2o}/\dot{N}_{GW} \leq 8.2\times10^{-9}$ for $0.5 \leq a_{BH} \leq 0.8$ and $4.5 \leq M_{BH}/M_{\odot} \leq 10.0$ in the region where $15.0a_{BH} - M_{BH} \geq 1.5$.
As in the case of radio follow-up of wind afterglows, the significantly small optical follow-up rates make the detection of wind afterglows especially challenging. 

The contour plots for the follow-up rates of the optical afterglow from the sGRB jet and cocoon for APR4/H4 NS EoS are shown in Figure \ref{fig11}.
For APR4 EoS, the relativistic jet afterglow has an optical follow-up rate $1.5\times10^{-3} \leq \dot{N}_{D3o}/\dot{N}_{GW} \leq 4.2\times10^{-3}$ for $0.6 \leq a_{BH} \leq 0.9$ and $5.0 \leq M_{BH}/M_{\odot} \leq 10.0$ in the region $22.5a_{BH} - M_{BH} \geq 8.5$. The corresponding rate is similar for H4 EoS with $1.8\times10^{-3} \leq \dot{N}_{D3o}/\dot{N}_{GW} \leq 4.7\times10^{-3}$ for $0.5 \leq a_{BH} \leq 0.8$ and $4.5 \leq M_{BH}/M_{\odot} \leq 10.0$ in the region where $15.0a_{BH} - M_{BH} \geq 1.5$. 
The cocoon afterglow has an optical follow-up rate $6.0\times10^{-3} \leq \dot{N}_{D4o}/\dot{N}_{GW} \leq 1.6\times10^{-2}$ with $0.7 \leq a_{BH} \leq 1.0$ and $5.5 \leq M_{BH}/M_{\odot} \leq 10.0$ for APR4 EoS. For H4 EoS, the optical follow-up rate is larger by a factor of $\sim$2 with $1.0\times10^{-2} \leq \dot{N}_{D4o}/\dot{N}_{GW} \leq 2.8\times10^{-2}$ for $0.6 \leq a_{BH} \leq 0.8$ and $4.5 \leq M_{BH}/M_{\odot} \leq 10.0$ in the region where $27.5a_{BH} - M_{BH} \geq 11.5$. 

\end{appendix}

\label{lastpage}

\end{document}